\documentclass[11pt,thmsb]{article}

\usepackage{setspace}
\usepackage{amssymb}
\usepackage{enumerate}
\usepackage{graphicx}
\usepackage{amsmath}
\usepackage{mathrsfs}
\usepackage{amsfonts}
\usepackage{harvard}
\usepackage{comment}
\usepackage{physics}
\usepackage{url}
\usepackage{autobreak}
\usepackage{amsthm}

\usepackage{geometry}
\usepackage[multiple]{footmisc}
\usepackage[all,2cell,ps]{xy}
\usepackage{titlesec}
\usepackage{tikz}
\usepackage{xr}
\usepackage{etoolbox}

\usepackage{titlesec}

\titleformat*{\section}{\normalsize\bfseries}
\titleformat*{\subsection}{\normalsize\bfseries}
\titleformat*{\subsubsection}{\normalsize\bfseries}
\titleformat*{\subparagraph}{\normalsize\itshape}

\theoremstyle{plain}

\titleformat{\paragraph}
{\normalfont\normalsize\bfseries}{\theparagraph}{1em}{}
\titlespacing*{\paragraph}{0pt}{3.25ex plus 1ex minus .2ex}{1.5ex plus .2ex}

\newtheorem{defn}{Definition}
\AfterEndEnvironment{defn}{\noindent\ignorespaces}
\newtheorem{thm}{Theorem}
\AfterEndEnvironment{thm}{\noindent\ignorespaces}
\newtheorem{prop}{Proposition}
\AfterEndEnvironment{prop}{\noindent\ignorespaces}

\newtheorem{obs}{Observation}
\newtheorem{cor}{Corollary}
\AfterEndEnvironment{cor}{\noindent\ignorespaces}
\newtheorem{fact}{Fact}

\newtheorem{lem}{Lemma}
\AfterEndEnvironment{lem}{\noindent\ignorespaces}

\newtheorem*{gen*}{Generalized Welfare Weights Inconsistency Theorem}
\AfterEndEnvironment{gen*}{\noindent\ignorespaces}
\newcommand{\edge}[1]{\ar@{-}[#1]}

\excludecomment{comment}

\def\l{\ensuremath\left}
\def\r{\ensuremath\right}
\allowdisplaybreaks

\begin{document}
\title{Generalized Social Marginal Welfare Weights\\ Imply Inconsistent Comparisons of Tax Policies}
\author{Itai Sher\thanks{email: isher@umass.edu. I am grateful for helpful comments and discussions with Matthew Adler, Eduardo Davila, 
Pawel Doligalski, Piotr Dworczak, Maya Eden, Wojciech Kopczuk, Maria Koumenta, Benjamin Lockwood, Juan Moreno-Cruz, Louis Perrault, Paolo Piacquadio, Peter Sher, and Matthew Weinzierl and to seminar audiences at UC Riverside, the Welfare Economics and Economic Policy virtual seminar, the University of Chicago Harris Public Policy's Political Economy Workshop, the PPE Society annual meeting, the Global Priorities Institute at Oxford University, Northwestern University, and King's College London, as well as to three anonymous referees.}\\University of Massachusetts Amherst}
\date{First version: February 15, 2021\\
This version: June 25, 2024}
\maketitle
\thispagestyle{empty}

\begin{abstract}
\noindent This paper concerns Saez and Stantcheva's (2016) \textit{generalized social marginal welfare weights}, which aggregate losses and gains due to tax policies while incorporating non-utilitarian ethical considerations. The approach evaluates local tax changes without a global social objective.  I show that local tax policy comparisons implicitly entail global comparisons.  Moreover, whenever welfare weights do not have a utilitarian structure, these implied global comparisons are inconsistent.  I argue that broader ethical values cannot in general be represented simply by modifying the weights placed on benefits to different people, and a more thoroughgoing modification of the utilitarian approach is required. (\textit{JEL} D60, D63, D71, H21, H23, I3)
\end{abstract}

\newpage

\setcounter{page}{1}
\section{Introduction}

The traditional optimal tax literature, building on the classic work of \citeasnoun{mirrlees1971exploration}, has adopted a broadly utilitarian normative framework.  As argued by several recent authors, including Weinzierl \citeyear{weinzierl2014promise,weinzierl2017popular} and \citeasnoun{fleurbaey2018optimal}, the omission of other ethical principles that people care about, such as libertarianism, equality of opportunity, and desert, is a serious problem for the classical approach.  \citeasnoun{saez2016generalized} have proposed a general, relatively simple, way of addressing these concerns: They argue that one can modify the  optimality conditions of the standard approach so that these can incorporate broader values while maintaining the structure of the standard optimal taxation theory.  According to Saez and Stantcheva's \textit{generalized social marginal welfare weights} (GSMWW) approach, all one has to do is substitute for the standard utilitarian welfare weights -- corresponding to the marginal utility of consumption -- other welfare weights reflecting broader values.  Such generalized welfare weights can effectively be used as a kind of ``get out of jail free" card that allows one to ignore normative issues on the assumption that they can be incorporated simply by appropriate selection of welfare weights.\footnote{I am grateful to an anonymous referee for suggesting the formulation of the problem in this paragraph as well as some of the wording.}   In this paper, I show formally that this solution to the problem of incorporating broader values into optimal tax does not work because it leads to inconsistencies.  It is not possible, in general, to capture broad ethical principles simply by means of welfare weights. Broadening the normative considerations that bear on taxation will require a more thoroughgoing revision of optimal tax theory.  

I now discuss the specific contributions of this paper. The GSMWW approach only claims to make \textit{local comparisons} among tax policies and accordingly to find local optima.  Indeed, Saez and Stantcheva write,  ``In our approach ... there is no social welfare objective primitive that the government maximizes".  The first contribution of the paper is to show how to collect local comparisons made by generalized welfare weights into implied global social comparisons. In particular, for any system of welfare weights $g$, I define  strict and weak rankings $\prec^g$ and $\sim^g$ over tax policies, which capture some of the global social comparisons implied by welfare weights (see Section \ref{rationalization section}).   Second, I define a critical property of welfare weights, \textit{structural utilitarianism} (see Section \ref{structural utilitarian section}), which is essential to the question of whether welfare weights  are consistent.  Third, I show that if welfare weights $g$ are structurally utilitarian, then welfare weights are consistent in the sense that there exists a social welfare function that generates those welfare weights (see Theorem \ref{necessity theorem} in Section \ref{structural utilitarian section}). More specifically, I show that the case in which welfare weights are structurally utilitarian is the case in which they can be generated by a generalized utilitarian social welfare function of the form $\int F_i\l(U_i\r) \dd i$, where $U_i$ represents agent $i$'s utility and $F_i\l(U_i\r)$ is an agent-specific monotonic transformation of this utility.\footnote{For utilitarianism to be meaningful and for generalized utilitarianism to be meaningfully different than utilitarianism, we must assume that we are given utilities $U_i$ that are cardinal and interpersonally comparable.}  Fourth, I show that when welfare weights are not structurally utilitarian, then they are inconsistent in the sense of the following theorem:
\newpage
\begin{gen*}
If welfare weights $g$ are not structurally utilitarian, then they
are inconsistent in the sense there exist tax policies $T_0,T_1,T_2, T_3$, each of which raises the same revenue, and such that welfare weights imply a social preference cycle of the form: $T_0 \prec^g T_1 \sim^g T_2 \prec^g T_3 \sim^g T_0.$
\end{gen*}
This is Theorem \ref{main theorem} in Section \ref{the main theorem section}.  Theorem \ref{individualized taxes necessity theorem} in Section \ref{preview section} is a simpler version of the result with a more accessible proof.  Putting together the third and fourth contributions, it follows that structural utilitarianism is necessary and sufficient for welfare weights to be consistent.  So the generalized welfare weights approach does not meaningfully add anything beyond what is already available by means of a generalized utilitarian social welfare function, a framework which is long established;\footnote{Indeed,  \citeasnoun{mirrlees1971exploration} posited a generalized utilitarian social welfare function, although with a common transformation $F\l(U_i\r)$ of utility for all agents $i$.} the additional possibilities offered by generalized welfare weights are inconsistent.  

Some ethical values can be captured in a generalized utilitarian framework by making the transformations $F_i$ suitably dependent on agent characteristics.   But Saez and Stancheva suggest that libertarian values can be captured by making welfare weights a function of total taxes paid or that a poverty alleviation imperative can be captured by making weights a function of consumption, and I show that such weights lead to inconsistent judgements (see Sections \ref{detailed example section} and \ref{poverty alleviation subsection}).  Section \ref{non-quasilinear preferences section} explains how my analysis generalizes when the assumption of quasilinear preferences, maintained through most of the paper, is dropped.  Section \ref{discussion section} continues the discussion of the significance of these results and their relation to the literature.

\section{\label{model section}Model}

This section presents the model of \citeasnoun{saez2016generalized}.   
 I assume all functions are \textbf{smooth} -- meaning infinitely differentiable -- unless their domain is discrete or explicitly stated otherwise.
   
\subsection{\label{standard aspects section}Standard aspects of the model}

There is a continuum of agents uniformly distributed on the interval $I=\l[0,1\r]$.  Each agent $i \in I$ has \textbf{observable characteristics} $x_i$ drawn from the set $X$ and \textbf{unobservable characteristics} $y_i$ drawn from the set $Y$.  I assume $X$ and $Y$ are either discrete or subsets of Euclidean spaces.  Let $c_i$ be agent $i$'s consumption and $z_i$ be agent $i$'s income.  Consumption belongs to the real line $\mathbb{R}$ and income to the nonnegative reals $Z = \mathbb{R}_+$.  Agent $i$ has the quasilinear utility function $U_i\l(c_i,z_i\r) = u\l(c_i -v_i\l(z_i\r)\r)$, where $v_i\l(z_i\r) = v\l(z_i,x_i,y_i\r)$ is the cost of earning income $z_i$ given characteristics $\l(x_i,y_i\r)$.  Assume $v_i'\l(z_i\r) > 0, v''_i\l(z_i\r) < 0$, for all $z_i$, so that $v_i$ is increasing and strictly convex in $z_i$, $v_i'\l(z_i\r) >1$ for sufficiently large $z_i$, and that $u$ is increasing and strictly concave.  Assume for simplicity that, for all $i$, $v'_i\l(0\r) < 1$, so that in the absence of taxes all agents earn a positive income.  A \textbf{tax policy} is a function $T\colon Z \times X \rightarrow \mathbb{R}$, where $T\l(z,x\r)$ is the tax paid by agents with income $z$ given observable characteristics $x$.   I write $T_i\l(z_i\r) = T\l(z_i,x_i\r)$, so that $T_i$ gives $i$'s personalized tax on the basis of $i$'s observable characteristics.  I assume  tax policies       have the formal structure requisite to support the exposition that follows.  Section \ref{parameterized families subsection} makes more precise assumptions about the set of tax policies for my formal results.  Given a tax policy $T$, we have $c_i = z_i - T_i\l(z_i\r)$.  Define $z_i\l(T\r)$ to be $i$'s optimal income when facing tax policy $T$, and $c_i\l(T\r)=z_i\l(T\r)-T_i\l(z_i\l(T\r)\r)$; formally, $z_i\l(T\r) \in  \arg\max_{z_i} U_i\l(z_i-T_i\l(z_i\r),z_i\r).$  The agent's indirect utility from tax policy $T$ is then $U_i\l(T\r)=U_i\l(c_i\l(T\r),z_i\l(T\r)\r)$.  Let  $R\l(T\r)=\int T_i\l(z_i\l(T\r)\r) \dd i$ be the revenue generated by $T$.  
\subsection{\label{generalized welfare weights introduction subsection}Generalized welfare weights}
The novelty in the GSMWW approach is the way that tax systems are evaluated.  We assume a system $g\l(c_i,z_i;x_i,y_i\r)$ of \textbf{generalized social marginal welfare weights}.  Thus, we assign a certain weight to each agent depending on their consumption $c_i$, income $z_i$, and characteristics $x_i, y_i$.  Formally, a system of generalized social welfare weights is a function $g: \mathbb{R} \times Z \times X \times Y \rightarrow \mathbb{R}$ such that $g\l(c_i,z_i;x_i,y_i\r) > 0, \forall c_i, z_i, x_i, y_i.$  Define $g_i\l(c_i,z_i\r) = g\l(c_i,z_i;x_i,y_i\r).$  The intuitive interpretation of generalized social marginal welfare weights is that they measure the marginal social value of consumption for each person $i$, and ratios of welfare weights $g_i\l(c_i,z_i\r)/g_j\l(c_j,z_j\r)$ measure social marginal rates of substitution of consumption for agents $i$ and $j$.    Given a tax system $T$, the local marginal welfare weight 
$g_i\l(T\r) = g_i\l(c_i\l(T\r),z_i\l(T\r)\r)$ is endogenously determined.  The key innovation of the approach is to assess small tax reforms via local marginal welfare weights rather than by reference to a global objective.   

I now present some illustrative examples from \citeasnoun{saez2016generalized}.  \textit{Utilitarian weights}: $g_i\l(c_i,z_i\r)= \pdv{c_i}U_i\l(c_i,z_i\r)= u'\l(c_i-v_i\l(z_i\r)\r)$.  These are the standard utilitarian weights that prioritize benefits according to the marginal utility of consumption. \textit{Libertarian weights}: $g_i\l(c_i,z_i\r)= \hat{g}\l(z_i-c_i\r) = \hat{g}\l(t_i\r)$,  where $t_i=z_i-c_i$ is the tax paid and we assume that $\hat{g}'\l(t_i\r) > 0$.  That is, the more tax a person has already paid, the greater the weight placed on that person.  \textit{Libertarian-utilitarian mix}: $g_i\l(c_i,z_i\r) = \hat{g}\l(c_i-v_i\l(z_i\r),z_i-c_i\r)= \hat{g}\l(\hat{u}_i,t_i\r)$ where $\hat{u}_i= c_i-v_i\l(z_i\r)$ with $\pdv{\hat{g}}{\hat{u}_i} <0$ and $\pdv{\hat{g}}{t_i} >0$; the first inequality can be interpreted as saying that weights are increasing in marginal utility for consumption (since $u'\l(c_i-v_i\l(z_i\r)\r)$ is decreasing in $c_i-v_i\l(z_i\r)$) and the second says that they are also increasing in taxes paid.  \textit{Poverty alleviation}: $g\l(c_i,z_i\r) = 1$ if $c_i < \bar{c}$ where $\bar{c}$ is the poverty threshold and $g\l(c_i,z_i\r) =0$ otherwise; that is, we put positive and equal weight on those beneath the poverty line, and no weight on those above the poverty line.\footnote{Such weights are only assumed to be nonnegative but not positive everywhere, contrary to our assumption.}   \textit{Counterfactuals}: Welfare weights can be made to depend on how much someone would have worked in the absence of taxes (which depends on their type) in comparison to how much they work in the presence of taxes.   \textit{Equality of opportunity:} Weights can be made to depend on one's rank in the income distribution conditional on one's background conditions. Such weights go beyond the formal framework in that they depend on the entire income distribution and not just on $c_i, z_i, x_i$, and $y_i$; Saez and Stantcheva present several examples that go beyond the basic formal framework they present.

\subsection{\label{local opt imp section}Local optimality and local improvements}

A \textbf{tax reform} is a function $\Delta T:Z \times X \rightarrow \mathbb{R}$, satisfying appropriate regularity conditions,\footnote{See Section \ref{parameterized families subsection}, which imposes formal regularity conditions on parameterized families of tax policies $\l(T^\theta\r)$; tax policies modified by small reforms $T+\theta \Delta T$ are a special case of parameterized tax policies $T^\theta$.} whose interpretation is that it represents some change to the status quo tax policy. Define $\Delta T_i\l(z_i\r)= \Delta T\l(z_i,x_i\r)$.  Say tax reform $\Delta T$ is \textbf{locally budget neutral} at tax policy $T$ if $\l.\dv{\varepsilon}\r|_{\varepsilon=0}R\l(T+\varepsilon \Delta T\r)=0$.  Say that a locally budget neutral tax reform $\Delta T$ is \textbf{locally desirable} at $T$ if 
\begin{align}\label{local desirability original statement}
 \int g_i\l(T\r)\Delta T_i\l(z_i\l(T\r)\r) \dd i <0.
\end{align}  
In words, $\Delta T$ is locally desirable at $T$ if the cost of the tax change to different individuals due to a small version of the reform $\varepsilon \Delta T$, weighted by the local welfare weights, is negative.  Say that tax system $T$ satisfies the \textbf{local optimal tax criterion} if, for all locally budget neutral tax reforms $\Delta T$, $\int g_i\l(T\r)\Delta T_i\l(z_i\l(T\r)\r) \dd i = 0$.  \citeasnoun{saez2016generalized} say that this criterion gives a necessary condition for local optimality of a tax system $T$, and use it to derive optimal tax formulas for generalized welfare weights that are analogous to the standard optimal tax formulas.  

In the traditional utilitarian framework, the goal is to choose a tax policy $T$ to maximize the utilitarian objective $\int U_i\l(T\r) \dd i$ subject to a revenue requirement.  Given this formulation, employing utilitarian weights $g_i\l(c_i,z_i\r)=\pdv{c_i}U_i\l(c_i,z_i\r)$, and using the envelope theorem, the local optimal tax criterion is a necessary condition for $T$ to be an optimum and (\ref{local desirability original statement}) is a sufficient condition for a small version of the reform $\Delta T$ to be a local improvement.  However, in the GSMWW framework, there is no global objective from which to derive these conditions; so the conditions for a locally desirable reform and for a local optimum are posited by analogy to the utilitarian case. 

\section{\label{rationalization section}Global social comparisons implied by welfare weights}

Generalized social marginal welfare weights provide \textit{local} comparisons: conditions for a local improvement and for local optimality of tax policies.  This section shows how to derive \textit{global} social comparisons implicit in welfare weights.  

\subsection{\label{parameterized families subsection}Modifying tax policies}

To derive global comparisons, I need to smoothly vary tax policies in a parametric way.  To do so, I append a (real-valued) parameter $\theta$ to tax policies, writing $T^\theta$.  Varying $\theta$ corresponds to changing tax policy in some way.  For example, if $T^\theta = T+\theta \Delta T$, then $\theta$ measures the size of the tax reform $\Delta T$ to tax policy $T$.  Alternatively, consider a (non-individualized) linear tax $T^\theta\l(z\r)=\theta z + \kappa\l(\theta\r)$, where,  when we vary $\theta$, we vary both the marginal tax rate and the lumpsum tax $\kappa\l(\theta\r)$.  In general, let $\Theta=\l[\underline{\theta},\overline{\theta}\r]$ be an interval in the real line, where $\underline{\theta} < \overline{\theta}$.  Consider a parameterized collection $\l(T^{\theta}\r)_{\theta \in \Theta}$ of tax policies.  Below I sometimes use the abbreviated notation $\l(T^\theta\r)$ rather than $\l(T^{\theta}\r)_{\theta \in \Theta}$. Given $\l(T^\theta\r)$, define $T_i\l(z,\theta\r)=T_i^{\theta}\l(z\r)$, so that $T_i\l(z,\theta\r)$ can be regarded as a real-valued function with domain $Z\times \Theta$.  A family of tax policies $\l(T^{\theta}\r)_{\theta \in \Theta}$ is \textbf{well-behaved} if (1) for each $i$ and $\theta$, $i$'s optimal income in response to $T^\theta$ exists, is unique, and positive, and the second order condition for $i$'s optimization problem, when facing $T^\theta$, holds with strict inequality at the optimum, and (2) for all $i$, the map $\l(z,\theta\r) \mapsto T_i\l(z,\theta\r)$ is smooth, and, except for at most at finitely many values of $i$, the map $\l(i,z,\theta\r) \mapsto T_i\l(z,\theta\r)$ is smooth.  Note that the second condition allows that, when taxes are individualized, there may be finitely many $i$ such that tax policy is discontinuous at $i$.  Say a tax policy $T$ is \textbf{regular} if there exists a well-behaved family $\l(T^\theta\r)_{\theta \in \Theta}$ and $\theta' \in \Theta$ such that $T^{\theta'}=T$.  Regular tax policies are characterized by conditions similar to (1) and (2) above (see Appendix \ref{wb appendix}).  Given a family $\l(T^\theta\r)$, write $z_i\l(\theta\r)=z_i\l(T^\theta\r), c_i\l(\theta\r)=c_i\l(T^\theta\r), U_i\l(\theta\r)=U_i\l(T^\theta\r), g_i\l(\theta\r)=g_i\l(T^\theta\r)$ for, respectively, $i$'s optimal income, optimal consumption, indirect utility, and welfare weight at $T^\theta$.
  
Sometimes I introduce a second parameter $\epsilon$ in  $E=\l[\underline{\epsilon},\overline{\epsilon}\r]$, where $\underline{\epsilon} < \overline{\epsilon}$, and consider a doubly parameterized family $\l(T^{\theta,\epsilon}\r)_{\theta \in \Theta,\epsilon \in E}$ (abbreviated as $\l(T^{\theta,\epsilon}\r)$).  As above, I write $T_i\l(z,\theta,\epsilon\r)= T_i^{\theta,\epsilon}\l(z\r)$. $\l(T^{\theta,\epsilon}\r)$ is \textbf{well-behaved} if it satisfies conditions analogous to (1) and (2) above, with $\l(\theta,\epsilon\r)$ playing the role of $\theta$, so that, for example, the first part of (2) becomes: for all $i$, the map $\l(z,\theta,\epsilon\r) \mapsto T_i\l(z,\theta,\epsilon\r)$ is smooth.  For a complete definition, see Appendix \ref{wb appendix}. Given family $\l(T^{\theta,\epsilon}\r)$, write $z_i\l(\theta,\epsilon\r), c_i\l(\theta,\epsilon\r), U_i\l(\theta,\epsilon\r),$ and $ g_i\l(\theta,\epsilon\r)$ for $i$'s optimal income, optimal consumption, indirect utility, and welfare weight at $T^{\theta,\epsilon}$.  

\subsection{\label{improvement and indifference principles subsection}The global improvement and indifference principles}

Consider a system of generalized social welfare weights $g$.  I now define a relation $\prec^g$, which captures some of the strict social preferences implied by $g$, and a relation $\sim^g$, which captures some of the social indifferences implied by $g$.\footnote{\label{not all implications footnote} I do not claim that $\prec^g$ and $\sim^g$ capture \textit{all} social preferences implicit in welfare weights $g$.} For any pair of tax policies $T_0$ and $T_1$, whenever $T_0 \prec^g T_1$, this indicates that welfare weights $g$ imply that $T_1$ is strictly socially preferred to $T_0$, and whenever $T_0 \sim^g T_1$, this indicates that welfare weights $g$ imply that $T_1$ is socially indifferent to $T_0$.

Let $\l(T^\theta\r)_{\theta \in \Theta}$ be a well-behaved parameterized collection of tax policies, and let $\theta_0, \theta_1 \in \Theta$ be such that $\theta_0 < \theta_1$.  Consider the following principles:    

\begin{itemize}
\item \textbf{Global improvement principle.} Suppose that, for all $\hat{\theta} \in \l[\theta_0,\theta_1\r]$, 
\begin{align}\label{improvement implication}
 \int  g_i\l(\hat{\theta}\r)\l.\pdv{\theta}\r|_{\theta =\hat{\theta}}T_i\l(z_i\l(\hat{\theta}\r),\theta\r)\dd i < 0
 \end{align}
-- that is, increasing $\theta$ is locally socially desirable at $\hat{\theta}$.  Then $T^{\theta_0} \prec^g T^{\theta_1}$: $T^{\theta_1}$ is socially preferred to $T^{\theta_0}$.
\item \textbf{Global indifference principle.} Suppose that, for all $\hat{\theta} \in \l[\theta_0,\theta_1\r]$, \begin{align}\label{indifference implication}
  \int  g_i\l(\hat{\theta}\r)\l.\pdv{\theta}\r|_{\theta =\hat{\theta}}T_i\l(z_i\l(\hat{\theta}\r),\theta\r)\dd i = 0
 \end{align}
 -- that is, at $\hat{\theta}$, welfare weights don't detect any change in social welfare as $\theta$ changes.  Then $T^{\theta_0} \sim^g T^{\theta_1}$: $T^{\theta_0}$ and $T^{\theta_1}$ are socially indifferent. 
\end{itemize}
We can think of these two principles as axioms that allow us to draw inferences about social preferences from welfare weights. Henceforth, I shall assume that $\prec^g$ and $\sim^g$ satisfy these principles.   

To understand these principles, consider first the standard utilitarian case. The utilitarian social welfare of tax policy $T^\theta$ is $W_{\textup{util}}\l(\theta\r)=\int U_i\l(\theta\r) \dd i$.  Because 
$U_i\l(\theta\r)= U_i(\underbrace{z_i\l(\theta\r)-T_i\l(z_i\l(\theta\r),\theta\r)}_{c_i\l(\theta\r)},z_i\l(\theta\r))$, it follows from the envelope theorem that, for any $\hat{\theta} \in \Theta$, 
\begin{align}\label{key envelope}
\dv{\theta} U_i\l(\hat{\theta}\r)= -\underbrace{\pdv{c_i} U_i\l(c_i\l(\hat{\theta}\r),z_i\l(\hat{\theta}\r)\r)}_{\textup{utilitarian welfare weight}}\underbrace{\l.\pdv{\theta}\r|_{\theta =\hat{\theta}}T_i\l(z_i\l(\hat{\theta}\r),\theta\r)}_{\textup{direct effect on taxes}}.
\end{align} 
That is, the envelope theorem tells us that, the marginal effect of a change in tax policy on an agent's utility is the product of the agent's marginal utility of consumption and the marginal direct effect of the change in $\theta$ on the agent's tax bill, and we can ignore the indirect effects due to changes in behavior -- the choices of consumption and income -- as taxes change.  So in the utilitarian case,
\begin{align*}
\dv{\theta}W_{\textup{util}}\l(\hat{\theta}\r) = \int \dv{\theta}U_i\l(\hat{\theta}\r) \dd i &= -\int  \pdv{c_i} U_i\l(c_i\l(\hat{\theta}\r),z_i\l(\hat{\theta}\r)\r)\l.\pdv{\theta}\r|_{\theta =\hat{\theta}}T_i\l(z_i\l(\hat{\theta}\r),\theta\r)\dd i\\
&= -\int  g_i\l(\hat{\theta}\r)\l.\pdv{\theta}\r|_{\theta =\hat{\theta}}T_i\l(z_i\l(\hat{\theta}\r),\theta\r)\dd i.  
\end{align*}
Given this equation, (\ref{improvement implication}) becomes $\dv{\theta}W_{\textup{util}}\l(\hat{\theta}\r)>0$ and (\ref{indifference implication}) becomes $\dv{\theta}W_{\textup{util}}\l(\hat{\theta}\r)=0$, so that the global improvement principle says that if utilitarian welfare is increasing as we vary $\theta$ from $\theta_0$ to $\theta_1$, then utilitarian welfare is greater at $\theta_1$ than at $\theta_0$, and the global indifference principle says that if utilitarian welfare is unchanging as we vary $\theta$, then utilitarian welfare is the same at $\theta_1$ as at $\theta_0$.  In the utilitarian case, these principles are obviously valid.   

In the case of generalized welfare weights, the global improvement and indifference principles are posited by analogy with the utilitarian case.  This is the same as the justification for Saez and Stantcheva's definitions for a local desirability of a tax reform and local optimality of a tax policy, which substitute generalized welfare weights $g_i\l(\hat{\theta}\r)$ for utilitarian welfare weights  $\pdv{c_i} U_i\l(c_i\l(\hat{\theta}\r),z_i\l(\hat{\theta}\r)\r)$ in principles that are valid for utilitarianism.  Indeed, when the parameterized family of tax policies has the form $T^\theta = T + \theta \Delta T$ and $\hat{\theta} =0$, (\ref{improvement implication}) simplifies to (\ref{local desirability original statement}) in Section \ref{local opt imp section}, Saez and Stantcheva's condition for a locally desirable tax reform.\footnote{Saez and Stantcheva apply this condition to locally revenue neutral tax reforms; in my main theorem, I use the global improvement and indifference principles to construct a cycle when revenue remains constant.} 

The following useful result assumes the global improvement principle and follows from our smoothness assumptions  -- see the Appendix for the proof.  
\begin{prop}\label{local improvement principle proposition}
\textbf{Local improvement principle.} \\ Let $g$ be a system of welfare weights, let $\l(T^\theta\r)_{\theta \in \l[\underline{\theta},\overline{\theta}\r]}$ be a well-behaved parameterized family of tax policies, and let $\theta_0 \in \l[\underline{\theta},\overline{\theta}\r)$. If 
$\int g_i\l(\theta_0\r) \l.\pdv{\theta}\r|_{\theta=\theta_0}T_i\l(z_i\l(\theta_0\r),\theta\r) \dd i < 0$, then there exists $\theta_1 \in \l(\theta_0,\overline{\theta}\r]$ such that, for all $\theta \in \l(\theta_0,\theta_1\r)$, $T^{\theta_0} \prec^g T^\theta$.  Similarly, if $
\int g_i\l(\theta_0\r) \l.\pdv{\theta}\r|_{\theta=\theta_0}T_i\l(z_i\l(\theta_0\r),\theta\r) \dd i > 0$, then there exists $\theta_1 \in \l(\theta_0,\overline{\theta}\r]$ such that, for all $\theta \in \l(\theta_0,\theta_1\r)$, $T^{\theta_0} \succ^g T^\theta$.
\end{prop}

\subsection{Pareto} 

Certain Pareto conditions, which are useful below, are implicit in the welfare weights framework.  In particular, it follows from (\ref{key envelope}), which was derived using the envelope theorem, and the fact that the marginal utility of consumption is positive that the following relation holds:
\begin{align}\label{equivalent inequalities U T 0}
\forall i,\forall \hat{\theta},\;\;\; \dv{\theta}U_i\l(\hat{\theta}\r)\gtreqqless 0 \Leftrightarrow \l.\pdv{\theta}\r|_{\theta=\hat{\theta}}T_i\l(z_i\l(\hat{\theta}\r),\theta\r) \lesseqqgtr 0. 
\end{align}
That is, $\dv{\theta}U_i\l(\hat{\theta}\r)$ and $\l.\pdv{\theta}\r|_{\theta=\hat{\theta}}T_i\l(z_i\l(\hat{\theta}\r),\theta\r)$ always have the opposite sign when nonzero, and otherwise both are zero.  This shows that the term $-\l.\pdv{\theta}\r|_{\theta=\hat{\theta}}T_i\l(z_i\l(\hat{\theta}\r),\theta\r)$ captures preferences in the sense that it points in the same direction as preferences do in response to a change in $\theta$; and it also shows why the global improvement and indifference principles respect preferences.  If, at $\hat{\theta}$, an increase in $\theta$ makes all agents better off, the terms $\l.\pdv{\theta}\r|_{\theta=\hat{\theta}}T_i\l(z_i\l(\hat{\theta}\r),\theta\r)$ will be negative for all agents, and so $\int g_i\l(\hat{\theta}\r) \l.\pdv{\theta}\r|_{\theta=\hat{\theta}}T_i\l(z_i\l(\hat{\theta}\r),\theta\r) \dd i <0$, no matter what (positive) welfare weights $g$ are used.  This is formalized by the following proposition, which is proved in the Appendix and assumes, as above, that welfare weights are always positive, and also assumes the global improvement and indifference principles.    
  
\begin{prop}\label{indifference Pareto coro}
Let $\l(T^\theta\r)_{\theta \in \Theta}$ be a well-behaved family of tax policies, and let $\theta_0, \theta_1 \in \Theta$ be such that $\theta_0 < \theta_1$.
\begin{enumerate}
\item\label{indifference along paths} \textbf{Pareto indifference along paths.}  Suppose that all agents are indifferent among all tax policies $T^\theta$ for $\theta \in \l[\theta_0,\theta_1\r]$.  Then, for all systems of welfare weights $g$,  $T^{\theta_0} \sim^g T^{\theta_1}$. 
\item \textbf{Weak Pareto along paths.} Suppose that for all $\hat{\theta} \in \l[\theta_0,\theta_1\r]$ and all agents $i$, $\dv{\theta}U_i\l(\hat{\theta}\r) > 0$ so that, for all agents, tax policies become more preferred as $\theta$ increases within $\l[\theta_0,\theta_1\r]$. Then, for all systems of welfare weighs $g$, $T^{\theta_0} \prec^g T^{\theta_1}$.
\end{enumerate}
\end{prop}
The Pareto principles stated above are weaker than the standard principles because they only apply to paths of smoothly changing tax policies along which the direction of preferences is constant.  Say that a social welfare function is \textbf{Paretian along paths} if it satisfies a weakened version of the Pareto principle, analogous to the properties that the above proposition shows to be satisfied by all systems of welfare weights.  A formal statement of this property of social welfare functions, as well as of what it means for a system of welfare weights to implement a social welfare function and a proof of the following corollary is in the Appendix.   
\begin{cor}\label{non Pareto coro}
Any social welfare function that is not Paretian along paths cannot be implemented by any system of generalized social welfare weights.
\end{cor}
 The corollary shows that the expressive power of welfare weights is limited in the sense that non-Paretian (in a weak sense of Paretian) objectives cannot be implemented by welfare weights.

\section{\label{structural utilitarian section}Structural utilitarianism}

The key condition for generalized welfare weights to be consistent is \textit{structural utilitarianism}.
\begin{defn}\label{depends only on definition}
A system of welfare weights $g$ is \textbf{structurally utilitarian} if and only if $\forall i \in I,\forall z_i,z'_i \in Z, \forall c_i, c'_i \in \mathbb{R}$,
\begin{align}\label{depends only on utility condition}
c_i-v_i\l(z_i\r) = c'_i - v_i\l(z'_i\r) \Rightarrow g_i\l(c_i,z_i\r)=g_i\l(c'_i,z'_i\r). 
\end{align}
\end{defn}
To interpret this definition, observe that, given quasilinear utility $U_i\l(c_i,z_i\r) = u\l(c_i-v_i\l(z_i\r)\r)$, we have $\pdv{c_i} U_i\l(c_i,z_i\r) = u'\l(c_i-v_i\l(z_i\r)\r)$.  Thus the marginal utility of consumption $\pdv{c_i} U_i\l(c_i,z_i\r)$ is determined by the quantity $c_i-v_i\l(z_i\r)$, and given our assumption that the outer utility function $u\l(\cdot\r)$ is strictly concave, the condition (\ref{depends only on utility condition}) for structural utilitarianism is equivalent to:
\begin{align}\label{first version structural utilitarianism}
\pdv{c_i}U_i\l(c_i,z_i\r) = \pdv{c_i} U_i\l(c'_i,z'_i\r) \Rightarrow g_i\l(c_i,z_i\r)=g_i\l(c'_i,z'_i\r).  
\end{align} 
Thus, structural utilitarianism allows that welfare weights are not necessarily \textit{equal to} the marginal utility of consumption $\pdv{c_i}U_i\l(c_i,z_i\r)$, the utilitarian welfare weight, but it requires that welfare weights are \textit{determined by} the marginal utility of consumption in the sense that, if $i$'s marginal utility of consumption does not change, then $i$'s welfare weight does not change.  Note that the condition is imposed separately on each agent $i$; it is a condition on how that agent's welfare weight changes as their allocation $\l(c_i,z_i\r)$ changes, and no relation is posited between the welfare weights of different agents $i$ and $j$.  So structural utilitarianism is consistent with welfare weights being dependent on agents' characteristics $\l(x_i,y_i\r)$.  Recalling that utility is given by $U_i\l(c_i,z_i\r) = u\l(c_i-v_i\l(z_i\r)\r)$, utility is also determined by the quantity $c_i-v_i\l(z_i\r)$.  When the outer utility function $u\l(\cdot\r)$ is both strictly increasing and strictly concave,  $\pdv{c_i}U_i\l(c_i,z_i\r) = \pdv{c_i} U_i\l(c'_i,z'_i\r)$ if and only if $U_i\l(c_i,z_i\r) = U_i\l(c'_i,z'_i\r)$, and the condition (\ref{depends only on utility condition}) for structural utilitarianism is also equivalent to:
\begin{align}\label{second version structural utilitarianism}
U_i\l(c_i,z_i\r) =  U_i\l(c'_i,z'_i\r) \Rightarrow g_i\l(c_i,z_i\r)=g_i\l(c'_i,z'_i\r).  
\end{align}
Thus, structural utilitarianism can also be interpreted as saying that $i$'s welfare weight doesn't change when $i$'s utility doesn't change.  The coincidence of (\ref{first version structural utilitarianism}) and (\ref{second version structural utilitarianism}) depends on the assumption of quasilinear utility, and, indeed, in Section \ref{non-quasilinear preferences section}, I show  how to generalize structural utilitarianism when utility is no longer assumed quasilinear.  

Define $\hat{U}_i\l(c_i,z_i\r)= c_i-v_i\l(z_i\r)$.  $\hat{U}_i\l(c_i,z_i\r)$ is a utility function over $\l(c_i,z_i\r)$ pairs that is ordinally equivalent to $U_i\l(c_i,z_i\r)$.    Define the variable $\hat{u}_i$ by $\hat{u}_i = \hat{U}_i\l(c_i,z_i\r)$.  We can then re-express welfare weights as a function $\hat{g}_i\l(\hat{u}_i,z_i\r)$ of utility and income $\l(\hat{u}_i,z_i\r)$ rather than as a function $g_i\l(c_i,z_i\r)$ of consumption and income $\l(c_i,z_i\r)$.  The relationship between the two expressions is as follows:  
\begin{align}\label{g g hat relation}
\hat{g}_i\l(\hat{u}_i,z_i\r) = g_i\l(\hat{u}_i+v_i\l(z_i\r),z_i\r), \;\;\; \forall \hat{u}_i \in \mathbb{R}, \forall z_i \in Z.
\end{align}     
The following result is useful. (The straightforward proof is in the Appendix.)
\begin{prop}\label{g observation}
Let $g$ and $\hat{g}$ be related as in (\ref{g g hat relation}).  Then welfare weights $g$ are structurally utilitarian if and only if $\forall i \in I, \forall \hat{u}_i \in \mathbb{R},\forall z_i \in Z, \pdv{z_i}\hat{g}_i\l(\hat{u}_i,z_i\r) =0.$
\end{prop}

We now come to a theorem that shows that when welfare weights are structurally utilitarian, they correspond to a global social ranking.  Say that a real valued function $W\l(T\r)$, whose domain is the set of regular tax policies, is a \textbf{generalized utilitarian social welfare function} if there exists a real-valued function $F_i\l(u_i\r)=F\l(u_i,x_i,y_i\r)$, which is (i) smooth in $u_i$ and smooth in $\l(u_i,x_i,y_i\r)$ unless $\l(x_i,y_i\r)$ are discrete and (ii) strictly increasing in $u_i$, such that $W\l(T\r) = \int F_i\l(U_i\l(c_i\l(T\r),z_i\l(T\r)\r)\r) \dd i$.\footnote{Note that I build smoothness into the definition of a generalized utilitarian social welfare function because I assumed similar smoothness properties on welfare weights.  If the smoothness requirements on welfare weights were relaxed somewhat, one could correspondingly weaken the smoothness requirements for a generalized utilitarian social welfare function and still prove a corresponding version of Theorem \ref{necessity theorem} below.}  It follows from the envelope theorem that, 
for all well-behaved families $\l(T^\theta\r)_{\theta \in \Theta}$ and $\theta_0 \in \Theta$, \begin{align*}\l.\dv{\theta}\r|_{\theta=\theta_0} W\l(T^\theta\r) = -\int F'_i\l(U_i\l(c_i\l(\theta_0\r),z_i\l(\theta_0\r)\r)\r) \pdv{c_i}U_i\l(c_i\l(\theta_0\r),z_i\l(\theta_0\r)\r) \l.\pdv{\theta}\r|_{\theta =\theta_0} T_i\l(z_i\l(\theta_0\r),\theta\r) \dd i.\end{align*} 
Hence, $F'_i\l(U_i\l(c_i,z_i\r)\r) \pdv{c_i}U_i\l(c_i,z_i\r)$ are the social welfare weights arising from a generalized utilitarian social welfare function.  Formally, say that a system of welfare weights $g$ \textbf{arise from a generalized utilitarian social welfare function} if there exists $F_i\l(u_i\r)=F\l(u_i,x_i,y_i\r)$ satisfying properties (i) and (ii) above such that for all $i, c_i,$ and $z_i$, $g_i\l(c_i,z_i\r)= F'_i\l(U_i\l(c_i,z_i\r)\r) \pdv{c_i}U_i\l(c_i,z_i\r)$.  
 \begin{thm}\label{necessity theorem}
Welfare weights $g$ are structurally utilitarian if and only if they arise from a generalized utilitarian social welfare function.\footnote{One might wonder why the social welfare function in the above theorem is additively separable; the answer is that $i$'s welfare weight is assumed to depend only on $i$'s consumption, income, and characteristics, and not on the distribution of these in society.} \end{thm} 
The theorem has the following important corollary:
\begin{cor}\label{noncycle corollary}
If welfare weights $g$ are structurally utilitarian, then there exists a generalized utilitarian social welfare function $W$ from which the welfare weights can be derived in the sense that for all well-behaved families  $\l(T^\theta\r)_{\theta \in \Theta}$  and $\theta_0 \in \Theta$, $\l.\dv{\theta}\r|_{\theta=\theta_0} W\l(T^\theta\r) = -\int g_i\l(T^{\theta_0}\r) \l.\pdv{\theta}\r|_{\theta =\theta_0} T_i\l(z_i\l(T^{\theta_0}\r),\theta\r) \dd i$, so that the welfare weights correspond to a consistent social ranking.   
\end{cor}
Both the theorem and the corollary are proved in the Appendix.  It should be clear that if welfare weights arise from a social welfare function, then it is not possible to use them to construct a social preference cycle.  A proof sketch of Theorem \ref{necessity theorem} is as follows.  First, if welfare weights arise from a generalized utilitarian social welfare function, then they are of the form  $g_i\l(c_i,z_i\r)=F'_i\l(U_i\l(c_i,z_i\r)\r) \pdv{c_i}U_i\l(c_i,z_i\r)$. These weights are structurally utilitarian because, given quasilinearity, both $U_i\l(c_i,z_i\r)$ and $\pdv{c_i}U_i\l(c_i,z_i\r)$ are determined by   $c_i-v_i\l(z_i\r)$.   Going in the other direction,  by Proposition \ref{g observation}, structural utilitarianism is equivalent to the requirement that welfare weights are a function of $\hat{u}_i = c_i-v_i\l(z_i\r)$, so that, assuming structural utilitarianism, we can write $g_i\l(c_i,z_i\r)= \hat{g}_i\l(c_i-v_i\l(z_i\r)\r)=\hat{g}_i\l(\hat{u}_i\r)$. Define the function $w_i\l(\hat{u}_i\r)$ by $w_i\l(\hat{u}^0_i\r) = \int_0^{\hat{u}^0_i} \hat{g}_i\l(\hat{u}_i\r) \dd \hat{u}_i.$   Then define the utility function $W_i\l(c_i,z_i\r)= w_i\l(c_i-v_i\l(z_i\r)\r)$.  Observe that the utility function $W_i\l(c_i,z_i\r)$ is ordinally equivalent to $U_i\l(c_i,z_i\r)$ in the sense that the two represent the same preferences over consumption and income. It follows that there exists a strictly increasing function $F_i$ such that $W_i\l(c_i,z_i\r)=F_i\l(U_i\l(c_i,z_i\r)\r)$.  Since $g_i\l(c_i,z_i\r)= g\l(c_i,z_i,x_i,y_i\r)$, there exists some function $F$ such that  $F_i\l(u_i\r)=F\l(u_i,x_i,y_i\r)$.   By construction, $g_i\l(c_i,z_i\r)= \hat{g}_i\l(c_i-v_i\l(z_i\r)\r) = w'_i\l(c_i-v_i\l(z_i\r)\r)=
\pdv{c_i} W_i\l(c_i,z_i\r)= F'_i\l(U_i\l(c_i,z_i\r)\r) \pdv{c_i}U_i\l(c_i,z_i\r)$, which is what we need to show.  

\section{\label{preview section}A simple version of the main theorem}

\subsection{\label{individualized version section}The special case when taxes can be completely individualized}
I now prove a simplified version of my main result.  A stronger version is in Section \ref{the main theorem section}. Consider the special case in which taxes can be completely individualized so that each agent $i$ faces an individualized tax schedule $T^{\theta}_i$ that can differ from the tax schedule faced by other agents.  In our framework, this is possible if each agent's observable characteristics uniquely identify them: formally, for all $i, j\in I$, $i \neq j \Rightarrow x_i \neq x_j$.  I assume that the map $i \mapsto x_i$ is smooth, that there are no unobservable characteristics $y_i$, and that the functions $u\l(\cdot\r)$, $\l(z_i,x_i\r) \mapsto v\l(z_i,x_i\r)$, $\l(c_i,z_i,x_i\r) \mapsto g\l(c_i,z_i,x_i\r)$ are smooth.   This case is not interesting from an optimal tax perspective because we can simply set the marginal tax rate equal to zero for each agent, so that all agents earn the efficient level of income and we can meet the revenue requirement and achieve any redistribution we wish via individualized lumpsum taxes.  However, the assumption of completely individualized taxes does allow us to illustrate the problems with welfare weights in a simple way.

\begin{thm}\label{individualized taxes necessity theorem}
Suppose that taxes can be completely individualized. If welfare weights $g$ are not structurally utilitarian, then they
are inconsistent in the sense there exist tax policies $T_0,T_1,T_2, T_3$, each of which raises the same revenue, and such that welfare weights imply a social preference cycle of the form: $T_0 \prec^g T_1 \sim^g T_2 \prec^g T_3 \sim^g T_0.$
\end{thm}
A proof sketch follows.  Assume that welfare weights are not structurally utilitarian.  Then it is possible to construct a completely individualized family $\l(T^\theta\r)$ of tax policies such that for some set $S$ of agents, where both $S$ and the set of agents not in $S$ have positive measure, we have that
\begin{enumerate}
\item for agents not in $S$, taxes are completely unchanged as $\theta$ varies; 
\item for agents in $S$, the optimal response $\l(c_i,z_i\r)$ to taxes changes as $\theta$ changes in such a way that the aggregate welfare weight on $S$, $g_S=\int_S g_i\l(c_i,z_i\r) \dd i$, changes but the utility of each agent $i$ is unchanged, so that agents in $S$ are indifferent about the value of $\theta$. 
\end{enumerate}
That it is possible to construct a family with the second property follows from the assumption that welfare weights are not structurally utilitarian. The characterization (\ref{second version structural utilitarianism}) of structural utilitarianism implies that, when welfare weights are not structurally utilitarian, for some agent $i$, it is possible to vary $\l(c_i,z_i\r)$ in such a way that utility $U_i\l(c_i,z_i\r)$ does not change, but the welfare weight $g_i\l(c_i,z_i\r)$ changes.  By the smoothness of welfare weights and utility functions, this holds for all agents in a neighborhood $S$ of $i$, and we may choose the neighborhood so that $g_i\l(c_i,z_i\r)$ changes in the same direction for all agents in $S$ as $\theta$ changes, and hence the aggregate welfare weight $g_S$ changes as well.  These changes can be brought about as optimal responses to a linear tax individualized policy (for agents in $S$), $T^\theta_i\l(z_i\r) =\tau_i\l(\theta\r) z_i- \kappa_i\l(\theta\r)$, where the marginal tax rate $\tau_i\l(\theta\r)$ controls the choice pretax income $z_i$ and consumption $c_i$ is brought to desired level by the lumpsum tax $\kappa_i\l(\theta\r)$. In the above construction, all agents are indifferent as $\theta$ changes.  So it follows from part \ref{indifference along paths} of Proposition \ref{indifference Pareto coro} -- Pareto indifference along paths -- that, letting $\theta$ vary from $\theta_0$ to $\theta_1$,  welfare weights will imply that the resulting change is socially indifferent: 
\begin{align}\label{indifference theta 0 theta 1}
T^{\theta_0} \sim^g T^{\theta_1}.
\end{align}
Let $O$ be a positive measure set of agents that is disjoint from $S$, and such that the set of agents outside of both $S$ and $O$ has positive measure.  By our assumptions, the aggregate welfare weight $g_S$ on agents in $S$ changes as $\theta$ varies between $\theta_0$ and $\theta_1$, while the aggregate welfare weight $g_O=\int_O g_i\l(c_i,z_i\r) \dd i$ on agents in $O$ does not change.  It follows that the social marginal rate of substitution $g_S/g_O$ of consumption of agents in $S$ for consumption of agents in $O$ changes as $\theta$ moves from $\theta_0$ to $\theta_1$.  Assume without loss of generality that $g_S$ increases as $\theta$ increases.  It follows that there exists some pair of payments $t_S$ and $t_O$, such that, for sufficiently small $\epsilon >0$, increasing taxes for agents in $S$ by $\epsilon t_S$ lumpsum, while reducing the taxes of agents in $O$ by $\epsilon t_O$ lumpsum is desirable at $\theta_0$ and undesirable at $\theta_1$. Formally, if we define $T^{\theta,\epsilon}$ by:
\begin{align}\label{T theta epsilon definition}
T^{\theta,\epsilon}_i\l(z_i\r)=\begin{cases} T^\theta_i\l(z_i\r)+ \epsilon t_S, & \textup{if $i \in S$;}\\
   T^\theta_i\l(z_i\r)-\epsilon t_O, & \textup{if $i \in O$;}\\
    T^\theta_i\l(z_i\r), & \textup{otherwise.} \end{cases} 
\end{align}
It then follows that if $t_S$ and $t_O$ are chosen as described above, then for sufficiently small $\epsilon >0$,
\begin{align}\label{conflicting inequalities +}
\begin{split}
T^{\theta_0} \prec^g T^{\theta_0,\epsilon},\\
T^{\theta_1} \succ^g T^{\theta_1,\epsilon}.
\end{split}
\end{align}
Formally this part of the argument appeals to Proposition \ref{local improvement principle proposition} -- the local improvement principle.  $T^{\theta,\epsilon}$ differs from $T^\theta$ for each agent $i$ at most by a change in the lumpsum payment that is independent of $\theta$.  Because utility is quasilinear, $T^{\theta,\epsilon}$ then inherits from $T^\theta$ the property that each agent is indifferent as $\theta$ changes, so that again by Pareto indifference along paths (Proposition \ref{indifference Pareto coro}), 
\begin{align}\label{indifference theta 0 theta 1+}   
T^{\theta_0,\epsilon} \sim^g T^{\theta_1,\epsilon}
\end{align}
Putting (\ref{indifference theta 0 theta 1}),(\ref{conflicting inequalities +}), and (\ref{indifference theta 0 theta 1+}) together, we have that for sufficiently small $\epsilon >0$, 
\begin{align}\label{very first preference cycle}
T^{\theta_0} \prec^g T^{\theta_0,\epsilon} \sim T^{\theta_1,\epsilon} \prec^g T^{\theta_1} \sim^g T^{\theta_0}.
\end{align} 
So on the assumption that welfare weights are not structurally utilitarian, we have constructed a social preference cycle.  

The last step is to show that revenue can be held fixed across the tax policies in the cycle.  This requires a modification of the tax policies $T^\theta$ and $T^{\theta,\epsilon}$.  Observe that $T^\theta = T^{\theta,\epsilon}$ when $\epsilon=0$, so we can identify $T^\theta$ and $T^{\theta,0}$.  Now consider a positive measure set of agents $Q$, which is disjoint from both $S$ and $O$. We modify the tax policies $T^{\theta,\epsilon}_i$ only for agents $i$ in $Q$, and otherwise these policies are not altered.  We assume that, for $i \in Q$, $T^{\theta,\epsilon}_i\l(z_i\r) = \bar{\tau}\l(\theta,\epsilon\r) z_i + \bar{\kappa}_i\l(\theta,\epsilon\r)$, where $\bar{\tau}\l(\theta,\epsilon\r)$ is a marginal tax rate, common to agents in $Q$, and $\bar{\kappa}_i\l(\theta,\epsilon\r)$ is a lumpsum tax.  We may assume that, for each agent $i \in Q$, the lumpsum tax $\bar{\kappa}_i\l(\theta,\epsilon\r)$ is chosen so as to offset any utility change as the marginal tax rate $\bar{\tau}\l(\theta,\epsilon\r)$ changes, so that agents in $Q$ are indifferent among tax policies $T^{\theta,\epsilon}$ as $\theta$ and $\epsilon$ vary.  Note however that if the marginal tax rate changes, and the lumpsum tax adjusts to keep agents' utility constant, this will change the revenue raised by the tax policy.  We may then also assume that $\bar{\tau}\l(\theta,\epsilon\r)$  (which determines $\kappa_i\l(\theta,\epsilon\r)$ for each $i$ in $Q$ up to a constant) is chosen so that the change in revenue among agents in $Q$ just offsets any change in revenue among agents in $S$ and $O$ as $\theta$ and $\epsilon$ change.  In this way, we keep revenue constant as we create the social preference cycle.  The above arguments establishing the cycle are unaltered because agents in $Q$ are indifferent as $\theta$ and $\epsilon$ change.  A formal version of the proof of Theorem \ref{individualized taxes necessity theorem} is in the Appendix and Appendix \ref{well behaved appendix individualized} shows how to fill in details when constructing $\l(T^{\theta,\epsilon}\r)$ so that it is well-behaved.  

\subsection{\label{detailed example section}A detailed example: libertarian weights}

I now present a detailed example.  The argument is parallel to that in the previous section, although some of the details differ.  In particular, in this example, I no longer assume that taxes can be completely individualized.  Instead, I assume that agents have a single observable binary characteristic $x_i$ that takes values $A$ and $B$.  For $i\in \l[0,\frac{1}{2}\r], x_i=A$ and for $i \in \l(\frac{1}{2},1\r], x_i=B$, so half of the population has each characteristic.  I assume it is possible to condition taxes on the characteristic, but the characteristic is not relevant to payoffs or welfare weights.  In particular, all types share the same cost  of earning income $v\l(z_i,A\r)=v\l(z_i,B\r)=v\l(z_i\r) = \frac{1}{2} z_i^2$.  I assume that welfare weights are libertarian and identical across agents, so that, for all $i \in \l[0,1\r]$, welfare weights are of the form $ g_i\l(c_i,z_i\r) =\tilde{g}\l(t_i\r)$, where $\tilde{g}$ is increasing in the tax $t_i =z_i-c_i$ paid by the agent.  
\begin{prop}\label{simple contradiction proposition}
In the model of the preceding paragraph, welfare weights 
are inconsistent in the sense there exist tax policies $T_0,T_1,T_2, T_3$, each of which raises the same revenue, and such that welfare weights imply a social preference cycle of the form: $T_0 \prec^g T_1 \sim^g T_2 \prec^g T_3 \sim^g T_0.$
 \end{prop}
This result resembles Theorem \ref{individualized taxes necessity theorem}.  Libertarian weights are not structurally utilitarian.  This can be seen in the argument below, in which we construct a tax policy such that utility is held fixed but the total tax paid by specific agents, and hence also their libertarian welfare weight, varies.

I now establish the proposition.  Consider linear taxes of the form $T\l(z\r)= \tau z+\kappa $, where $\tau$ is the marginal tax rate and $\kappa$ is a lumpsum payment.  Agents facing marginal tax rate $\tau$ solve the problem $\max_z \l[z\l(1-\tau\r) - \frac{1}{2}z^2-\kappa\r]$, and the optimal income is $z\l(\tau\r)= 1-\tau$.  Because utility is quasilinear, $z\l(\tau\r)$ does not depend on the lumpsum tax.  Define $\kappa\l(\tau\r)$ to be the lumpsum tax that makes agents' utility equal to zero when facing marginal tax rate $\tau$ (using the utility representation $\hat{U}_i\l(c_i,z_i\r)=c_i - v_i\l(z_i\r)$ that omits the outer utility function $u\l(\cdot\r)$). Formally, $\kappa\l(\tau\r)$ solves:
$z\l(\tau\r)\l(1-\tau\r)- \kappa\l(\tau\r) - v\l(z\l(\tau\r)\r)=0$.   Given our assumptions, $\kappa\l(\tau\r)=\frac{1}{2}\l(1-\tau\r)^2$.\footnote{We have $\kappa\l(\tau\r)=z\l(\tau\r)\l(1-\tau\r)-v\l(z\l(\tau\r)\r)= \l(1-\tau\r)\l(1-\tau\r)-\frac{1}{2}\l[\l(1-\tau\r)\r]^2=\frac{1}{2}\l(1-\tau\r)^2$.}  Consider the doubly parameterized family of tax policies $\l(T^{\theta,\epsilon}\r)$, where $\theta \in \l[\theta_0,\theta_1\r]$, with $\theta_0 = \sqrt{\frac{1}{3}}, \theta_1 = \sqrt{\frac{2}{3}}\;$:
 \begin{align*}
T^{\theta,\epsilon}_i\l(z_i\r) &=\begin{cases} \theta z_i+\kappa\l(\theta\r) +  \epsilon, &\textup{if $x_i= A$,}\\
 \l(\sqrt{1-\theta^2}\r) z_i+\kappa\l(\sqrt{1-\theta^2}\r) -\epsilon, & \textup{if $x_i= B$.}\end{cases}
\end{align*}
Observe first that $\epsilon$ just parameterizes a transfer from agents with characteristic $A$ to agents with characteristic $B$; since utility is quasilinear, such a transfer does not lead to a behavioral response, and hence, because there is an equal mass of type $A$ and type $B$ agents, the transfer is revenue neutral.  Agents with characteristic $A$ face a marginal tax rate of $\theta$, and agents with characteristic $B$ face a marginal tax rate of $\sqrt{1-\theta^2}$.  As $\theta$ rises from $\theta_0$ to $\theta_1$, the marginal tax rate of type $A$ agents rises from $\sqrt{\frac{1}{3}}$ to $\sqrt{\frac{2}{3}}$ while the marginal tax rate of type $B$ agents falls from $\sqrt{\frac{2}{3}}$ to $\sqrt{\frac{1}{3}}$.  Moreover, as $\theta$ rises from $\theta_0$ to $\theta_1$, the per agent revenue raised from type $A$ agents falls from $\frac{1}{3}+\epsilon$ to $\frac{1}{6}+\epsilon$, and, the per agent revenue raised from type $B$ agents rises from $\frac{1}{6}-\epsilon$ to $\frac{1}{3}-\epsilon$.\footnote{These numbers are derived in  Appendix \ref{libertarian calculations appendix}.}  The formula $\sqrt{1-\theta^2}$ was chosen for type $B$ agents' marginal tax rate because this is the formula required for the revenue effects from type $A$ and type $B$ agents to exactly offset one another so that the total revenue of the tax policy remains equal to $\frac{1}{4}$ for all $\theta$ and $\epsilon$.\footnote{\label{1 4 calculation footnote}This calculation is verified in Appendix \ref{libertarian calculations appendix}.}     

When $\epsilon=0$, observe that the lumpsum tax $\kappa\l(\theta\r)$ is chosen so as to keep type $A$ agents' utility equal to zero as $\theta$ varies.  So type $A$ agents are indifferent among all tax policies of the form $T^{\theta,0}$.  Likewise the lumpsum tax $\kappa\l(\sqrt{1-\theta^2}\r)$ makes type $B$ agents indifferent among all tax policies of the form $T^{\theta,0}$.  Because utility is quasilinear, these indifference conditions continue to hold if, in addition,  there is a fixed transfer from type $A$ to type $B$ agents.  So for any fixed $\epsilon$, all agents are indifferent among tax policies $T^{\theta,\epsilon}$ as $\theta$ varies.   So by part \ref{indifference along paths} of Proposition \ref{indifference Pareto coro} -- Pareto indifference along paths -- it follows that varying $\theta$ from $\theta_0$ to $\theta_1$ is socially indifferent: $ \forall \epsilon, T^{\theta_0,\epsilon} \sim^g T^{\theta_1,\epsilon}$.

As mentioned above, when $\theta =\theta_0$ and $\epsilon =0$, type $A$ agents pay a per person tax of $\frac{1}{3}$ and while type $B$ agents pay $\frac{1}{6}$.  Since libertarian weights $\tilde{g}\l(t\r)$ are increasing in taxes paid $t$, half of the agents fall into each category $A$ and $B$, and type $A$ agents pay more in tax than type $B$ agents, a small transfer $\epsilon >0$ from type $A$ to type $B$ agents is \textit{bad} at $T^{\theta_0,0}$ according to libertarian weights.  That is, $T^{\theta_0,0}  \succ^g T^{\theta_0,\epsilon}$ for sufficiently small $\epsilon>0$.  When $\theta=\theta_1$ and $\epsilon=0$, the situation is exactly reversed, so that type $A$ agents pay a tax of $\frac{1}{6}$ while type $B$ agents pay $\frac{1}{3}$.  So, at $T^{\theta_1,0}$, a small transfer from type $A$ to type $B$ is \textit{good}.  That is, $T^{\theta_1,0} \prec^g T^{\theta_1,\epsilon}$ for sufficiently small $\epsilon>0$.\footnote{A formal derivation, appealing to Proposition \ref{local improvement principle proposition}, is in Appendix \ref{libertarian calculations appendix}.}
  
Putting together the social preferences and indifferences derived in the preceding paragraphs, for sufficiently small $\epsilon > 0$, we have $T^{\theta_1,0} \prec^g T^{\theta_1,\epsilon}   \sim^g T^{\theta_0,\epsilon} \prec^g T^{\theta_0,0}  \sim^g T^{\theta_1,0}$.  This establishes that the libertarian welfare weights imply a cycle. As I show in the next section, the fact that in this example, taxes depend on characteristics, specifically ones that do not affect utility, is inessential to the argument.  The problem arises because endogenously chosen quantities (in this case $t=z-c$) can affect welfare weights without affecting utility.

\section{\label{the main theorem section}The main theorem without individualized taxes}

In this section, I show that it is possible to generate social preference cycles when all agents face the same tax schedule.  The argument then becomes more complicated but its overall structure is similar.  One of the reasons that the proof becomes more complicated is that if taxes are not individualized, it will no longer be possible to hold all agents indifferent as we modify taxes in a nontrivial way.  So the proof of the theorem in the general case no longer appeals to the Pareto indifference principle inherent in the welfare weights approach (Proposition \ref{indifference Pareto coro}).  The step in the preceding argument in which all agents are kept indifferent as the parameter $\theta$ varies is replaced by a step in which if benefits and costs to different agents are aggregated according to the system of social welfare weights $g$, then the change as $\theta$ varies is socially indifferent. 

\subsection{\label{additional structure section}Additional assumptions}
For the main result, I assume that there are no observable characteristics, but there is a single one-dimensional real valued unobservable characteristic $y$.  Because there are no observable characteristics on which to condition taxes, I omit the subscript $i$ on taxes and write $T\l(z_i\r)$ rather than $T_i\l(z_i\r)$.  This also simplifies the definition of a well-behaved family of tax policies in Section \ref{parameterized families subsection};  condition (2) in the definition of a well-behaved family $\l(T^\theta\r)$ simplifies to: the map $\l(z,\theta\r)\mapsto T\l(z,\theta\r)$ is smooth (and similarly to $\l(z,\theta,\epsilon\r)\mapsto T\l(z,\theta,\epsilon\r)$ is smooth for a doubly parameterized family $\l(T^{\theta,\epsilon}\r)$, see Appendix \ref{wb appendix non-individualized} for a complete defintion).  I assume that the function $i \mapsto y_i$ assigning to each $i$ their characteristic $y_i$ is smooth, strictly increasing in $i$, and, more specifically, the derivative of $y_i$ with respect to $i$ is positive at all values of $i$ in $I=\l[0,1\r]$.  In this case we can write $v_i\l(z_i\r) = v\l(z_i,y_i\r)$ and $g_i\l(c_i,z_i\r)=g\l(c_i,z_i,y_i\r)$.  Moreover, I assume that a higher value of $y$ corresponds to the ability to earn income at a lower cost, so that $\forall z, \forall y, \pdv[2]{}{y}{z} v\l(z,y\r) < 0.$  This implies that, in response to any regular tax policy,\footnote{Assuming that taxes are not individualized also simplifies the characterization of regular tax policies; see Appendix \ref{wb appendix non-individualized}.} agents with a higher index $i$ -- hence a higher value of $y_i$ -- earn higher income. 

\subsection{\label{statement of main theorem section}Statement of the theorem}
\begin{thm}\label{main theorem}
Under the supplementary assumptions of Section \ref{additional structure section}, if welfare weights  $g$ are not structurally utilitarian, then there exist tax policies $T_0,T_1,T_2, T_3$, each of which raises the same revenue, and such that welfare weights imply a social preference cycle of the form $T_0 \prec^g T_1 \sim^g T_2 \prec^g T_3 \sim^g T_0$.
\end{thm}
Together, Theorems \ref{necessity theorem} and \ref{main theorem} characterize the exact property on welfare weights -- structural utilitarianism -- that is required for welfare weights to be consistent.  If welfare weights are structurally utilitarian, they are compatible with a social welfare function and hence with a consistent social preference, and if welfare weights are not structurally utilitarian, they imply a social preference cycle.  This means that to acquire a \textit{consistent} method of evaluating tax policies from welfare weights, generalized welfare weights must be quite similar to traditional welfare weights, and the promise of the GSMWW approach that one can represent very general values with generalized welfare weights is not fulfilled.  To really represent broader values, we need to seek more general approaches that differ more fundamentally from the traditional utilitarian approach. 

\subsection{\label{proof sketch subsection}Proof sketch}

Here I sketch the proof of the main theorem; the missing details can be found in the Appendix.  Like in the proof of the simpler version of the theorem in Section \ref{individualized version section}, we construct a doubly parameterized family of tax policies, $\l(T^{\theta,\epsilon}\r)_{\theta \in \Theta, \epsilon \in E}$, where $\Theta =\l[\underline{\theta},\overline{\theta}\r]$ and $E=\l[\underline{\epsilon},\overline{\epsilon}\r]$.   Heuristically, we can think of $\epsilon$ as parameterizing a redistribution from some a set of higher income agents $S$ to a set of lower income agents $O$ -- as $\epsilon$ rises, taxes on agents in $S$ rise while those in $O$ fall.  The specific construction of $T^{\theta,\epsilon}$ in the Appendix bears out this interpretation (see the proof of Lemma \ref{lemma without assumptions on welfare weights}), and, in this way, the argument resembles the argument in Section \ref{individualized version section}.

\subsubsection{\label{sufficient conditions subsubsection}Sufficient conditions for a social preference cycle}
Now suppose that we construct such a family $\l(T^{\theta,\epsilon}\r)$ with the following two properties:\begin{enumerate}
\item \textbf{Indifference to $\theta$.}  Holding fixed $\epsilon$, the value of $\theta$ is socially indifferent:
\begin{align}
\label{2} \forall \epsilon \in E,\forall \theta' \in \Theta,\;\;\; \int g_i\l(\theta',\epsilon\r) \l.\pdv{\theta}\r|_{\theta = \theta'} T\l(z_i\l(\theta',\epsilon\r),\theta,\epsilon\r) \dd i = 0.  
\end{align}
\item \textbf{Changing desirability of redisribution $\epsilon$.} There exist $\theta_0 \in \l(\underline{\theta},\overline{\theta}\r)$ and $\epsilon_0 \in \l(\underline{\epsilon},\overline{\epsilon}\r)$ such that at $\l(\theta_0,\epsilon_0\r)$, as $\theta$ crosses $\theta_0$, a change in $\epsilon$ goes from being undesirable to being desirable: 
\begin{align}
\label{3}\int g_i\l(\theta_0,\epsilon_0\r) \l.\pdv{\epsilon}\r|_{\epsilon = \epsilon_0} T\l(z_i\l(\theta_0,\epsilon_0\r),\theta_0,\epsilon\r)\dd i =\; &0,\\
\label{4} \l.\dv{\theta}\r|_{\theta = \theta_0} \int g_i\l(\theta,\epsilon_0\r) \l.\pdv{\epsilon}\r|_{\epsilon = \epsilon_0} T\l(z_i\l(\theta,\epsilon_0\r),\theta,\epsilon\r) \dd i <\;& 0.  
\end{align}
\end{enumerate}
The following lemma shows that, if we can construct a family $\l(T^{\theta,\epsilon}\r)$ with the above properties, that is sufficient to construct a social preference cycle. 
\begin{lem}\label{cycle structure lemma}
Suppose that welfare weights $g$ are such that there exists a well-behaved family $\l(T^{\theta,\epsilon}\r)$ that satisfies (\ref{2})-(\ref{4}).  Then there exist parameter values $\theta_-, \theta_+ \in \Theta$ and $\epsilon_0,\epsilon_+ \in E$ for which there exists a social preference cycle of the form
$T^{\theta_+,\epsilon_0} \prec^g T^{\theta_+,\epsilon_+} \sim^g T^{\theta_-,\epsilon_+} \prec^g  T^{\theta_-,\epsilon_0} \sim^g T^{\theta_+,\epsilon_0}.$
\end{lem}
\textit{Proof.} Suppose there is a family $\l(T^{\theta,\epsilon}\r) $ satisfying (\ref{2})-(\ref{4}). Then (\ref{3}) and (\ref{4}) imply that for $\theta_- \in \Theta$ such that $\theta_- <\theta_0$ and $\theta_-$ is sufficiently close to $\theta_0$,  $\int g_i\l(\theta_-,\epsilon_0\r) \l.\pdv{\epsilon}\r|_{\epsilon = \epsilon_0} T\l(z_i\l(\theta_-,\epsilon_0\r),\theta_-,\epsilon\r)\dd i >0$, while at the same time for $\theta_+ \in \Theta$ such that $\theta_+ >\theta_0$ and $\theta_+$ is sufficiently close to $\theta_0$, $\int g_i\l(\theta_+,\epsilon_0\r) \l.\pdv{\epsilon}\r|_{\epsilon = \epsilon_0} T\l(z_i\l(\theta_+,\epsilon_0\r),\theta_+,\epsilon\r)\dd i <0.$ 
It follows from these two inequalities and the local improvement principle (Proposition \ref{local improvement principle proposition}) that for $\epsilon_+ \in E$ such that $\epsilon_+ > \epsilon_0$ and $\epsilon_+$ sufficiently close to $\epsilon_0$, $T^{\theta_-,\epsilon_0} \succ^g T^{\theta_-,\epsilon_+}$ and $T^{\theta_+,\epsilon_0} \prec^g T^{\theta_+,\epsilon_+}$.  It follows from (\ref{2}) and the global indifference principle (Section \ref{improvement and indifference principles subsection}) that $T^{\theta_-,\epsilon_0} \sim^g T^{\theta^+,\epsilon_0}$ and  $T^{\theta_-,\epsilon_+} \sim^g T^{\theta_+, \epsilon_+}$.\footnote{Observe that when $\l(T^{\theta,\epsilon}\r)$ is well-behaved, then, for each fixed $\epsilon \in E$, the family $\l(T^{\theta,\epsilon}\r)_{\theta \in \Theta}$ is well-behaved, and for each $\theta \in \Theta$, $\l(T^{\theta,\epsilon}\r)_{\epsilon \in E}$ is well-behaved.  So the improvement and indifference principles can be applied to one of the parameters $\theta$ or $\epsilon$ at a time, holding the other fixed.}  Putting the just derived relations together, we derive the cycle promised by the lemma. $\square$

\subsubsection{\label{jointly satisfiable section}Non-structurally utilitarian weights allow a family $\l(T^{\theta,\epsilon}\r)$ satisfying the sufficient conditions for a social preference cycle}

I now show that the sufficient conditions for a social preference cycle (\ref{2})-(\ref{4}) are jointly satisfiable if (and only if) welfare weights are not structurally utilitarian.  It is convenient to define $\hat{U}_i\l(T\r)=\hat{U}_i\l(c_i\l(T\r),z_i\l(T\r)\r)$ and $\hat{U}_i\l(\theta,\epsilon\r)=\hat{U}_i\l(T^{\theta,\epsilon}\r)$.  I begin by stating a fairly immediate corollary of Proposition \ref{g observation}, which is proved in the Appendix:
\begin{cor}\label{convex tax policy corollary} 
If $g$ is not structurally utilitarian, then there exists a regular tax policy $T $ for which there exist agents $i_a, i_b \in \l(0,1\r)$ with $i_a < i_b$ such that either
\begin{align}\label{negative derivative integral}
\forall i \in \l(i_a,i_b\r),\;\;\;\pdv{z_i}\hat{g}_i\l(\hat{U}_i\l(T\r),z_i\l(T\r)\r)<0
\end{align}
or 
\begin{align}\label{positive derivative integral}
\forall i \in \l(i_a,i_b\r),\;\;\; \pdv{z_i}\hat{g}_i\l(\hat{U}_i\l(T\r),z_i\l(T\r)\r)>0.
\end{align}
\end{cor}
Next I show that in the presence of condition (\ref{2}), condition (\ref{4}) takes a more convenient form.   

\begin{lem}\label{alternative condition lemma}
\label{simplification contradiction lemma}Assume that $\l(T^{\theta,\epsilon}\r)$ is well-behaved and satisfies (\ref{2}).  Then (\ref{4}) holds if and only if 
\begin{align}\label{alternative essential condition}
\begin{split}
\int \pdv{z_i}\hat{g}_i\l(\hat{U}_i\l(\theta_0,\epsilon_0\r),z_i\l(\theta_0,\epsilon_0\r)\r)
&\l[\l.\pdv{\theta}\r|_{\theta =\theta_0} z_i\l(\theta,\epsilon_0\r) \l.\pdv{\epsilon}\r|_{\epsilon =\epsilon_0} T\l(z_i\l(\theta_0,\epsilon_0\r),\theta_0,\epsilon\r)\r.\\& \l. -  \l.\pdv{\epsilon}\r|_{\epsilon =\epsilon_0} z_i\l(\theta_0,\epsilon\r)\l.\pdv{\theta}\r|_{\theta =\theta_0} T\l(z_i\l(\theta_0,\epsilon_0\r),\theta,\epsilon_0\r)\r] \dd i < 0.
\end{split}
\end{align}
\end{lem}
Since, by Proposition \ref{g observation}, for structurally utilitarian weights, $\pdv{z_i}\hat{g}_i\l(\hat{U}_i\l(\theta_0,\epsilon_0\r),z_i\l(\theta_0,\epsilon_0\r)\r)=0$ everywhere, the integral in the left-hand side of (\ref{alternative essential condition}) is always equal to zero when welfare weights are structurally utilitarian.  Hence, it follows immediately from Lemma \ref{alternative condition lemma} that a necessary condition for (\ref{2})-(\ref{4}) to be satisfied is for welfare weights \textit{not} to be structurally utilitarian.  However, what we need to show here is that not being structurally utilitarian is a \textit{sufficient} condition for the ability to construct a family of tax policies for which  (\ref{2})-(\ref{4}) to hold.   

\textit{Proof outline of Lemma \ref{alternative condition lemma}.}  The key is to show that, when expanded, the expression in the left-hand side of (\ref{4}) and the $\epsilon$-derivative of the expression on the left-hand side of (\ref{2}), evaluated at $\l(\theta_0,\epsilon_0\r)$, have overlapping terms.  In particular, I will define terms $A, B,$ and $C$ such that:
\begin{align}
\label{A+C}\l. \dv{\epsilon}\r|_{\epsilon=\epsilon_0}\int g_i\l(\theta_0,\epsilon\r) \l.\pdv{\theta}\r|_{\theta = \theta_0} T\l(z_i\l(\theta_0,\epsilon\r),\theta,\epsilon\r)\dd i &= A+C,\\
\label{B+C}\l.\dv{\theta}\r|_{\theta = \theta_0} \int g_i\l(\theta,\epsilon_0\r) \l.\pdv{\epsilon}\r|_{\epsilon = \epsilon_0} T\l(z_i\l(\theta,\epsilon_0\r),\theta,\epsilon\r)\dd i &=B+C.
\end{align}
Above, $A= \int \pdv{z_i}\hat{g}_i\l(\hat{U}_i\l(\theta_0,\epsilon_0\r),z_i\l(\theta_0,\epsilon_0\r)\r)\l.\pdv{\epsilon}\r|_{\epsilon=\epsilon_0}z_i\l(\theta_0,\epsilon\r)
\l.\pdv{\theta}\r|_{\theta = \theta_0} T\l(z_i\l(\theta_0,\epsilon_0\r),\theta,\epsilon_0\r)\dd i$, and $B= \int \pdv{z_i}\hat{g}_i\l(\hat{U}_i\l(\theta_0,\epsilon_0\r),z_i\l(\theta_0,\epsilon_0\r)\r)\l.\pdv{\theta}\r|_{\theta=\theta_0}z_i\l(\theta,\epsilon_0\r)
\l.\pdv{\epsilon}\r|_{\epsilon = \epsilon_0} T\l(z_i\l(\theta_0,\epsilon_0\r),\theta_0,\epsilon\r)\dd i.$  The term $C$, as well as the derivation of (\ref{A+C}) and (\ref{B+C}), are in the Appendix.  Note that (\ref{2}) implies that the left-hand side of (\ref{A+C}) is equal to zero, which implies that the right-hand side is equal to zero as well.  It follows that $C=-A$.  So $B+C = B-A$.  It follows that the left-hand side of (\ref{B+C}) is less than zero--which is what (\ref{4}) says--if and only if $B-A <0$.  But $B-A < 0$ is equivalent to (\ref{alternative essential condition}).  This completes the proof of Lemma \ref{alternative condition lemma}.  $\square$

The next Lemma shows that in order to able to construct a family $\l(T^{\theta,\epsilon}\r)$ that satisfies (\ref{2}), (\ref{3}), and (\ref{alternative essential condition}), it is sufficient to find a tax policy $T$ and $i_a,i_b$ for which (\ref{positive derivative integral}) holds.  (The Online Appendix presents an analogous lemma -- Lemma \ref{lemma without assumptions on welfare weights variant} -- corresponding to condition (\ref{negative derivative integral}).)

\begin{lem}\label{lemma without assumptions on welfare weights}
Let $T$ be a regular tax policy and let $i_a, i_b \in \l(0,1\r)$ be such that $i_a < i_b$.  Then there exists a well-behaved family $\l(T^{\theta,\epsilon}\r)$ with $T^{\theta_0,\epsilon_0}=T$ for some interior parameter values $\theta_0,\epsilon_0$ and that satisfies (\ref{2}),  (\ref{3}), and 
\begin{align}\label{wanted key inequality}
\begin{split}
& \l.\pdv{\theta}\r|_{\theta =\theta_0} z_i\l(\theta,\epsilon_0\r) \l.\pdv{\epsilon}\r|_{\epsilon =\epsilon_0} T\l(z_i\l(\theta_0,\epsilon_0\r),\theta_0,\epsilon\r) \\& -  \l.\pdv{\epsilon}\r|_{\epsilon =\epsilon_0} z_i\l(\theta_0,\epsilon\r)\l.\pdv{\theta}\r|_{\theta =\theta_0} T\l(z_i\l(\theta_0,\epsilon_0\r),\theta,\epsilon_0\r)\end{split}\;\;\begin{cases} < 0, &\textup{ if } i \in \l(i_a,i_b\r),\\
= 0,&\textup{ if } i \not\in \l(i_a,i_b\r).
\end{cases}
\end{align}
\end{lem}
The lemma is proven in the Online Appendix.  This lemma does not depend on any assumptions on welfare weights, but just on the broad flexibility that is available in constructing tax policies. 

The following lemma puts together the previous results derived in this section.

\begin{lem}\label{key lemma minus revenue}
If $g$ is not structurally utilitarian, then there exists a well-behaved family of tax policies $\l(T^{\theta,\epsilon}\r) $ satisfying (\ref{2})-(\ref{4}). 
\end{lem} 
\textit{Proof.}  Assume that $g$ is not structurally utilitarian.  It then follows from Corollary \ref{convex tax policy corollary} that there exists a regular tax policy $T$ and $i_a, i_b \in \l(0,1\r)$ such that $i_a < i_b$ and either (\ref{negative derivative integral}) or (\ref{positive derivative integral}) hold. First assume that (\ref{positive derivative integral}) holds.  It follows from Lemma \ref{lemma without assumptions on welfare weights} that there exists a well-behaved family of tax policies  $\l(T^{\theta,\epsilon}\r)_{\theta \in \Theta,\epsilon \in E}$ with $T^{\theta_0,\epsilon_0}=T$ satisfying (\ref{2}), (\ref{3}), and (\ref{wanted key inequality}), where, in (\ref{wanted key inequality}), $i_a$ and $i_b$ are chosen to be the same values for which (\ref{positive derivative integral}) holds.  Moreover, (\ref{positive derivative integral}) and (\ref{wanted key inequality}) together imply (\ref{alternative essential condition}).  So in this case, we can construct well-behaved family $\l(T^{\theta,\epsilon}\r)$ satisfying (\ref{2}), (\ref{3}), and (\ref{alternative essential condition}).  A similar argument -- invoking a variant of Lemma \ref{lemma without assumptions on welfare weights} (Lemma \ref{lemma without assumptions on welfare weights variant} in Section \ref{lemma without assumptions on welfare weights variant section} of the Online Appendix) shows that, when (\ref{negative derivative integral}) rather than (\ref{positive derivative integral}) holds, we can still construct a well behaved family $\l(T^{\theta,\epsilon}\r)$ satisfying  (\ref{2}), (\ref{3}), and (\ref{alternative essential condition}).  It now follows from Lemma \ref{alternative condition lemma} that whenever welfare weights are not structurally utilitarian, it is possible to construct a tax policy satisfying (\ref{2})-(\ref{4}).   $\square$

\subsubsection{\label{holding revenue constant subsubsection}Holding revenue constant}

The construction of the previous section can be extended so that the family $\l(T^{\theta,\epsilon}\r)$ is be chosen so that revenue is held constant, as stated by the following lemma.
\begin{lem}\label{key lemma with revenue}
If $g$ is not structurally utilitarian, then there exists a well-behaved constant revenue family of tax policies $\l(T^{\theta,\epsilon}\r)$ satisfying (\ref{2})-(\ref{4}).
\end{lem}
Lemma \ref{key lemma with revenue} is a strengthening of Lemma \ref{key lemma minus revenue} that differs from Lemma \ref{key lemma minus revenue} only in that family $\l(T^{\theta,\epsilon}\r)$ is required to be a constant revenue family in the sense that all tax policies $T^{\theta,\epsilon}$ raise the same revenue.  I have separated this additional requirement into a separate lemma because the argument that revenue can be held constant appeals to different principles than the proof of the other properties.  The basic idea is similar to that described in Section \ref{individualized version section} for holding revenue constant.  In particular, once we construct a family $\l(T^{\theta,\epsilon}\r)$ satisfying (\ref{2})-(\ref{4}), as we know we can do from Lemma \ref{key lemma minus revenue}, we consider a positive measure set $Q$ of agents at a different income level than agents in $S$ and $O$, and vary the revenue raised from agents in $Q$ as $\theta$ and $\epsilon$ vary exactly so as to offset revenue changes elsewhere in the tax schedule in such a way that there is no detectable welfare change in $Q$ according to welfare weights; this is analogous to moving along a social indifference curve for agents in $Q$ along which the revenue raised from those agents varies.  The details are in the Online Appendix.

\subsubsection{Putting it all together} 

Putting Lemmas \ref{cycle structure lemma} and \ref{key lemma with revenue} together yields Theorem \ref{main theorem}, the main result.  

\subsection{\label{poverty alleviation subsection}An application: Poverty alleviation}

I now present an application to illustrate the main result.  Maintain all of the assumptions of Section \ref{additional structure section}.      
Let $\bar{c}$ be the poverty line; that is, $\bar{c}$ is the level of consumption below which agents are considered to be poor.  Now consider welfare weights which capture the goal of poverty alleviation by concentrating weight on agents beneath the poverty line.  Saez and Stantacheva presented such an example.\footnote{\citeasnoun{besley1992workfare} and \citeasnoun{kanbur1994optimal} incorporate poverty alleviation in optimal tax.}  I modify their example slightly to make welfare weights smooth.  Suppose that $g_i\l(c_i,z_i\r)= \tilde{g}\l(c_i\r)$, where $\tilde{g}\l(c_i\r)$ is decreasing in $c_i$ until $c_i$ gets to $\bar{c}$ and then remains constant at the value $\underline{g}$ thereafter, where $\underline{g} >0$.  I assume that $\underline{g} >0$ to be in conformity with my prior assumptions but we may assume that $\underline{g}$ is arbitrarily close to zero.  So agents below the poverty line have a higher welfare weight than agents above the poverty line, the welfare weight is greater the further below the poverty line the agent is, and constant for agents above the poverty line. 

Now consider a doubly parameterized family of tax policies $\l(T^{\theta,\epsilon}\r)$ of the form $T\l(z,\theta,\epsilon\r)=\theta f\l(z\r) + \l(\theta-\epsilon\r) z  +\alpha \epsilon -\kappa\l(\theta,\epsilon\r)$ where $f\l(z\r)$ is a smooth function and, for some $\theta_0$, $\kappa\l(\theta_0,\epsilon\r) =0, \forall \epsilon$. Assume that there exists $\epsilon_0$ and income level $\bar{z}$ (within the income distribution), such that, when facing tax schedule $T^{\theta_0,\epsilon_0}$, all agents earn positive income, all agents earning income $\bar{z}$ or above are strictly above the poverty line, and a positive measure of agents with income below $\bar{z}$ are beneath the poverty line.    I assume that $f\l(z\r) = 0$ for all $z$ with $z \leq \bar{z}$, and $f\l(z\r) > 0$ for all $z$ with $z> \bar{z}$, so that the $\theta f\l(z\r)$ term specifies taxes that only apply to agents above the poverty line when $\l(\theta,\epsilon\r)$ is close to $\l(\theta_0,\epsilon_0\r)$.  Noting that the optimal income for $i$, $z_i\l(\theta,\epsilon\r)$, is independent of $\alpha$, assume that $\alpha$ is chosen so that $\int g\l(\theta_0,\epsilon_0\r) \l[z_i\l(\theta_0,\epsilon_0\r)-\alpha\r] \dd i =0$, which says that, at $\hat{T}^{\theta_0,\epsilon_0}$, the positive welfare effect of increasing $\epsilon$ due to decreasing marginal tax rates through the term $-\epsilon z$ is just offset by the negative welfare effect of the increase in the lumpsum tax $\alpha \epsilon$.  (Note that, by our assumptions, $\l.\pdv{\epsilon}\r|_{\epsilon=\epsilon_0} \kappa\l(\theta_0,\epsilon\r) =0$.)         Finally, we assume that $\kappa\l(\theta,\epsilon\r)$ satisfies the following set of differential equations (note that $g_i\l(\theta,\epsilon\r)$ depends on $\kappa\l(\theta,\epsilon\r)$):
\begin{align}\label{kappa differential equation}
\l.\pdv{\theta}\r|_{\theta=\theta'} \kappa\l(\theta,\epsilon\r)= \int \frac{g_i\l(\theta',\epsilon\r)}{\int g_j \l(\theta',\epsilon\r) \dd j}\l[z_i\l(\theta',\epsilon\r)+f\l(z_i\l(\theta',\epsilon\r)\r)\r] \dd i, \;\;\;\; \forall \theta', \forall \epsilon. \end{align} 
Rearranging terms, one can see that (\ref{kappa differential equation}) says that for any fixed value of $\epsilon$, when changing $\theta$, the welfare effect due to increasing marginal tax rates through the term $\theta f\l(z\r)+\theta z$ is just offset by the welfare effect of the change in the lumpsum tax $\kappa\l(\theta,\epsilon\r)$.  Note that the differential equations (\ref{kappa differential equation}) and the conditions $\kappa\l(\theta_0,\epsilon\r) =0, \forall \epsilon$ uniquely determine $\kappa\l(\theta,\epsilon\r)$. 

\begin{prop}\label{poverty cycle proposition}
With poverty alleviation welfare weights, if $\l(T^{\theta,\epsilon}\r)$ has the properties assumed in this section, $\l(T^{\theta,\epsilon}\r)$ satisfies (\ref{2})-(\ref{4}), the sufficient conditions for a preference cycle in Lemma \ref{cycle structure lemma}.   
\end{prop} 
The proof is in the Online Appendix. Condition (\ref{2}) corresponds to $\int g\l(\theta_0,\epsilon_0\r) \l[z_i\l(\theta_0,\epsilon_0\r)-\alpha\r] \dd i =0$ (and $\l.\pdv{\epsilon}\r|_{\epsilon=\epsilon_0} \kappa\l(\theta_0,\epsilon\r) =0$), and (\ref{3}) corresponds to (\ref{kappa differential equation}).  The key calculation that drives the argument is that:
$\l.\dv{\theta}\r|_{\theta = \theta_0} \int g_i\l(\theta,\epsilon_0\r) \l.\pdv{\epsilon}\r|_{\epsilon = \epsilon_0} T\l(z_i\l(\theta,\epsilon_0\r),\theta,\epsilon\r)= \int  \tilde{g}'\l(c_i\l(\theta_0,\epsilon_0\r)\r) \frac{v'_i\l(z_i\l(\theta_0,\epsilon_0\r)\r)}{v''_i\l(z_i\l(\theta_0,\epsilon_0\r)\r)} \dd i \times \int \frac{\tilde{g}\l(c_i\l(\theta_0,\epsilon_0\r)\r)}{\int \tilde{g}\l(c_j\l(\theta_0,\epsilon_0\r)\r)\dd j} f\l(z_i\l(\theta_0,\epsilon_0\r)\r) \dd i < 0$, which establishes (\ref{4}).  It follows from Proposition \ref{poverty cycle proposition} that, in the poverty alleviation example, with tax policies as described above, we can construct a social preference cycle exactly as in the proof of Lemma \ref{cycle structure lemma} (see Section \ref{sufficient conditions subsubsection} above).  We have not worried about holding revenue constant, but Lemma \ref{key lemma with revenue} tells us that we can modify the construction of $\l(T^{\theta,\epsilon}\r)$ so as to hold revenue constant as well.  Of course, the reason we could construct a cycle is that poverty reduction welfare weights are not structurally utilitarian.  In particular, by increasing both consumption $c_i$ and income $z_i$ so as to hold total utility $u\l(c_i-v_i\l(z_i\r)\r)$ fixed, it is possible to bring an agent above the poverty line, and, in this way, we can change their welfare weight; this is not consistent with structural utilitarianism.  In general welfare weights that respond to changes in consumption but do not take into account labor supply costs will not be structurally utilitarian, and hence will lead to social preference cycles. More generally, welfare weights that respond to a only subset of the endogenously chosen arguments that determine utility will be vulnerable to inconsistency.

\section{\label{non-quasilinear preferences section}Generalization to non-quasilinear preferences}

Throughout the paper, I assumed quasilinear utility, which rules out income effects.   This section discusses how the results generalize without quasilinearity.  For more general utility functions $U_i\l(c_i,z_i\r)$ that are not necessarily quasilinear, structural utilitarianism can be defined as follows.
\begin{defn}\label{nonquasilinear welfare weights definition} \textbf{Structural utilitarianism without quasilinearity.}
A system of welfare weights $g$ is \textbf{structurally utilitarian} if and only if $\forall i \in I,\forall z_i,z'_i \in Z, \forall c_i, c'_i \in \mathbb{R}$,
\begin{align}\label{nonquasilinear structural utilitarianism}
U_i\l(c_i,z_i\r)=U_i\l(c'_i,z'_i\r) \Rightarrow \frac{\pdv{c_i}U_i\l(c_i,z_i\r)}{\pdv{c_i}U_i\l(c'_i,z'_i\r)}=\frac{g_i\l(c_i,z_i\r)}{g_i\l(c'_i,z'_i\r)}.
\end{align}
\end{defn} 
This condition says that, as we move along a fixed $\l(c_i,z_i\r)$-indifference curve for agent $i$, $i$'s marginal welfare weight must be proportional to the marginal utility of consumption. Of course, utilitarian weights must satisfy this condition as they are equal to the marginal utility of consumption.  Section \ref{structural utilitarian section} provided several equivalent conditions characterizing structural utilitarianism for the quasilinear case, (\ref{depends only on utility condition}), (\ref{first version structural utilitarianism}), and (\ref{second version structural utilitarianism}).  To see that (\ref{nonquasilinear structural utilitarianism}) is indeed a generalization of these conditions, it is easiest to compare with  (\ref{second  version structural utilitarianism}).   As discussed in Section \ref{structural utilitarian section}, for quasilinear utility, $U_i\l(c_i,z_i\r) = U_i\l(c'_i,z'_i\r)$ implies $\pdv{c_i}U_i\l(c_i,z_i\r) = \pdv{c_i} U_i\l(c'_i,z'_i\r)$, 
or equivalently, if $U_i\l(c_i,z_i\r)=U_i\l(c'_i,z'_i\r)$, then  $\frac{\pdv{c_i}U_i\l(c_i,z_i\r)}{\pdv{c_i}U_i\l(c'_i,z'_i\r)}=1$.  So, with quasilinearity, (\ref{nonquasilinear structural utilitarianism}) reduces to $U_i\l(c_i,z_i\r)=U_i\l(c'_i,z'_i\r) \Rightarrow 1=\frac{g_i\l(c_i,z_i\r)}{g_i\l(c'_i,z'_i\r)}$, which is equivalent to (\ref{second  version structural utilitarianism}).  In other words, for quasilinear utility,  the marginal utility of consumption is constant along any $\l(c_i,z_i\r)$-indifference curve, and hence (\ref{nonquasilinear structural utilitarianism}) says that  structually utilitarian welfare weights must be constant too. So Definition \ref{nonquasilinear welfare weights definition} indeed generalizes the previous definition of structural utilitarianism.

Our results also generalize.  Even without quasilinearity,  welfare weights are structurally utilitarian if and only if they arise from a generalized utilitarian social welfare function -- so that Theorem \ref{necessity theorem} still holds -- and  if welfare weights are not structurally utilitarian, then it is possible to construct a social preference cycle -- so that Theorem \ref{main theorem} holds as well.  These results assume some regularity conditions on the utility functions $U_i\l(c_i,z_i\r)$.  Section \ref{nonquasilinear appendix} of the Online Appendix presents these conditions and explains how to modify the proofs of the theorems when quasilinearity is no longer assumed.

\section{\label{discussion section}Discussion}

The motivation for generalized social marginal welfare weights was as a means of addressing the omission of broader values in economic analysis.  I have argued in this paper that this solution does not work because generalized welfare weights, once they stray too far from traditional utilitarian weights, are inconsistent.  In this closing section, I will discuss some related literature and how the current contribution differs, as well as ways forward on the problem of incorporating broader normative values in economic analysis.

\subsection{\label{related literature section}The Pareto principle and broader values: related literature}

\citeasnoun{saez2016generalized} write ``if the weights are nonnegative, then our theory respects the Pareto principle in the sense that, around the local optimum, there is no Pareto improving small reform."  It may appear that Saez and Stantcheva have uncovered a way of incorporating broader values into economic analysis compatibly with the Pareto principle.  Several authors, including Sen \citeyear{sen1970impossibility,sen1979personal,sen1979utilitarianism} and Kaplow and Shavell \citeyear{kaplow2001any,kaplow2009fairness}, have argued that incorporating broader moral considerations into economic evaluation is inconsistent with the Pareto principle.   Sen interprets this as an argument against insisting on the Pareto principle, whereas \citeasnoun{kaplow2001any} interpret it as an argument against including non-welfarist considerations in normative economic evaluation.\footnote{See also \citeasnoun{weymark2017conundrums}.} As they say, one philosopher's modus ponens is another philosopher's modus tollens. \citeasnoun{fleurbaey2003any} are critical of \citeasnoun{kaplow2001any}, and take a more positive view of incorporating broader values compatibly with the Pareto principle.  In discussing the Saez and Stantcheva approach critically,  \citeasnoun{fleurbaey2018optimal}, who also take a more positive view of incorporating broader values, compatibly with Pareto, write, 
\begin{quote}
... the social welfare function approach has been introduced by \citeasnoun{bergson1938reformulation} and \citeasnoun{samuelson1947foundations} not out of a taste for elegance, but because it is the only way to define social preferences that are both transitive and Paretian. Therefore, a method that directly weights tax changes at the various earning levels is compatible with transitive and Paretian social preferences, and then extendable to the study of nonlocal reforms, only if it relies on the classical framework of the social welfare function. (p. 1059)
\end{quote} 
This informal passage is closely related to the results developed formally in the current paper.     

\subsection{\label{contribution subsection}The contribution of this paper}
My result differs from the  \citeasnoun{kaplow2001any} result in three ways: (1) Kaplow and Shavell are concerned with \textit{social welfare functions}, which give global rankings, while I am concerned with \textit{systems of generalized welfare weights}, which give local marginal rates of substitution.\footnote{I do however bridge the gap between social welfare functions and marginal welfare weights to some extent by showing how to derive some of the global comparisons implied by welfare weights.} (2)  The key property for Kaplow and Shavell is whether a social welfare function is \textit{individualistic}, meaning that changes in states that do not affect individual utility cannot affect social welfare, whereas the key property for me  is \textit{structural utilitarianism}.  Structural utilitarianism, at least under the assumption of quasilinear preferences, is thematically similar to individualism in that both say that some aspect of social evaluation cannot change in response to certain types of changes that do not affect individual utility, but formally the two properties are quite different, imposing different restrictions on different types of formal objects.\footnote{Individualism says that changes in \textit{social states} that do not affect utility do not affect \textit{social welfare}, whereas, under the assumption of quasilinearity, structural utilitarianism says that changes in agent's \textit{decisions} (specifically of consumption and income) that do not affect an individual's utility do not affect affect \textit{that individual's welfare weight}, but structural utilitarianism allows that exogenous characteristics contained in $\l(x_i,y_i\r)$ may affect welfare weights without affecting utility.  Moreover, in the non-quasilinear case, structural utilitarianism generalizes to the property that, along any $\l(c_i,z_i\r)$-indifference curve, an agent's welfare weight is proportional to their marginal utility of consumption, which does not seem to be analogous to individualism in the same way as in the quasilinear case.}  This difference is perhaps easiest to see by observing that individualism is equivalent to the property of Pareto indifference and, by Proposition \ref{indifference Pareto coro}, the social preferences induced by welfare weights satisfy a version of Pareto indifference regardless of whether they are structurally utilitarian.  (3) For Kaplow and Shavell, the penalty for violating their key property is that the social ranking \textit{violates weak Pareto}, whereas, for me, the penalty is that the implied social ranking contains a preference cycle, and hence is \textit{inconsistent}.  My paper shows that eschewing social welfare functions in favor of the local comparisons of generalized marginal welfare weights is not a successful approach to avoiding Kaplow-Shavell type impossibilities because it leads to inconsistencies.

\citeasnoun{fleurbaey2018optimal} only discuss the potential intransitivity of welfare weights briefly and they do not present a formal result characterizing when generalized social welfare weights are consistent.   Nor do they provide a methodology for collecting the local judgements of the generalized social welfare weights into implicit global comparisons.  In this paper, I do both of these things.  I show how to collect the local judgements of generalized social welfare weights into global social judgements (see Section \ref{improvement and indifference principles subsection}) and that the precise property that is necessary and sufficient for welfare weights to be consistent is structural utilitarianism (see Theorems \ref{necessity theorem} and \ref{main theorem}). Unlike Fleurbaey and Maniquet, I also construct specific examples of cases in which generalized welfare weights are inconsistent. Moreover, my result is stronger than the point made by Fleurbaey and Maniquet in another way.  I show that when welfare weights are not structurally utilitarian, they are not consistent with \textit{any} social welfare function, \textit{Paretian or not}. Notice, in this regard, that Theorem \ref{main theorem} does not mention any Pareto principle; it simply says that if welfare weights are not structurally utilitarian, then they are inconsistent.  

\subsection{Two ways forward}

I now highlight two ways forward if broader values are to be incorporated into normative economic analysis and specifically optimal tax.  \citeasnoun{fleurbaey2018optimal} write that ``the classical social welfare function framework is more flexible than commonly thought, and can accommodate a very large set of nonutilitarian values. More specifically, fairness concepts can help solve the interpersonal comparison difficulties that the utilitarian approach faces when agents have different preferences by providing useful selections of suitable individual utility indexes," and their paper shows that \textit{Paretian} social welfare functions can capture a broad set of values in an optimal tax context.\footnote{Other work representing broader values with Paretian social welfare functions includes \citeasnoun{fleurbaey2011theory},  \citeasnoun{piacquadio2017fairness} and \citeasnoun{berg2020equal}.}  In the setting of the current paper, Theorem \ref{necessity theorem} shows that structurally utilitarian welfare weights are compatible with a generalized utilitarian social welfare function of the form $\int F\l(U_i\l(c_i,z_i\r),x_i,y_i\r) \dd i$.  We may think of the function $F\l(u_i, x_i, y_i\r)$ as reweighting utilities $u_i$ -- and hence also reweighting the social value we assign to tax changes -- on the basis of certain moral considerations which are responsive to the characteristics $\l(x_i,y_i\r)$.  The welfare weights induced by such a social welfare function must be consistent because they are derived from a consistent social ranking to begin with.

Not all values can be captured with Paretian approaches.\footnote{ \citeasnoun{fleurbaey2018optimal} recognize this,  writing ``we highlight another way in which at least some fairness principles can remain compatible with the Pareto principle ... Not all fairness principles fall in this category, obviously, and the socialist and libertarian principles mentioned two paragraphs earlier provide examples of non-Paretian approaches." (p. 1040)  For criticisms of the Pareto principle, see \citeasnoun{sen1979utilitarianism}, \citeasnoun{mongin2016spurious} and \citeasnoun{sher2020perspective}.}  The second way forward embraces this point. Consider libertarianism as an example.\footnote{For approaches to libertarian taxation, see \citeasnoun{nozick1974anarchy}, \citeasnoun{feldstein1976theory}, \citeasnoun{young1987progressive}, \citeasnoun{weinzierl2014promise}, and \citeasnoun{v2018taxation}. For an approach to non-welfarist optimal taxation, see \citeasnoun{kanbur2006non}.}  Suppose that one thinks that people are entitled to their pre-tax incomes and that in some way taxation is like theft.  This view is not faithfully rendered as saying that additional income to people who have been taxed more should be given additional weight in comparison to those who have been taxed less; rather it is the view that it is wrong to tax, or at least, if not absolutely wrong, that it is bad to tax, and that this bad is tolerated, to the extent that it is, because of the other important purposes of taxation.  On a rights-based version of libertarianism, taxing people is bad not because it reduces their utility but because it violates their entitlements.   Imagine there is a function $s\l(t_i\r)$ for each agent $i$, that measures how bad it is to violate $i$'s entitlements.  We might then minimize the non-Paretian social welfare function  $W\l(T\r) = -\int_i s\l(T_i\l(z_i\l(T\r)\r)\r) \dd i$ subject to a revenue requirement.  Such an approach will not be Paretian, even in the sense of Proposition \ref{indifference Pareto coro}, and so it follows from Corollary \ref{non Pareto coro} that this approach cannot be captured by welfare weights.  Alternatively we may trade off rights based concerns as captured by $s\l(t_i\r)$ against utilitarian concerns.  Or we may want to go  farther, and consider more thoroughly procedural approaches that do not appeal to a social objective (or even a local social objective).  Whatever the right approach, it seems unlikely that we can capture the richness of  broader ethical values by means of conservative modifications, such as by modifications of welfare weights, in a way that strongly preserves the structure of traditional optimal tax theory; we should expect that incorporation of broader values will require a more thorough change in the way that we normatively evaluate taxes and other economic policies.

\bibliographystyle{agsm}
\bibliography{gswwbib}

\newpage
\appendix
\numberwithin{obs}{section}
\numberwithin{equation}{section}
\numberwithin{lem}{section}
\numberwithin{prop}{section}
\numberwithin{cor}{section}
\numberwithin{thm}{section}

\section*{Appendix}
\section{\label{main appendix}Definitions and proofs of results stated in main text}

\subsection{\label{wb appendix}Well-behaved families of tax policies}

In this section, I spell out the requirements for a well-behaved families of tax policies introduced in Section \ref{parameterized families subsection} more formally and completely, both for individualized tax policies that can depend on $i$, and for non-individualized tax policies that do not depend on $i$, as in Section \ref{additional structure section}.

\subsubsection{Individualized tax policies}

A family of tax policies $\l(T^{\theta}\r)_{\theta \in \Theta}$ is \textbf{well-behaved} if 
\begin{enumerate}
\item\label{first formal well-behaved}  for each $i$ and $\theta$, $i$'s optimal income in response to $T^\theta$, $z_i\l(\theta\r)$ exists, is unique, and $z_i\l(\theta\r)>0$, and the second order condition for $i$'s optimization problem, when facing $T^\theta$, holds with strict inequality at the optimum: $\l.\dv[2]{z}\r|_{z=z_i\l(\theta\r)} U_i\l(z-T_i\l(z,\theta\r),z\r)< 0$, and
\item \begin{enumerate}
\item for all $i$, the map $\l(z,\theta\r) \mapsto T_i\l(z,\theta\r)$ is smooth, and
\item\label{second b formal well-behaved}  there exists a finite set subset of $\l\{i_0, i_1,\ldots, i_n\r\}$ of $I$, with $n \geq 1$ and $i_0=0 < i_1 < i_2 < \ldots < i_n=1$ such that the map $\l(i,z,\theta\r) \mapsto T_i\l(z,\theta\r)$ is smooth on $\l(i_{k-1},i_k\r) \times Z \times \Theta$, for $k=1,\ldots n$.
\end{enumerate}
\end{enumerate}
To eliminate any possible ambiguity, $\l.\dv[2]{z}\r|_{z=z_i\l(\theta\r)} U_i\l(z-T_i\l(z,\theta\r),z\r)$ is the second derivative of the function $z \mapsto U_i\l(z-T_i\l(z,\theta\r),z\r)$.  Assuming quasilinear utility, $\l.\dv[2]{z}\r|_{z=z_i\l(\theta\r)} U_i\l(z-T_i\l(z,\theta\r),z\r)=\l.\dv[2]{z}\r|_{z=z_i\l(T\r)}u\l(z-T_i\l(z,\theta\r)-v_i\l(z\r)\r)$.  As mentioned in the main text, condition \ref{second b formal well-behaved} allows for a finite number of discontinuities in $i$.     

A tax policy is \textbf{regular} if there exists a well-behaved family $\l(T^\theta\r)_{\theta \in \Theta}$ and $\theta' \in \Theta$ such that $T^{\theta'}=T$.  Given this definition, it is easy to see that a tax policy $T$ is regular if and only if
\begin{enumerate}
\item  for each $i$, $z_i\l(T\r)$ exists and is unique, $z_i\l(T\r)>0$, and $\l.\dv[2]{z_i}\r|_{z_i=z_i\l(T\r)} U_i\l(z_i-T_i\l(z_i\r),z_i\r)< 0$, and
\item \begin{enumerate}
\item for all $i$, the map $z \mapsto T_i\l(z\r)$ is smooth, and,
\item there exists a finite set subset of $\l\{i_0, i_1,\ldots, i_n\r\}$ of $I$, with $n \geq 1$ and $i_0=0 < i_1 < i_2 < \ldots < i_n=1$ such that the map $\l(i,z\r) \mapsto T_i\l(z\r)$ is smooth on $\l(i_{k-1},i_k\r) \times Z $, for $k=1,\ldots n$.
\end{enumerate}
\end{enumerate}
It follows immediately from the definitions of well-behaved families of tax policies and regular tax policies that any regular tax policy must satisfy the above conditions.  Going in the other direction, if $T$ satisfies the above conditions then the family $\l(T^\theta\r)$, defined by $T^\theta = T, \forall \theta$ is well-behaved.  So the above conditions are sufficient for a tax policy to be regular as well.

A doubly parameterized family $\l(T^{\theta,\epsilon}\r)_{\theta \in \Theta,\epsilon \in E}$ is \textbf{well-behaved} if
\begin{enumerate}
\item\label{first formal well-behaved}  for each $i, \theta$, and $\epsilon$, $z_i\l(\theta,\epsilon\r)$ exists and is unique, $z_i\l(\theta,\epsilon\r)>0$, and the second order condition holds with strict inequality: $\l.\dv[2]{z}\r|_{z=z_i\l(\theta,\epsilon\r)} U_i\l(z-T_i\l(z,\theta,\epsilon\r),z\r)< 0$, and
\item \begin{enumerate}
\item for all $i$, the map $\l(z,\theta,\epsilon\r) \mapsto T_i\l(z,\theta,\epsilon\r)$ is smooth, and
\item\label{second b formal well-behaved}  there exists a finite set subset of $\l\{i_0, i_1,\ldots, i_n\r\}$ of $I$, with $n \geq 1$ and $i_0=0 < i_1 < i_2 < \ldots < i_n=1$ such that the map $\l(i,z,\theta,\epsilon\r) \mapsto T_i\l(z,\theta,\epsilon\r)$ is smooth on $\l(i_{k-1},i_k\r) \times Z \times \Theta \times E$, for $k=1,\ldots n$.
\end{enumerate}
\end{enumerate}
\subsubsection{\label{wb appendix non-individualized}Non-individualized tax policies}
When taxes are not individualized, and hence are the same for all agents and do not depend on $i$, the requirements for well-behavedness simplify.  In particular, in this case, a tax policy $T$ is regular if and only if
\begin{enumerate}
\item  for each $i$, $z_i\l(T\r)$ exists and is unique, $z_i\l(T\r)>0$, and $\l.\dv[2]{z_i}\r|_{z_i=z_i\l(T\r)} U_i\l(z_i-T\l(z_i\r),z_i\r)< 0$, and
\item the map $z \mapsto T\l(z\r)$ is smooth. 
\end{enumerate}
Likewise, when taxes are not individualized, a family $\l(T^{\theta,\epsilon}\r)$ is well behaved if 
\begin{enumerate}
\item\label{first formal well-behaved}  for each $i, \theta$, and $\epsilon$, $z_i\l(\theta,\epsilon\r)$ exists and is unique, $z_i\l(\theta,\epsilon\r)>0$, and the second order condition holds with strict inequality: $\l.\dv[2]{z}\r|_{z=z_i\l(\theta,\epsilon\r)} U_i\l(z-T\l(z,\theta,\epsilon\r),z\r)< 0$, and
\item  the map $\l(z,\theta,\epsilon\r) \mapsto T\l(z,\theta,\epsilon\r)$ is smooth.
\end{enumerate}
The following observation is useful
\begin{obs}\label{regularity observation}
A family of non-individualized tax policies $\l(T^{\theta,\epsilon}\r)$ is well behaved if and only if (i) for all $\theta$ and $\epsilon$, $T^{\theta,\epsilon}$ is regular and (ii) the map $\l(z,\theta,\epsilon\r) \mapsto T\l(z,\theta,\epsilon\r)$ is smooth.
\end{obs}

\subsection{Proof of Proposition \ref{local improvement principle proposition}}
Assume that $g$, $\l(T^\theta\r)$ and $\theta_0$ are as in the hypothesis of the proposition.  Now, first assume that $\int g_i\l(\theta_0\r) \l.\pdv{\theta}\r|_{\theta=\theta_0}T_i\l(z_i\l(\theta_0\r),\theta\r) \dd i < 0$.  It follows from the smoothness of welfare weights, utility functions and parameterized families of tax policies that if $\theta_1$ is such that $\theta_1 > \theta_0$ and $\theta_1$ is sufficiently close to $\theta_0$, then for all $\theta' \in \l[\theta_0,\theta_1\r]$, $\int g_i\l(T^{\theta'}\r) \l.\pdv{\theta}\r|_{\theta=\theta'}T_i\l(z_i\l(T^{\theta'}\r),\theta\r) \dd i < 0.$
It follows from the global improvement principle (in Section \ref{improvement and indifference principles subsection}) that  
 for all $\theta' \in \l(\theta_0,\theta_1\r)$, $T^{\theta_0} \prec^g T^{\theta'}$.  This establishes the first claim in Proposition \ref{local improvement principle proposition}. 

Next assume that $
\int g_i\l(\theta_0\r) \l.\pdv{\theta}\r|_{\theta=\theta_0}T_i\l(z_i\l(\theta_0\r),\theta\r) \dd i > 0$.  Now define the parameterized family of tax policies, $\l(\tilde{T}^\theta\r)_{\theta \in \l[-\overline{\theta},-\underline{\theta}\r]}$ by $\tilde{T}^\theta = T^{-\theta}, \forall \theta \in \l[-\overline{\theta},-\underline{\theta}\r]$, and, using notation analogous to that introduced in Section \ref{parameterized families subsection}, let $\tilde{T}_i\l(z,\theta\r)=\tilde{T}^\theta_i\l(z\r)$.  Then we have:
\begin{align*}
  \int g_i\l(\tilde{T}^{-\theta_0}\r) \l.\pdv{\theta}\r|_{\theta=-\theta_0}\tilde{T}_i\l(z_i\l(\tilde{T}^{-\theta_0}\r),\theta\r) \dd i &= \int g_i\l(T^{\theta_0}\r)\times \l(- \l.\pdv{\theta}\r|_{\theta=\theta_0}T_i\l(z_i\l(T^{\theta_0}\r),\theta\r) \r)\dd i \\ &=- \int g_i\l(T^{\theta_0}\r) \l.\pdv{\theta}\r|_{\theta=\theta_0}T_i\l(z_i\l(T^{\theta_0}\r),\theta\r)\dd i <0,
  \end{align*}
where the inequality follows from the assumption made at the beginning of the paragraph.  It follows from the smoothness of welfare weights, utility functions and parameterized families of tax policies that if $-\theta_1 \in \l(-\overline{\theta},-\theta_0\r)$ is sufficiently close to $-\theta_0$, then for all $\theta' \in \l[-\theta_1,-\theta_0\r]$,  $\int g_i\l(\tilde{T}^{\theta'}\r) \l.\pdv{\theta}\r|_{\theta=\theta'}\tilde{T}_i\l(z_i\l(\tilde{T}^{\theta'}\r),\theta\r) \dd i < 0.  $ So the global improvement principle implies that, for all $-\theta' \in \l(-\theta_1,-\theta_0\r)$, $\tilde{T}^{-\theta'} \prec^g \tilde{T}^{-\theta_0}$.  So for all $\theta' \in \l(\theta_0,\theta_1\r), T^{\theta_0} \succ^g T^{\theta'}$.  This establishes the second claim of Proposition \ref{local improvement principle proposition}.   
 $\square$  
 
 \subsection{Proof of Proposition \ref{indifference Pareto coro}}
First assume that all agents are indifferent as $\theta$ varies in the interval $\l[\theta_0,\theta_1\r]$.   Then, for all $\theta' \in \l[\theta_0,\theta_1\r]$ and agents $i$, $\l.\dv{\theta}\r|_{\theta'=\theta}U_i\l(T^{\theta'}\r) =0$.  Hence, by (\ref{equivalent inequalities U T 0}), for all $\theta' \in \l[\theta_0,\theta_1\r]$ and agents $i$, $\l.\pdv{\theta}\r|_{\theta=\theta'}T_i\l(z_i\l(T^{\theta'}\r),\theta\r) =0$. So, for all $\theta' \in \l[\theta_0,\theta_1\r]$, $\int g_i\l(T\r)  \l.\pdv{\theta}\r|_{\theta=\theta'}T_i\l(z_i\l(T^{\theta'}\r),\theta\r) \dd i =0$.  So by the global indifference principle (in Section \ref{improvement and indifference principles subsection}), $T^{\theta_0} \sim^g T^{\theta_1}$.  This establishes Pareto indifference along paths.  Weak Pareto along paths is similar, appealing again to (\ref{equivalent inequalities U T 0}), and using the global improvement principle (also in Section \ref{improvement and indifference principles subsection}) instead of the global indifference principle. $\square$  

 \subsection{Definitions for and proof of Corollary \ref{non Pareto coro}}
 Consider a real-valued social welfare function $W\l(T\r)$, whose domain is the set of regular tax policies. Say the social welfare function is \textbf{sufficiently differentiable} if for all well-behaved families $\l(T^\theta\r)_{\theta \in \Theta}$ and $\theta_0 \in \Theta$, the derivative $\l.\dv{\theta}\r|_{\theta=\theta_0} W\l(T^\theta\r)$ exists.  Say that a social welfare function $W$ is \textbf{Paretian along paths} if for all well-behaved $\l(T^\theta\r)_{\theta \in \Theta}$ and all $\theta_0, \theta_1 \in \Theta$ with $\theta_0 < \theta_1$, $W$ satisfies the following properties:  
  \begin{enumerate}
\item \textbf{Pareto indifference along a path.}  Suppose that all agents are indifferent among all tax policies $T^\theta$ for $\theta \in \l[\theta_0,\theta_1\r]$.  Then $W\l(T^{\theta_0}\r) = W\l(T^{\theta_1}\r)$. 
\item \textbf{Weak Pareto along paths.} Suppose that, for all $\hat{\theta} \in \l[\theta_0,\theta_1\r]$ and all agents $i$, $\dv{\theta}U_i\l(\hat{\theta}\r) > 0$. Then $W\l(T^{\theta_0}\r) < W\l(T^{\theta_1}\r)$.  
\end{enumerate}
Say that a system of welfare weights $g$ \textbf{implements  social welfare function $W$} if $W$ is sufficiently differentiable and for all well-behaved families $\l(T^\theta\r)_{\theta \in \Theta}$ and all $\theta' \in \Theta$, 
\begin{align}
\label{W g equiv >}\l.\dv{\theta}\r|_{\theta=\theta'}W\l(T^\theta\r) > 0 &\Leftrightarrow \int g_i\l(T^{\theta'}\r) \l.\pdv{\theta}\r|_{\theta =\theta'} T\l(z_i\l(T^{\theta'}\r),\theta\r) \dd i <0 \textup{ and }\\
\label{W g equiv =}\l.\dv{\theta}\r|_{\theta=\theta'}W\l(T^\theta\r) = 0 &\Leftrightarrow \int g_i\l(T^{\theta'}\r) \l.\pdv{\theta}\r|_{\theta =\theta'} T\l(z_i\l(T^{\theta'}\r),\theta\r) \dd i =0.
\end{align}  
The first condition says that increasing $\theta$ is good according to the social welfare function $W$ and this is detected by the $\theta$-derivative of  $W\l(T^\theta\r)$ if and only if increasing $\theta$ is desirable according welfare weights $g$.  The second condition says that the $\theta$-derivative of $W\l(T^\theta\r)$ does not detect any change in social welfare if and only if welfare weights do not detect any change in social welfare.  

Having made the terms in the corollary precise, I now prove the corollary.  Assume that the system of welfare weights $g$ implements social welfare function $W$. Let $\l(T^\theta\r)_{\theta \in \Theta}$ be well-behaved and let $\theta_0, \theta_1 \in \Theta$ with $\theta_0 < \theta_1$, and suppose that all agents are indifferent among all tax policies $T^{\theta'}$ for $\theta' \in \l[\theta_0,\theta_1\r]$.  Then arguing as in the proof of Proposition \ref{indifference Pareto coro}, it follows that $\int g_i\l(T^{\theta'}\r) \l.\pdv{\theta}\r|_{\theta =\theta'} T\l(z_i\l(T^{\theta'}\r),\theta\r) \dd i =0$.  So by (\ref{W g equiv =}), $\l.\dv{\theta}\r|_{\theta=\theta'}W\l(T^\theta\r) = 0$, for all $\theta' \in \l[\theta_0,\theta_1\r]$.  So $W\l(T^{\theta_0}\r) = W\l(T^{\theta_1}\r)$.  So any social welfare function implemented by $g$ satisfies Pareto indifference along paths.  The argument that any social welfare function $W$ implemented by welfare weights satisfies Weak Pareto along paths, proceeds similarly, using (\ref{W g equiv >}) in the place of (\ref{W g equiv =}) to derive $\l.\dv{\theta}\r|_{\theta=\theta'}W\l(T^\theta\r)> 0$, for all $\theta' \in \l[\theta_0,\theta_1\r]$, and hence $W\l(T^{\theta_0}\r)  < W\l(T^{\theta_1}\r)$. $\square$  

\subsection{Proof of Proposition \ref{g observation}}
It is convenient to prove a stronger version of Proposition \ref{g observation}, which adds a third equivalent condition -- condition \ref{measurability 2} in Proposition \ref{g observation more detailed} below -- to conditions \ref{measurability 1} and \ref{measurability 3}. Recall that we have assumed that $g_i\l(c_i,z_i\r)$ is a smooth function of  $\l(c_i,z_i\r)$.
\begin{prop}\label{g observation more detailed}
Let $g$ and $\hat{g}$ be related as in (\ref{g g hat relation}).  Then the following conditions are equivalent:
\begin{enumerate}
\item\label{measurability 1} $g$ is structurally utilitarian.
\item\label{measurability 2}
 $\forall i \in I, \forall \hat{u}_i \in \mathbb{R}, \;\forall z_i, z_i'  \in Z,\;\;\hat{g}_i\l(\hat{u}_i,z_i\r) = \hat{g}_i\l(\hat{u}_i,z_i'\r).$
\item\label{measurability 3} $\forall i \in I, \forall \hat{u}_i \in \mathbb{R},\forall z_i \in Z, \pdv{z_i}\hat{g}_i\l(\hat{u}_i,z_i\r) =0.$ 
\end{enumerate}
\end{prop}
Proof.  First I argue that condition \ref{measurability 1} of the proposition implies condition \ref{measurability 2}.  Assume that $g$ is structurally utilitarian.   Now choose $i \in I, z_i,z_i' \in Z, \textup{ and } \hat{u}_i \in \mathbb{R}$.   Define $c_i = \hat{u}_i + v_i\l(z_i\r)$ and $c_i'= \hat{u}_i+v_i\l(z_i'\r)$.  Then observe that
\begin{align}\label{for structural utilitarianism}c_i- v_i\l(z_i\r) =\hat{u}_i= c_i'-v_i\l(z_i'\r).\end{align}  Then $\hat{g}_i\l(\hat{u}_i,z_i\r)= g_i\l(c_i,z_i\r) = g_i\l(c_i',z_i'\r) = \hat{g}_i\l(\hat{u}_i,z_i'\r)$, where the first and last equalities follow from (\ref{g g hat relation}), and the middle equality follows from (\ref{for structural utilitarianism}) and the assumption that $g$ is structurally utilitarian. It follows that condition \ref{measurability 2} of the proposition holds. 

Next I argue that condition \ref{measurability 2} implies condition \ref{measurability 1}.  So assume condition \ref{measurability 2}.  Choose $i \in I, c_i, c_i' \in \mathbb{R}, z_i,z_i' \in Z$ and $\hat{u}_i \in \mathbb{R}$ such that $\hat{u}_i=c_i-v_i\l(z_i\r)=c_i'-v_i\l(z_i'\r)$.  It follows that $
g_i\l(c_i,z_i\r)= \hat{g}_i\l(\hat{u}_i,z_i\r)=\hat{g}_i\l(\hat{u}_i,z_i'\r)= g_i\l(c_i',z_i'\r)$, where the first and last equalities follow from (\ref{g g hat relation}), and the middle equality follows from condition \ref{measurability 2} of the proposition.  This establishes condition \ref{measurability 1}.

Finally, consider the equivalence of conditions \ref{measurability 2} and \ref{measurability 3}.  First observe that our smoothness assumptions imply that condition \ref{measurability 2} implies: $\forall i \in I, \forall \hat{u}_i \in \mathbb{R},\forall z_i \in Z, \frac{\partial}{\partial  z_i}\hat{g}_i\l(\hat{u}_i,z_i\r) =0.$  Going in the other direction, the equivalence now follows from the fundamental theorem of calculus.  $\square$

\subsection{Proof of Theorem \ref{necessity theorem}}
First assume welfare weights arise from a generalized utilitarian social welfare function, meaning that they are of the form $g_i\l(c_i,z_i\r)=F'_i\l(U_i\l(c_i,z_i\r)\r) \pdv{c_i}U_i\l(c_i,z_i\r)$.  These weights are structurally utilitarian because, if, for all $c_i, c'_i, z_i, z'_i$, if $c_i-v_i\l(z_i\r)= c'_i-v_i\l(z'_i\r)$, then $U_i\l(c_i,z_i\r)= u\l(c_i-v_i\l(z_i\r)\r)=u\l(c'_i-v_i\l(z'_i\r)\r)=U_i\l(c'_i,z'_i\r)$ and $\pdv{c_i}U_i\l(c_i,z_i\r) = u'\l(c_i-v_i\l(z_i\r)\r)=u'\l(c'_i-v_i\l(z'_i\r)\r)=\pdv{c_i}U_i\l(c'_i,z'_i\r)$.  So if $c_i-v_i\l(z_i\r)= c'_i-v_i\l(z'_i\r)$, then $g_i\l(c_i,z_i\r)=g_i\l(c'_i,z'_i\r)$.    

Going in the other direction,  by Proposition \ref{g observation}, structural utilitarianism is equivalent to the requirement that, holding fixed agent characteristics $\l(x_i,y_i\r)$, welfare weights are a function of $\hat{u}_i = c_i-v_i\l(z_i\r)$, so that, assuming structural utilitarianism, we can write $g_i\l(c_i,z_i\r)= g\l(c_i,z_i,x_i,y_i\r) = \hat{g}\l(\hat{u}_i,x_i,y_i\r)=\hat{g}_i\l(\hat{u}_i\r)$.  Define the function $w_i\l(\hat{u}_i\r)= w\l(\hat{u}_i,x_i,y_i\r)$ by $w_i\l(\hat{u}^0_i\r) = \int_0^{\hat{u}^0_i} \hat{g}_i\l(\hat{u}_i\r) \dd \hat{u}_i.$  Now define the Function $F: \mathbb{R} \times X \times Y\rightarrow \mathbb{R}$ by $F\l(v_i,x_i,y_i\r) = w\l(u^{-1}\l(v_i\r),x_i,y_i\r)$, where $u^{-1}\l(\cdot\r)$ is the inverse of $u\l(\cdot\r)$.  If $x_i$ and $y_i$ are not discrete, the smoothness of $w$ and $u$ imply that $F$ is smooth.  If $x_i$ and $y_i$ are discrete, $w$ is smooth in its first argument and hence $F$ is smooth in $v_i$.  Let $F_i\l(v_i\r)=F\l(v_i,x_i,y_i\r)$ and define $W_i\l(c_i,z_i\r)=F_i\l(U_i\l(c_i,z_i\r)\r)$.  We have $W_i\l(c_i,z_i\r)=F_i\l(U_i\l(c_i,z_i\r)\r)= w_i\l(u^{-1}\l(u\l(c_i-v_i\l(z_i\r)\r)\r)\r)=w_i\l(c_i-v_i\l(z_i\r)\r)$.  Note that, from the above, we have $g_i\l(c_i,z_i\r)= \hat{g}_i\l(c_i-v_i\l(z_i\r)\r) = w'_i\l(c_i-v_i\l(z_i\r)\r)= \pdv{c_i} W_i\l(c_i,z_i\r)=F'_i\l(U_i\l(c_i,z_i\r)\r) \pdv{c_i}U_i\l(c_i,z_i\r)$.  So the weights arise from a generalized utilitarian social welfare function.  $\square$

\subsection{Proof of Corollary \ref{noncycle corollary}}
Suppose that welfare weights $g$ are structurally utilitarian.  It follows from Theorem \ref{necessity theorem} that welfare weights are of the form $g_i\l(c_i,z_i\r) = F'_i\l(U_i\l(c_i,z_i\r)\r) \pdv{c_i} U_i\l(c_i,z_i\r)$ for $F_i\l(u_i\r) = F\l(u_i,x_i,y_i\r)$ for some $F$.  So for the social welfare function $W\l(T\r)=- \int F_i\l(U_i\l(c_i\l(T\r),z_i\l(T\r)\r)\r) \dd i$, the envelope theorem implies that, for all well-behaved families $\l(T^\theta\r)_{\theta \in \Theta}$ and $\theta_0 \in \Theta$, $\l.\dv{\theta}\r|_{\theta=\theta_0} W\l(T^\theta\r) = -\int g_i\l(T^{\theta_0}\r) \l.\pdv{\theta}\r|_{\theta =\theta_0} T_i\l(z_i\l(T^{\theta_0}\r),\theta\r) \dd i$. $\square$

\subsection{\label{individualized taxes necessity theorem appendix}Proof of Theorem \ref{individualized taxes necessity theorem}}

\subsubsection{\label{individualized main argument appendix}Main argument}

What follows is a more formal version of the argument in the main text.  Assume that welfare weights are not structurally utilitarian.  It follows from Proposition \ref{g observation} that there exists $j \in I, \hat{u}^* \in \mathbb{R}, z^* \in Z,$ such that $\pdv{z_j}\hat{g}_j\l(\hat{u}^*,z^*\r) \neq0$.  Smoothness of the primitives implies that we can choose $z^*$ so that $z^* >0$.  Assume that  $\pdv{z_j}\hat{g}_j\l(\hat{u}^*,z^*\r) <0$.  (The argument would be similar if we assumed instead that $\pdv{z_j}\hat{g}_j\l(\hat{u}^*,z^*\r) >0$.)  Our smoothness assumptions then imply that there exists a non-degenerate\footnote{By a non-degenerate closed interval, I mean a closed interval which is not equal to a single point.} closed interval of agents $S$, which is a proper subset of $I=\l[0,1\r]$, such that, for all agents $i \in S$, $\pdv{z_i}\hat{g}_i\l(\hat{u}^*,z^*\r) <0$.\footnote{Of course, it is possible that $\pdv{z_i}\hat{g}_i\l(\hat{u}^*,z^*\r) <0$ for all $i \in \l[0,1\r]$, but in this case there is also a closed interval $S$, which is a proper subset of $\l[0,1\r]$, on which this property holds.}  Let $O$ and $Q$ be two other non-degenerate closed intervals contained in $\l[0,1\r]$, such that $S, O,$ and $Q$ are pairwise disjoint.  Now consider a doubly parameterized family of tax policies $\l(T^{\theta,\epsilon}\r)_{\theta \in \Theta, \epsilon \in E}$, where $\Theta = \l[\underline{\theta},\overline{\theta}\r]$ for some $\underline{\theta} < \overline{\theta}$ and $E=\l[-\bar{\epsilon},\bar{\epsilon}\r]$ for some $\bar{\epsilon} >0$, and  which takes the following form:  
\begin{align}\label{cases theta epsilon tax policy}
T^{\theta,\epsilon}_i\l(z_i\r) =\begin{cases}\tau_i\l(\theta\r) z_i + \kappa_i\l(\theta\r) +\epsilon t_S, & \textup{if $i \in S$},\\
- \epsilon t_O,& \textup{if $i \in O$},\\
\bar{\tau}\l(\theta,\epsilon\r)z_i + \bar{\kappa}_i\l(\theta,\epsilon\r), & \textup{if $i \in Q$}, \\
0, & \textup{otherwise.}
\end{cases}
\end{align}
Above $\tau_i\l(\theta\r)$ is a personalized marginal tax rate for agents in $i$ in $S$, and 
 $\bar{\tau}\l(\theta,\epsilon\r)$ is a marginal tax rate which is not personalized on $Q$; both $\tau_i\l(\theta\r)$ and $\bar{\tau}\l(\theta,\epsilon\r)$  depend on parameter values. $\kappa_i\l(\theta\r)$ and $\bar{\kappa}_i\l(\theta,\epsilon\r)$ are personalized lumpsum taxes that depend on parameters. $t_S$ and $t_O$ are positive real numbers, so that $\epsilon t_S$ and $-\epsilon t_O$ are lumpsum taxes as well. I assume that the map $\l(i,\theta\r) \mapsto \tau_i\l(\theta\r)$ is smooth on the domain $S \times \Theta$ and that the map $\l(\theta,\epsilon\r) \mapsto \bar{\tau}\l(\theta,\epsilon\r)$ is smooth on the domain $\Theta \times E$.  Moreover, I assume that there exists $\theta_0 \in \l(\underline{\theta},\overline{\theta}\r)$ such that, for all $i \in S$, $\tau_i\l(\theta_0\r)= 1-v'_i\l(z^*\r)$ and, for all $\theta \in \Theta$, $\tau_i'\l(\theta\r) >0$.\footnote{We allow for the possibility that $\tau_i\l(\theta_0\r) < 0$.} In what follows, let $\hat{U}_i\l(\theta,\epsilon\r)=\hat{U}_i\l(T^{\theta,\epsilon}\r)=z_i\l(\theta,\epsilon\r)-T^{\theta,\epsilon}_i\l(z_i\l(\theta,\epsilon\r)\r)-v_i\l(z_i\l(\theta,\epsilon\r)\r)$ be $i$'s utility in response to $T^{\theta,\epsilon}$, using the representation that omits the outer utility function $u\l(\cdot\r)$, and note that $g_i\l(\theta,\epsilon\r)=g_i\l(T^{\theta,\epsilon}\r)=\hat{g}_i\l(\hat{U}_i\l(\theta,\epsilon\r),z_i\l(\theta,\epsilon\r)\r)$.  When an agent $i$ in $S$ faces tax policy $T^{\theta_0,0}$, they will solve the problem $\max_{z_i} \l(1-\tau_i\l(\theta_0\r)\r) z_i - \kappa_i\l(\theta_0\r) - v_i\l(z_i\r)$.  It follows from the construction of $\tau_i\l(\theta_0\r)$ and the fact that $v_i\l(z_i\r)$ is strictly convex that $z_i=z^*$ uniquely satisfies the agent's first order condition when $\l(\theta,\epsilon\r) = \l(\theta_0,0\r)$, namely, $\l(1-\tau\l(\theta_0\r)\r) - v'_i\l(z_i\r)=0$.   Because agents' objective is strictly concave, it follows that $z_i=z^*$ is the unique optimum for all agents $i \in S$ when facing tax policy $T^{\theta_0,0}$, so that $z_i\l(\theta_0,0\r)=z^*$ for all $i \in S$.  For all $i \in S$, define the function $\kappa_i\l(\theta\r)$ in (\ref{cases theta epsilon tax policy}) to solve:  
\begin{align}\label{kappa i definition theta}
\l(1-\tau_i\l(\theta\r)\r)z_i\l(\theta,0\r)- v_i\l(z_i\l(\theta,0\r)\r) - \kappa_i\l(\theta\r) =\hat{u}^*, \;\;\;\; \forall \theta \in \Theta.
\end{align}   
That is, the lumpsum tax $\kappa_i\l(\theta\r)$ is chosen so as the keep the agents' (in $S$) utility fixed at $\hat{u}^*$ when the agent faces tax policies of the form $T^{\theta,0}$ as $\theta$ changes -- where we measure utility via the representation $\hat{U}_i\l(T^{\theta,0}\r)$ that excludes the outer utility function $u\l(\cdot\r)$.  Note that we can freely define $\kappa_i\l(\theta\r)$ in this way because the optimal income $z_i\l(\theta,0\r)$ depends only on the marginal tax rate $\tau_i\l(\theta\r)$ and not on the lumpsum tax $\kappa_i\l(\theta\r)$.  Note, moreover, that,  for any $\epsilon \in E$, $\theta \in \Theta$, and $i \in S$, $i$'s utility, when facing $T^{\theta,\epsilon}$, is $\hat{U}_i\l(\theta,\epsilon\r) = \hat{u}^*-\epsilon t_S$, which does not depend on $\theta$.  So, holding $\epsilon$ fixed, each agent $i \in S$ is indifferent as $\theta$ varies.  Likewise, for all $i \in Q$, define $\bar{\kappa}_i\l(\theta,\epsilon\r)$ to satisfy the following equation:  
\begin{align}\label{bar kappa definition}
\l(1-\bar{\tau}\l(\theta,\epsilon\r)\r)z_i\l(\theta,\epsilon\r)- v_i\l(z_i\l(\theta,\epsilon\r)\r) - \bar{\kappa}_i\l(\theta,\epsilon\r) =0, \;\;\;\; \forall \theta \in \Theta, \forall \epsilon \in E.
\end{align}  
That is, the lumpsum tax $\bar{\kappa}_i\l(\theta,\epsilon\r)$ is selected to keep the utility $\hat{U}_i\l(\theta,\epsilon\r)$ of all agents $i \in Q$ equal to zero as $\theta$ and $\epsilon$ vary.  Again, observe that $z_i\l(\theta,\epsilon\r)$ only depends on the marginal tax rate $\bar{\tau}\l(\theta,\epsilon\r)$ and not on the lumpsum tax $\bar{\kappa}_i\l(\theta,\epsilon\r)$.  Given the above, it follows by construction that, holding $\epsilon$ fixed, \textit{all} agents are indifferent, as $\theta$ varies in $T^{\theta,\epsilon}$.  So, it follows from part \ref{indifference along paths} of Proposition \ref{indifference Pareto coro} -- Pareto indifference along paths -- that 
\begin{align}\label{indifference theta 0 theta 1 appendix}
T^{\theta_0,\epsilon} \sim^g T^{\theta_1,\epsilon}, \;\;\;\; \forall \epsilon \in E,
\end{align}
where $\theta_1$, satisfying $\theta_0 < \theta_1$, is a value of $\theta$ that we now select.  In particular, it follows from the facts that $\pdv{z_i} \hat{g}_i\l(z^*,\hat{u}^*\r)<0$ and $z_i\l(\theta_0,0\r)=z^*$ for all $i \in S$ and the smoothness of the primitives of the model that if we choose $\theta_1$ sufficiently close to $\theta_0$, 
\begin{align}\label{change of g in theta}
\pdv{z_i} \hat{g}_i\l(\hat{U}_i\l(\theta,0\r),z_i\l(\theta,0\r)\r)= \pdv{z_i} \hat{g}_i\l(\hat{u}^*,z_i\l(\theta,0\r)\r) <0, \;\;\; \forall \theta \in \l[\theta_0,\theta_1\r], \forall i \in S.\end{align}
So let us choose $\theta_1$ so that (\ref{change of g in theta}) is satisfied.  Moreover, since $z_i\l(\theta_0,0\r)=z^* > 0, \forall i \in S$, we may assume that $\theta_1$ is chosen sufficiently close to $\theta_0$ that, for all $i \in S$ and $\theta \in \l[\theta_0,\theta_1\r], z_i\l(\theta,0\r) >0$.  

For any $\theta \in \Theta$, define $g_S\l(\theta,0\r)= \int_S g_i\l(\theta,0\r) \dd i$ and $g_O\l(\theta,0\r)= \int_O g_i\l(\theta,0\r) \dd i$.  It follows from the fact that $\hat{U}_i\l(\theta,0\r)=\hat{u}^*, \forall \theta \in \Theta, \forall i \in S$,  (\ref{change of g in theta}), and the assumption that $\tau_i'\l(\theta\r) > 0, \forall \theta \in \Theta, \forall i \in S$, which, given that $z_i\l(\theta,0\r) > 0, \forall \theta \in \l[\theta_0,\theta_1\r], \forall i \in S$, implies that $\pdv{\theta}z_i\l(\theta,0\r) < 0, \forall \theta \in \l[\theta_0,\theta_1\r], \forall i \in S$, that
\begin{align}\label{g S is increasing}
\pdv{\theta}g_S\l(\theta,0\r) >0,\;\;\; \forall \theta \in \l[\theta_0,\theta_1\r].\end{align}
Choose $\theta' \in \l(\theta_0,\theta_1\r)$ and suppose that the positive numbers $t_S$ and $t_O$ in (\ref{cases theta epsilon tax policy}) were selected to satisfy 
\begin{align}\label{gStSgOtO}
g_S\l(\theta',0\r) t_S = g_O\l(\theta',0\r)t_O. 
\end{align}
Then, writing $T\l(z_i,\theta,\epsilon\r)=T^{\theta,\epsilon}\l(z_i\r)$, we have:
\begin{align}\label{welfare inequalities for the individualized case}
\begin{split}
&\int g_i\l(\theta_0,0\r) \l.\pdv{\epsilon}\r|_{\epsilon=0} T_i\l(z_i\l(\theta_0,0\r),\theta_0,\epsilon\r) \dd i\\ =&\; g_S\l(\theta_0,0\r)t_S- g_O\l(\theta_0,0\r)t_O 
 + \int_Q \l( \l.\l[\pdv{\epsilon}\r|_{\epsilon=0} \bar{\tau}\l(\theta_0,\epsilon\r)\r] z_i\l(\theta_0,0\r)+ \l.\pdv{\epsilon}\r|_{\epsilon=0}\bar{\kappa}_i\l(\theta_0,\epsilon\r)\r)\dd i\\
 =&\; g_S\l(\theta_0,0\r)t_S- g_O\l(\theta_0,0\r)t_O 
 + \int_Q -\l.\pdv{\epsilon}\r|_{\epsilon=0} \hat{U}_i\l(\theta_0,\epsilon\r)\dd i\\
  =&\; g_S\l(\theta_0,0\r)t_S- g_O\l(\theta_0,0\r)t_O<0,  
  \end{split}
\end{align}
where the second equality follows from the envelope theorem, and the third equality follows from the fact that, by (\ref{bar kappa definition}), the utility of all agents in $Q$ is held fixed as $\epsilon$ varies in $T^{\theta_0,\epsilon}$, so that, for all $i \in Q$, $\l.\pdv{\epsilon}\r|_{\epsilon=0} \hat{U}_i\l(\theta_0,\epsilon\r)=0$.  The inequality follows from (\ref{gStSgOtO}), and the facts that $g_O\l(\theta,0\r)$ is constant in $\theta$, that, by (\ref{g S is increasing}), $g_S\l(\theta,0\r)$ is increasing in $\theta$, and that $\theta_0 < \theta'$.  Using similar arguments, 
\begin{align}\label{welfare inequalities for the individualized case 2}
\int g_i\l(\theta_1,0\r) \l.\pdv{\epsilon}\r|_{\epsilon=0} T_i\l(z_i\l(\theta_1,0\r),\theta_1,\epsilon\r) \dd i &= g_S\l(\theta_1,0\r)t_S- g_O\l(\theta_1,0\r)t_O>0,
\end{align}
The reason that the the inequality in (\ref{welfare inequalities for the individualized case 2}) points in the opposite direction of the inequality in (\ref{welfare inequalities for the individualized case}) is that, whereas $\theta_0 < \theta'$, $\theta_1 > \theta'$.  
It follows from (\ref{welfare inequalities for the individualized case}), (\ref{welfare inequalities for the individualized case 2}),  and the local improvement principle -- Proposition \ref{local improvement principle proposition} -- that\begin{align}\label{conflicting inequalities + appendix}
\begin{split}
T^{\theta_0,0} \prec^g T^{\theta_0,\epsilon},\\
T^{\theta_1,0} \succ^g T^{\theta_1,\epsilon},
\end{split} \;\;\; \textup{ for sufficiently small $\epsilon > 0$.}
\end{align}
Putting (\ref{indifference theta 0 theta 1 appendix}) and (\ref{conflicting inequalities + appendix}), together, we have that for sufficiently small $\epsilon >0$, 
\begin{align}\label{very first preference cycle}
T^{\theta_0,0} \prec^g T^{\theta_0,\epsilon} \sim T^{\theta_1,\epsilon} \prec^g T^{\theta_1,0} \sim^g T^{\theta_0,0}.
\end{align} 
So, on the assumption that welfare weights are not structurally utilitarian, we have constructed a social preference cycle.  

The last step is to show that revenue can be held fixed across the tax policies in the cycle.  This is achieved via the selection of $\bar{\tau}\l(\theta,\epsilon\r)$ in (\ref{cases theta epsilon tax policy}).  For any marginal tax rate $\tau$, write $z_i\l(\tau\r)$ to be the income that $i$ would earn, if $i$ faces the tax policy $T\l(z\r)=\tau z$, or, in other words, if $i$ faces a constant marginal tax rate of $\tau$.  It follows that, for all $i \in Q$, we can write $z_i\l(\bar{\tau}\l(\theta,\epsilon\r)\r) = z_i\l(\theta,\epsilon\r)$ because every agent $i \in Q$ faces the constant marginal tax rate $\bar{\tau}\l(\theta,\epsilon\r)$ under tax policy $T^{\theta,\epsilon}$.  Let $R_Q\l(\theta,\epsilon\r)$ be the revenue raised from agents in $Q$ by tax policy $T^{\theta,\epsilon}$.  Then we have
\begin{align}\label{revenue in Q}
\begin{split}
R_Q\l(\theta,\epsilon\r) =& \int_Q T\l(z_i\l(\theta,\epsilon\r),\theta,\epsilon\r) \dd i = \int_Q \l[\bar{\tau}\l(\theta,\epsilon\r)z_i\l(\theta,\epsilon\r) + \bar{\kappa}_i\l(\theta,\epsilon\r)\r]  \dd i \\
= &\int_Q \l[\bar{\tau}\l(\theta,\epsilon\r)z_i\l(\theta,\epsilon\r) + \l(1-\bar{\tau}\l(\theta,\epsilon\r)\r)z_i\l(\theta,\epsilon\r)- v_i\l(z_i\l(\theta,\epsilon\r)\r) \r]  \dd i\\
=&\int_Q \l[z_i\l(\theta,\epsilon\r) - v_i\l(z_i\l(\theta,\epsilon\r)\r) \r]  \dd i,
\end{split}
\end{align} 
where the third equality follows from (\ref{bar kappa definition}).    Next, for any marginal tax rate $\tau$, define $\tilde{R}_Q\l(\tau\r)$ by
\begin{align*}
\tilde{R}_Q\l(\tau\r) = \int_Q \l[z_i\l(\tau\r) - v_i\l(z_i\l(\tau\r)\r) \r]  \dd i. 
\end{align*}
Then it follows from (\ref{revenue in Q}) and the fact that $z_i\l(\bar{\tau}\l(\theta,\epsilon\r)\r) = z_i\l(\theta,\epsilon\r)$ that $\tilde{R}_Q\l(\bar{\tau}\l(\theta,\epsilon\r)\r)= R_Q \l(\theta,\epsilon\r)$.  Since we assume that, in the absence of taxes, all agents earn positive income (see Section \ref{model section}), there exists a positive marginal tax rate $\tau_0$, which is sufficiently small that, for all $i \in Q$, $z_i\l(\tau_0\r) >0$.\footnote{The assumption that, in the absence of taxes, all agents earn positive income, is not necessary for the proof.  In the absence of this assumption, we could instead select $\tau_0$ to be a sufficiently small negative marginal tax rate that, for all $i \in Q$, $z_i\l(\tau_0\r) >0$.  Then the proof would proceed in the same way as below except that $\tilde{R}_Q'\l(\bar{\tau}\l(\theta,\epsilon\r)\r) >0$ rather than $\tilde{R}_Q'\l(\bar{\tau}\l(\theta,\epsilon\r)\r) <0$.  However what matters for the argument is only that $\tilde{R}_Q'\l(\bar{\tau}\l(\theta,\epsilon\r)\r) \neq 0$.}  From agent $i$'s first order condition, when facing marginal tax rate $\tau_0$, we have that, for all $i \in Q$, $0=\l(1-\tau_0\r) - v'_i\l(z_i\l(\tau_0\r)\r) < 1-v'_i\l(z_i\l(\tau_0\r)\r)$.  Assume that $\bar{\tau}\l(\theta_0,0\r)=\tau_0$.  Define $R_{-Q}\l(\epsilon,\theta\r)= \int_{I \setminus Q} T_i\l(z_i\l(\theta,\epsilon\r),\theta,\epsilon\r) \dd i$ to be the revenue raised by tax policy $T^{\theta,\epsilon}$ from all agents not in $Q$.  Now consider the condition:
\begin{align}\label{implicit definition bar tau}
\tilde{R}_Q\l(\bar{\tau}\l(\theta,\epsilon\r)\r)+ R_{-Q}\l(\theta,\epsilon\r) = \tilde{R}_Q\l(\tau_0\r)+ R_{-Q}\l(\theta_0,0\r).
\end{align} 
Observe that $\tilde{R}_Q'\l(\bar{\tau}\l(\theta_0,0\r)\r) = \int_Q z_i'\l(\tau_0\r)\l[1-v'_i\l(z_i\l(\tau_0\r)\r)\r] \dd i < 0.$\footnote{This inequality follows from the facts that, by our assumptions above imply that, for all $i \in Q$ (i) $z_i\l(\tau_0\r) >0$, so that $z'_i\l(\tau_0\r) <0$, and that (ii) $1-v'_i\l(z_i\l(\tau_0\r)\r)>0$.} It follows from the implicit function theorem that the function $\bar{\tau}\l(\theta,\epsilon\r)$ is uniquely determined in a neighborhood of $\l(\theta_0,0\r)$ by $\bar{\tau}\l(\theta_0,0\r)=\tau_0$ and (\ref{implicit definition bar tau}).  Redefining $\bar{\epsilon}$ to be sufficiently small and $\overline{\theta}$ and $\underline{\theta}$ to be sufficiently close to $\theta_0$ if necessary, and assuming that $\theta_1$ was chosen sufficiently close to $\theta_0$ so that $\theta_0 < \theta_1 < \bar{\theta}$ still holds, we may assume that we have thus defined $\bar{\tau}\l(\theta,\epsilon\r)$ on all of $\Theta \times E$, and moreover such that $z_i\l(\theta,\epsilon\r) >0$ for all $i$ in $Q$, $\theta \in \Theta$, and $\epsilon \in E$ (since $z_i\l(\theta_0,0\r) = z_i\l(\tau_0\r) > 0, \forall i \in Q$ and $Q$ is compact).  Note now that (\ref{implicit definition bar tau}) implies that the revenue of $T^{\theta,\epsilon}$ is held constant as $\theta$ and $\epsilon$ vary.  This completes the proof. $\square$  

\subsubsection{\label{well behaved appendix individualized}Well-behavedness of $\l(T^{\theta,\epsilon}\r)$}

Here I verify that the family $\l(T^{\theta,\epsilon}\r)$ in (\ref{cases theta epsilon tax policy}) above is well-behaved (see Sections \ref{parameterized families subsection} and \ref{wb appendix}), as this is required for Propositions \ref{local improvement principle proposition} and \ref{indifference Pareto coro}.   I begin by verifying the first condition for well-behavedness.  Existence and uniqueness of $z_i\l(\theta,\epsilon\r)$ are straightforward to establish.\footnote{Existence and uniqueness follow from the assumptions of Section \ref{standard aspects section}, the fact that when facing a linear tax policy, agents' objectives are strictly concave, the selection of the marginal tax rates $\tau_i\l(\theta_0\r)$ and $\tau_0$, and the construction of $\tau_i\l(\theta\r)$ and $\bar{\tau}\l(\theta,\epsilon\r)$ using the implicit function theorem.}  That  $z_i\l(\theta,\epsilon\r) > 0$ for all $i$ in $S$ and $Q$ was established in the course of the proof (noting that $z_i\l(\theta,\epsilon\r)=   z_i\l(\theta,0\r), \forall i \in S$), and for $i$ not in $S$ or $Q$, $z_i\l(\theta,\epsilon\r) > 0$  follows from the assumption that, when facing a zero marginal tax rate, all agents select a positive income (see Section \ref{standard aspects section}).   That each agent's second order condition holds with a strict inequality follows from the fact that $v''_i > 0$ and $u' > 0$ hold everywhere and that all agents face a tax policy that is linear in $z$ (possibly with a zero marginal tax rate) under $T^{\theta,\epsilon}$.  This establishes that $\l(T^{\theta,\epsilon}\r)$ satisfies the first condition required for well-behavedness.   

To establish the second condition, I appeal to the following observation. 
\begin{obs}
\label{smoothness observation} The maps $\l(i,\theta\r) \mapsto \tau_i\l(\theta\r)$ and $\l(i,\theta\r) \mapsto \kappa_i\l(\theta\r)$ are smooth on $S \times \Theta$; $\l(\theta,\epsilon\r) \mapsto \bar{\tau}\l(\theta,\epsilon\r)$ is smooth on $\Theta \times E$; and the map $\l(i,\theta,\epsilon\r) \mapsto \kappa_i\l(\theta,\epsilon\r)$ is smooth on $Q \times \Theta \times E$.
\end{obs} 
The map $\l(i,\theta\r) \mapsto \tau_i\l(\theta\r)$ is smooth on $S \times \Theta$ by assumption.\footnote{This is consistent with the other assumptions made on $\tau_i\l(\theta\r)$. In particular, I assumed that, for all $i \in S$, $\tau_i\l(\theta_0\r)= 1-v'_i\l(z^*\r)$, and that, for all $\theta \in \Theta$, $\tau'_i\l(\theta\r) > 0$.  So for example, if I had specifically defined $\tau_i\l(\theta\r) = 1-v'_i\l(z^*\r) + \l(\theta-\theta_0\r)$ on $S \times \Theta$, $\l(i,\theta\r) \mapsto \tau_i\l(\theta\r)$ would have satisfied these properties, and, moreover, would be smooth on $S \times \Theta$, since the assumptions of Section \ref{individualized version section} imply that $i \mapsto v'_i\l(z^*\r)$ is smooth.} The map $\l(i,\theta\r) \mapsto \kappa_i\l(\theta\r)$ is smooth on $S\times \Theta$ because it is defined by (\ref{kappa i definition theta}) and all of the other functions in (\ref{kappa i definition theta}) are smooth.\footnote{\label{smooth z footnote} In particular, $\l(i, \theta\r) \mapsto z_i\l(\theta,0\r)$ is smooth because the latter is characterized by the implicit function theorem applied to $i$'s first order condition and the the functions that feature in the first order condition are smooth in $\l(i,\theta\r)$.}  The map $\l(\theta,\epsilon\r) \mapsto \bar{\tau}\l(\theta,\epsilon\r)$ is smooth because it is defined by the implicit function theorem via equation (\ref{implicit definition bar tau}) and the other functions in (\ref{implicit definition bar tau}) are smooth.  Finally, $\l(i,\theta,\epsilon\r) \mapsto \kappa_i\l(\theta,\epsilon\r)$ is smooth on $Q \times \Theta \times E$ because it is defined by (\ref{bar kappa definition}) and the other functions in (\ref{bar kappa definition}) are smooth.\footnote{Again, the map $\l(i,\theta,\epsilon\r) \mapsto z_i\l(\theta,\epsilon\r)$ is smooth for reasons similar to those explained in footnote \ref{smooth z footnote} of the appendix. }  

That, for all $i$, $\l(z,\theta,\epsilon\r) \mapsto T_i\l(z,\theta,\epsilon\r)$ is smooth follows from (\ref{cases theta epsilon tax policy}) and Observation \ref{smoothness observation}.  Recall that $S, O,$ and $Q$ are assumed in Section \ref{individualized main argument appendix} to be  pairwise disjoint closed intervals.  It then follows from (\ref{cases theta epsilon tax policy}) and Observation \ref{smoothness observation} that the map $\l(i,z,\theta,\epsilon\r) \mapsto T_i\l(z,\theta,\epsilon\r)$ only fails to be smooth when $i$ is one of the six endpoints of these three intervals.  This establishes the second condition required for the well-behavedness of $\l(T^{\theta,\epsilon}\r)$.

\subsection{\label{libertarian calculations appendix}Calculations from Section \ref{detailed example section}}
That the revenue of $T^{\theta,\epsilon}$ is $\frac{1}{4}$, for all $\theta$ and $\epsilon$, is verified by the following calculation: 
\begin{align*}
R\l(T^{\theta,\epsilon}\r)=& \underbrace{\frac{1}{2}\l[z\l(\theta\r)\theta+\kappa\l(\theta\r)+\epsilon\r]}_{\textup{revenue from type $A$ agents}} +\underbrace{\frac{1}{2}\l[z\l(\sqrt{1-\theta^2}\r)\sqrt{1-\theta^2}+\kappa\l(\theta\r)-\epsilon\r]}_{\textup{revenue from type $B$ agents.}}\\
=& \frac{1}{2}\l[\l(1-\theta\r)\theta + \frac{1}{2}\l(1-\theta\r)^2\r]+ \frac{1}{2}\l[\l(1-\sqrt{1-\theta^2}\r)\sqrt{1-\theta^2}+\frac{1}{2}\l(1-\sqrt{1-\theta^2}\r)^2
\r]\\
=& \frac{1}{2}\l[\frac{1}{2}\l(1-\theta\r)\l(1+\theta\r)\r]+ \frac{1}{2}\l[\frac{1}{2}\l(1-\sqrt{1-\theta^2}\r)\l(1+\sqrt{1-\theta^2}\r)\r]
= \frac{1}{4}\l(1-\theta^2\r)+ \frac{1}{4}\theta^2 =\frac{1}{4}.
\end{align*}

The above calculation also implies that, at $T^{\theta,\epsilon}$, the total tax paid by a type $A$ agent is $\frac{1}{2}\l(1-\theta^2\r) + \epsilon$ and the total tax paid by a type $B$ agent is $\frac{1}{2}\theta^2 -\epsilon$.  So as $\theta$ rises from $\theta_0=\sqrt{\frac{1}{3}}$ to $\theta_1=\sqrt{\frac{2}{3}}$, the total tax paid by a type $A$ agent falls from $\frac{1}{3} + \epsilon$ to $\frac{1}{6}+\epsilon$ while the total tax paid by a type $B$ agent rises from $\frac{1}{6}-\epsilon$ to $\frac{1}{3} - \epsilon$.

A formal derivation that $T^{\theta_0,0}  \succ^g T^{\theta_0,\epsilon}$ for sufficiently small $\epsilon>0$ is as follows.
\begin{align*}
 \int^1_0 g_i\l(\theta_0,0\r) \l.\pdv{\epsilon}\r|_{\epsilon =0}T_i\l(z_i\l(T\l(\theta_0,0\r)\r),\theta_0,\epsilon \r) \dd i  = &\underbrace{\int_0^{\frac{1}{2}}\l[\tilde{g}\l(\frac{1}{3}\r) \times 1\r] \dd i}_{\textup{type $A$ agents}} + \underbrace{\int_{\frac{1}{2}}^1 \l[\tilde{g}\l(\frac{1}{6}\r) \times \l(-1\r)\r] \dd i}_{\textup{type $B$ agents}} \\ =& \frac{1}{2} \tilde{g}\l(\frac{1}{3}\r) - \frac{1}{2} \tilde{g}\l(\frac{1}{6}\r) > 0. \end{align*} So by Proposition \ref{local improvement principle proposition} -- the local improvement principle -- it follows that $T^{\theta_0,0}  \succ^g T^{\theta_0,\epsilon}$ for sufficiently small $\epsilon>0$. 
 
Similarly, $\int^1_0 g_i\l(\theta_1,0\r) \l.\pdv{\epsilon}\r|_{\epsilon =0}T_i\l(z_i\l(\theta_1,0\r),\theta_1,\epsilon \r)\dd i = \frac{1}{2} \tilde{g}\l(\frac{1}{6}\r) - \frac{1}{2} \tilde{g}\l(\frac{1}{3}\r) < 0$, and, again by Proposition \ref{local improvement principle proposition}, $T^{\theta_1,0} \prec^g T^{\theta_1,\epsilon}$, for sufficiently small $\epsilon >0$.

\subsection{Proof of Corollary \ref{convex tax policy corollary}}
Assume that $g$ is not structurally utilitarian.  It follows from Proposition \ref{g observation} that there exists an agent $j \in \l(0,1\r)$, $z^* \in Z$ with $z^* > 0$ and $\hat{u}^* \in \mathbb{R}$ and such that 
\begin{align}\label{pd g not z}
\pdv{z_j}\hat{g}_j\l(\hat{u}^*,z^*\r) \neq 0.
\end{align}
We can assume that $j$ is in the interior of $I=\l[0,1\r]$ and $z^* >0$ because of the smoothness of the primitives.  Choose a smooth strictly convex tax policy $T$, with moreover $T''\l(z_i\r) > 0, \forall z_i$,  such that (i) $T'\l(z^*\r) = 1-v'_j\l(z^*\r)$, (ii) $T'\l(0\r)$ is sufficiently small (or negative if $1-v'_j\l(z^*\r)<0$) such that all agents would earn a positive income in response to $T$ -- recall that in the absence of taxes, all agents earn a positive income (see Section \ref{standard aspects section}) --, and (iii) $\lim_{z \rightarrow \infty} T'\l(z\r) > 0$.  These assumptions, together with the strict convexity of $v_i\l(z_i\r)$ and the assumption that $v'_i\l(z_i\r) >1$ for sufficiently large $z_i$ (see Section \ref{standard aspects section}), imply that $T$ is regular. (See Section \ref{wb appendix non-individualized} for the requirements for regularity.) It follows from property (i) that $z_j\l(T\r)=z^*$.   By the appropriate choice of a lumpsum transfer in $T$, we can ensure that $\hat{U}_j\l(T\r)=\hat{u}^*$.  (\ref{pd g not z}) together with the smoothness of the primitives and of $T$ now ensure that if we select a sufficiently small interval $\l(i_a,i_b\r)$ containing $j$, then either (\ref{negative derivative integral}) or (\ref{positive derivative integral}) holds.  $\square$ 

\subsection{Omitted details from the proof of Lemma \ref{alternative condition lemma}}
Here I present the details of the proof of Lemma \ref{alternative condition lemma} that were omitted in the main text: the expression for the overlapping term  $C$ discussed in the text, and the proof of conditions (\ref{A+C})-(\ref{B+C}).  First, I present the expression for the term $C$, which I will prove is the overlapping term below:
\begin{align}\label{definition of C}
\begin{split}
C=& \int \l(-\pdv{\hat{u}_i}\hat{g}_i\l(\hat{U}_i\l(\theta_0,\epsilon_0\r),z_i\l(\theta_0,\epsilon_0\r)\r)\l.\pdv{\epsilon}\r|_{\epsilon=\epsilon_0} T\l(z_i\l(\theta_0,\epsilon_0\r),\theta_0,\epsilon\r) \l.\pdv{\theta}\r|_{\theta = \theta_0} T\l(z_i\l(\theta_0,\epsilon_0\r),\theta,\epsilon_0\r) \r.\\
&+   g_i\l(\theta_0,\epsilon_0\r)\l[ -\frac{\l.\pdv[2]{}{\theta}{z_i}\r|_{\theta = \theta_0,z_i=z_i\l(\theta_0,\epsilon_0\r)} T\l(z_i,\theta,\epsilon_0\r)\l.\pdv[2]{}{\epsilon}{z_i}\r|_{\epsilon=\epsilon_0,z_i=z_i\l(\theta_0,\epsilon_0\r)}T\l(z_i,\theta_0,\epsilon\r)}{v''_i\l(z_i\l(\theta_0,\epsilon_0\r)\r)+\l.\pdv[2]{z_i}\r|_{z_i=z_i\l(\theta_0,\epsilon_0\r)}T\l(z_i,\theta_0,\epsilon_0\r)}\r.\\&\l.+\l. \l.\pdv[2]{}{\epsilon}{\theta}\r|_{\epsilon=\epsilon_0,\theta = \theta_0} T\l(z_i\l(\theta_0,\epsilon_0\r),\theta,\epsilon\r)\r]\r)\dd i.
\end{split}
\end{align}
Next, I present some useful preliminary facts, which I use to establish (\ref{A+C})-(\ref{B+C}).  Observe that at $\l(\theta_0,\epsilon_0\r)$, agent $i$'s optimization problem is: $\max_{z_i} \l[z_i - v_i\l(z_i\r) -T\l(z_i,\theta_0,\epsilon_0\r)\r]$. The first-order condition is: $1 -v'_i\l(z_i\r)-\pdv{z_i}T\l(z_i,\theta_0,\epsilon_0\r)=0$.  Applying the implicit function theorem to the first-order condition,\footnote{As $\l(T^{\theta,\epsilon}\r)_{\theta \in \Theta,\epsilon \in E}$ is well-behaved, it follows that the first-order condition uniquely characterizes agent $i$'s optimal income $z_i\l(\theta,\epsilon\r)$.} we have:
\begin{align}\label{implicit theta}
\l.\pdv{\theta}\r|_{\theta=\theta_0}z_i\l(\theta,\epsilon_0\r)&=-\frac{\l.\pdv[2]{}{\theta}{z_i}\r|_{\theta=\theta_0,z_i=z_i\l(\theta,\epsilon_0\r)}T\l(z_i,\theta,\epsilon_0\r)}{v''_i\l(z_i\l(\theta_0,\epsilon_0\r)\r)+\l.\pdv[2]{z_i}\r|_{z_i = z_i\l(\theta_0,\epsilon_0\r)}T\l(z_i,\theta_0,\epsilon_0\r)},\\
\label{implicit epsilon}\l.\pdv{\epsilon}\r|_{\epsilon=\epsilon_0}z_i\l(\theta_0,\epsilon\r)&=-\frac{\l.\pdv[2]{}{\epsilon}{z_i}\r|_{\epsilon=\epsilon_0,z_i=z_i\l(\theta_0,\epsilon_0\r)}T\l(z_i,\theta_0,\epsilon\r)}{v''_i\l(z_i\l(\theta_0,\epsilon_0\r)\r)+\l.\pdv[2]{z_i}\r|_{z_i=z_i\l(\theta_0,\epsilon_0\r)}T\l(z_i,\theta_0,\epsilon_0\r)}.
\end{align}
I am now ready to establish (\ref{A+C})-(\ref{B+C}).  First, I establish (\ref{A+C}):
 \begin{align}
 \nonumber  &\l. \dv{\epsilon}\r|_{\epsilon=\epsilon_0}\int g_i\l(\theta_0,\epsilon\r) \l.\pdv{\theta}\r|_{\theta = \theta_0} T\l(z_i\l(\theta_0,\epsilon\r),\theta,\epsilon\r)\dd i\\
\label{differentiating indifference derivation 2} =&\int \l.\pdv{\epsilon}\r|_{\epsilon = \epsilon_0}\l[g_i\l(\theta_0,\epsilon\r) \l.\pdv{\theta}\r|_{\theta = \theta_0} T\l(z_i\l(\theta_0,\epsilon\r),\theta,\epsilon\r)\r]\dd i\\ 
\begin{split}\label{differentiating indifference derivation 3}=&\int \l(\l[\l.\pdv{\epsilon}\r|_{\epsilon = \epsilon_0}g_i\l(\theta_0,\epsilon\r)\r] \l.\pdv{\theta}\r|_{\theta = \theta_0} T\l(z_i\l(\theta_0,\epsilon_0\r),\theta,\epsilon_0\r)\r.\\&\l.+g_i\l(\theta_0,\epsilon_0\r)\l.\dv{\epsilon}\r|_{\epsilon=\epsilon_0}\l.\pdv{\theta}\r|_{\theta = \theta_0} T\l(z_i\l(\theta_0,\epsilon\r),\theta,\epsilon\r)\r)\dd i 
\end{split}\\
\begin{split}
\label{differentiating indifference derivation 4}= &\int \l(\l[-\pdv{\hat{u}_i}\hat{g}_i\l(\hat{U}_i\l(\theta_0,\epsilon_0\r),z_i\l(\theta_0,\epsilon_0\r)\r)\l.\pdv{\epsilon}\r|_{\epsilon=\epsilon_0} T\l(z_i\l(\theta_0,\epsilon_0\r),\theta_0,\epsilon\r)\r.\r.\\&\l.+ \pdv{z_i}\hat{g}_i\l(\hat{U}_i\l(\theta_0,\epsilon_0\r),z_i\l(\theta_0,\epsilon_0\r)\r)\l.\pdv{\epsilon}\r|_{\epsilon=\epsilon_0}z_i\l(\theta_0,\epsilon\r)\r]
\l.\pdv{\theta}\r|_{\theta = \theta_0} T\l(z_i\l(\theta_0,\epsilon_0\r),\theta,\epsilon_0\r) \\
&\l.+ g_i\l(\theta_0,\epsilon_0\r)\l.\dv{\epsilon}\r|_{\epsilon=\epsilon_0}\l.\pdv{\theta}\r|_{\theta = \theta_0} T\l(z_i\l(\theta_0,\epsilon\r),\theta,\epsilon\r)\r)\dd i 
\end{split}\\
\begin{split}
\label{differentiating indifference derivation 5}=&\int \l(\l[-\pdv{\hat{u}_i}\hat{g}_i\l(\hat{U}_i\l(\theta_0,\epsilon_0\r),z_i\l(\theta_0,\epsilon_0\r)\r)\l.\pdv{\epsilon}\r|_{\epsilon=\epsilon_0} T\l(z_i\l(\theta_0,\epsilon_0\r),\theta_0,\epsilon\r)\r.\r.\\&\l.+ \pdv{z_i}\hat{g}_i\l(\hat{U}_i\l(\theta_0,\epsilon_0\r),z_i\l(\theta_0,\epsilon_0\r)\r)\l.\pdv{\epsilon}\r|_{\epsilon=\epsilon_0}z_i\l(\theta_0,\epsilon\r)\r]
\l.\pdv{\theta}\r|_{\theta = \theta_0} T\l(z_i\l(\theta_0,\epsilon_0\r),\theta,\epsilon_0\r) \\
&+ g_i\l(\theta_0,\epsilon_0\r)\l[ \l.\pdv[2]{}{z_i}{\theta}\r|_{z_i=z_i\l(\theta_0,\epsilon_0\r),\theta = \theta_0} T\l(z_i,\theta,\epsilon_0\r)
\l.\pdv{\epsilon}\r|_{\epsilon=\epsilon_0}z_i\l(\theta_0,\epsilon\r) \r.\\&\l.\l.+ \l.\pdv[2]{}{\epsilon}{\theta}\r|_{\epsilon=\epsilon_0,\theta = \theta_0} T\l(z_i\l(\theta_0,\epsilon_0\r),\theta,\epsilon\r)\r]\r)\dd i 
\end{split}\\
\begin{split}
\label{differentiating indifference derivation 6}=&\int \l(\l[-\pdv{\hat{u}_i}\hat{g}_i\l(\hat{U}_i\l(\theta_0,\epsilon_0\r),z_i\l(\theta_0,\epsilon_0\r)\r)\l.\pdv{\epsilon}\r|_{\epsilon=\epsilon_0} T\l(z_i\l(\theta_0,\epsilon_0\r),\theta_0,\epsilon\r)\r.\r.\\&\l.+ \pdv{z_i}\hat{g}_i\l(U_i\l(\theta_0,\epsilon_0\r),z_i\l(\theta_0,\epsilon_0\r)\r)\l.\pdv{\epsilon}\r|_{\epsilon=\epsilon_0}z_i\l(\theta_0,\epsilon\r)\r]
\l.\pdv{\theta}\r|_{\theta = \theta_0} T\l(z_i\l(\theta_0,\epsilon_0\r),\theta,\epsilon_0\r) \\
&+ g_i\l(\theta_0,\epsilon_0\r)\l[- \l.\pdv[2]{}{z_i}{\theta}\r|_{z_i=z_i\l(\theta_0,\epsilon_0\r),\theta = \theta_0} T\l(z_i,\theta,\epsilon_0\r)\frac{\l.\pdv[2]{}{\epsilon}{z_i}\r|_{\epsilon = \epsilon_0,z_i=z_i\l(\theta_0,\epsilon_0\r)} T\l(z_i,\theta_0,\epsilon\r)}{v''_i\l(z_i\l(\theta_0,\epsilon_0\r)\r)+\l.\pdv[2]{z}\r|_{z_i = z_i\l(\theta_0,\epsilon_0\r)}T\l(z_i,\theta_0,\epsilon_0\r)}\r.\\&\l.\l.+ \l.\pdv[2]{}{\epsilon}{\theta}\r|_{\epsilon=\epsilon_0,\theta = \theta_0} T\l(z_i\l(\theta_0,\epsilon_0\r),\theta,\epsilon\r)\r]\r)\dd i, 
\end{split}\\
\label{differentiating indifference derivation 7}=& \;A + C 
\end{align} 
where (\ref{differentiating indifference derivation 4}) analyzes the term $\l.\pdv{\epsilon}\r|_{\epsilon = \epsilon_0}g_i\l(\theta_0,\epsilon\r)$ and appeals to the fact that, by the envelope theorem, $\l.\pdv{\epsilon}\r|_{\epsilon=\epsilon_0}\hat{U}_i\l(\theta_0,\epsilon\r)=-\l.\pdv{\epsilon}\r|_{\epsilon=\epsilon_0} T\l(z_i\l(\theta_0,\epsilon_0\r),\theta_0,\epsilon\r)$, and  (\ref{differentiating indifference derivation 6}) follows from (\ref{implicit epsilon}), $A$ is defined as in the proof outline of Lemma \ref{alternative condition lemma} in the main text and $C$ is defined by (\ref{definition of C}).  This establishes (\ref{A+C}).

As the derivation of (\ref{B+C}) is similar, I present it in an abbreviated form:
\begin{align*}
 &\l.\dv{\theta}\r|_{\theta = \theta_0} \int g_i\l(\theta,\epsilon_0\r) \l.\pdv{\epsilon}\r|_{\epsilon = \epsilon_0} T\l(z_i\l(\theta,\epsilon_0\r),\theta,\epsilon\r)\dd i \\
=&\int \l(\l[-\pdv{\hat{u}_i}\hat{g}_i\l(\hat{U}_i\l(\theta_0,\epsilon_0\r),z_i\l(\theta_0,\epsilon_0\r)\r)\l.\pdv{\theta}\r|_{\theta=\theta_0} T\l(z_i\l(\theta_0,\epsilon_0\r),\theta,\epsilon_0\r)\r.\r.\\&\l.+ \pdv{z_i}\hat{g}_i\l(\hat{U}_i\l(\theta_0,\epsilon_0\r),z_i\l(\theta_0,\epsilon_0\r)\r)\l.\pdv{\theta}\r|_{\theta=\theta_0}z_i\l(\theta,\epsilon_0\r)\r]
\l.\pdv{\epsilon}\r|_{\epsilon = \epsilon_0} T\l(z_i\l(\theta_0,\epsilon_0\r),\theta_0,\epsilon\r)\\
&+ g_i\l(\theta_0,\epsilon_0\r)\l[ \l.\pdv[2]{}{z_i}{\epsilon}\r|_{z_i=z_i\l(\theta_0,\epsilon_0\r),\epsilon = \epsilon_0} T\l(z_i,\theta_0,\epsilon\r)\l.\pdv{\theta}\r|_{\theta=\theta_0}z_i\l(\theta,\epsilon_0\r)\r.\\&\l.\l.+ \l.\pdv[2]{}{\theta}{\epsilon}\r|_{\theta = \theta_0, \epsilon=\epsilon_0} T\l(z_i\l(\theta_0,\epsilon_0\r),\theta,\epsilon\r)\r]\r)\dd i
 \\
 =&\int \l(\l[-\pdv{\hat{u}_i}g_i\l(\hat{U}_i\l(\theta_0,\epsilon_0\r),z_i\l(\theta_0,\epsilon_0\r)\r)\l.\pdv{\theta}\r|_{\theta=\theta_0} T\l(z_i\l(\theta_0,\epsilon_0\r),\theta,\epsilon_0\r)\r.\r.\\&\l.+ \pdv{z}\hat{g}_i\l(\hat{U}_i\l(\theta_0,\epsilon_0\r),z_i\l(\theta_0,\epsilon_0\r)\r)\l.\pdv{\theta}\r|_{\theta=\theta_0}z_i\l(\theta,\epsilon_0\r)\r]
\l.\pdv{\epsilon}\r|_{\epsilon = \epsilon_0} T\l(z_i\l(\theta_0,\epsilon_0\r),\theta_0,\epsilon\r)\\
&+ \hat{g}_i\l(\theta_0,\epsilon_0\r)\l[- \l.\pdv[2]{}{z_i}{\epsilon}\r|_{z_i=z_i\l(\theta_0,\epsilon_0\r),\epsilon = \epsilon_0} T\l(z_i,\theta_0,\epsilon\r)\frac{\l.\pdv[2]{}{\theta}{z_i}\r|_{\theta = \theta_0,z=z_i\l(\theta_0,\epsilon_0\r)} T\l(z_i,\theta_0,\epsilon\r)}{v''_i\l(z_i\l(\theta_0,\epsilon_0\r)\r)+\l.\pdv[2]{z_i}\r|_{z_i=z_i\l(\theta,\epsilon\r)}T\l(z_i,\theta_0,\epsilon_0\r)}\r.\\
&\l.\l.+ \l.\pdv[2]{}{\theta}{\epsilon}\r|_{\theta = \theta_0, \epsilon=\epsilon_0} T\l(z_i\l(\theta_0,\epsilon_0\r),\theta,\epsilon\r)\r]\r)\dd i
\\
=&\; B+C.
 \end{align*}
 The justification is similar to the justification for (\ref{differentiating indifference derivation 2})-(\ref{differentiating indifference derivation 7}), using (\ref{implicit theta}) instead of (\ref{implicit epsilon}).  This establishes (\ref{B+C}). $\square$

\subsection{Proof of Lemma \ref{lemma without assumptions on welfare weights}}
The main argument proving Lemma \ref{lemma without assumptions on welfare weights} is presented in Section \ref{lemma without assumptions on welfare weights main argument subsubsection}.  The proofs of a supporting lemma and some related material are presented in the subsequent subsections.   
\subsubsection{\label{lemma without assumptions on welfare weights main argument subsubsection}Main argument}
Choose a regular tax policy $T$. (See Section \ref{wb appendix non-individualized} for the requirements for a regular tax policy when taxes are not individualized.)  To establish the lemma, I construct a well-behaved doubly parameterized family of tax policies $\l(T^{\theta,\epsilon}\r)_{\theta \in \Theta,\epsilon \in E}$ satisfying (\ref{2}), (\ref{3}) and (\ref{wanted key inequality}), and such that, for the $\theta_0 \in \l(\underline{\theta},\overline{\theta}\r), \epsilon_0 \in \l(\underline{\epsilon},\overline{\epsilon}\r)$ that feature in the preceding conditions, $T^{\theta_0,\epsilon_0}=T$.    

Recall that the \textit{support} of a function $h$ with argument $x$ is the closure of $\l\{x:h\l(x\r) \neq 0\r\}$.   

To construct $\l(T^{\theta,\epsilon}\r)_{\theta \in \Theta,\epsilon \in E}$, I consider four smooth tax reforms $ \mu_1, \mu_2, \eta_1, \eta_2$.  Let $ i_k, k=1,2,3, 5$ be elements of $\l(0,1\r)$ be such that  $ i_1 <  i_2 <  i_3= i_a <   i_5= i_b.$  The reader will notice that we have skipped $ i_4$; this term will be introduced below (see Lemma \ref{intermediate point lemma}).   If we let $\hat{z}_k = z_{i_k}\l(T\r)$ for $k=1,2, 3,5$, it follows from assumptions on $v$ and $i \mapsto y_i$ in Section \ref{additional structure section} that $\hat{z}_1 < \hat{z}_2< \hat{z}_3 < \hat{z}_5$.  I assume that $\mu_1\l(z\r)=0$ when $z \leq \hat{z}_3$, $\mu_1\l(z\r)$ is increasing in $z$ on the interval $\l(\hat{z}_3,\hat{z}_5\r)$, and $\mu_1\l(z\r)$ remains constant at some positive number thereafter.  I assume that $\mu_2\l(z\r)=0$ when $z \leq \hat{z}_2$, $\mu_2\l(z\r)$ is increasing in $z$ on the interval $\l(\hat{z}_2,\hat{z}_3\r)$ and $\mu_2\l(z\r) =1$ when $z \geq \hat{z}_3$.  Assume, moreover, that $\mu_1$ and $\mu_2$ are chosen such that:
\begin{align}\label{Xi definition}
\int_0^1 g_ i\l(T\r)\mu_1\l(z_ i\l(T\r)\r) \dd i=\int_0^1 g_ i\l(T\r)\mu_2\l(z_ i\l(T\r)\r) \dd  i.
\end{align}
That is, both tax reforms $\mu_1$ and $\mu_2$ have the same marginal effect on social welfare, when benefits are weighted by welfare weights. The above assumptions imply the following lemma, which is proved in Section \ref{intermediate point section} of the Appendix.
\begin{lem}\label{intermediate point lemma}
There exists $ i_4 \in \l( i_3, i_5\r)$ such that $\mu_1\l(z_{ i_4}\l(T\r)\r) =1$. 
\end{lem}
If we define $\hat{z}_4 = z_{ i_4}\l(T\r)$, it follows from the fact that $ i_3 <  i_4 <  i_5$ that $\hat{z}_3 < \hat{z}_4 < \hat{z}_5$.  So Lemma \ref{intermediate point lemma} says that there is some income level $\hat{z}_4$, between $\hat{z}_3$ and $\hat{z}_5$, such that $\mu_1\l(\hat{z}_4\r)=1$, and moreover income level $\hat{z}_4$ is chosen by some agent $i_4$ when facing tax policy $T$.  

Assume that $\eta_1$ has support $\l[\hat{z}_3,\hat{z}_5\r]$, and that $\eta_1$ is increasing on $\l(\hat{z}_3,\hat{z}_4\r)$ and decreasing on $\l(\hat{z}_4,\hat{z}_5\r)$, which implies that $\eta_1\l(z\r)> 0, \forall z \in \l(\hat{z}_3,\hat{z}_5\r)$.  Assume that the support of $\eta_2$ is $\l[\hat{z}_1,\hat{z}_2\r]$, that $\eta_2\l(z\r) < 0, \forall z \in \l(\hat{z}_1,\hat{z}_2\r)$, and that
\begin{align}\label{balancing condition}
  \int_0^1 g_ i\l(T\r) \eta_1\l(z_ i\l(T\r)\r)\dd i = -\int_0^1 g_{ i}\l(T\r)\eta_2\l(z_ i\l(T\r)\r)\dd i.
  \end{align} 
In other words the marginal welfare effect of reform $\eta_1$ is the negative of the marginal welfare effect of reform $\eta_2$, so that the two cancel out.

For any real numbers, $\theta$ and $\epsilon$, define $T_*^{\theta,\epsilon}$ by:
 \begin{align}\label{hat T theta epsilon definition}
 T_*^{\theta,\epsilon} = T + \theta \mu_1+\epsilon \l(\eta_1 + \eta_2\r). 
 \end{align} 

It follows from the Picard-Lindel\"{o}f  theorem (see Section \ref{T theta epsilon well behaved section} for a more explicit formulation) that there exist real numbers $\underline{\theta}, \overline{\theta}, \underline{\epsilon}, \overline{\epsilon}$ such that $\underline{\theta} < 0 < \overline{\theta}, \underline{\epsilon} < 0 < \overline{\epsilon}$, and such that we can define the real-valued function $\zeta\l(\theta,\epsilon\r)$ on $\Theta \times E$, where $\Theta =\l[\underline{\theta},\overline{\theta}\r]$ and $E= \l[\underline{\epsilon},\overline{\epsilon}\r]$, by
\begin{align}
\label{AXzeta equals zero}&\zeta\l(0, \epsilon\r) =0, \;\;\;\forall \epsilon \in E, \\
\label{AXintegral zeta condition}\begin{split}&\int g_ i\l(T_*^{\theta,\epsilon} -\zeta\l(\theta,\epsilon\r)\mu_2\r)\\&\times \l[\mu_1\l(z_ i\l(T_*^{\theta,\epsilon} -\zeta\l(\theta,\epsilon\r)\mu_2\r)\r)-\pdv{\theta}\zeta\l(\theta,\epsilon\r)\mu_2\l(z_ i\l(T_*^{\theta,\epsilon} -\zeta\l(\theta,\epsilon\r)\mu_2\r)\r)\r]  \dd i =0, \\ &\forall \theta \in \Theta, \forall \epsilon \in E.
\end{split}
\end{align}
Next, for all $\theta \in \Theta$ and $\epsilon \in E$, define 
\begin{align}\label{T theta epsilon def 1}
T^{\theta,\epsilon} &= T + \l[\theta \times \mu_1\r] -\l[\zeta\l(\theta,\epsilon\r) \times \mu_2\r]+ \l[\epsilon \times \l(\eta_1 + \eta_2\r)\r]\\
\label{T theta epsilon def 2}& =T_*^{\theta,\epsilon} -\zeta\l(\theta,\epsilon\r) \mu_2.
\end{align}
In Section \ref{well behaved section}, I establish that if $\underline{\theta},\overline{\theta},\underline{\epsilon},$ and $\overline{\epsilon}$ are all chosen sufficiently close to 0, then $\l(T^{\theta,\epsilon}\r)_{\theta \in \Theta, \epsilon \in E}$ is well-behaved.   

So now consider the parameterized family of tax policies $\l(T^{\theta,\epsilon}\r)_{\theta \in \Theta, \epsilon \in E}$, for which we will verify the properties required in the lemma.  Let $\theta_0=0$ and $\epsilon_0=0$.  Then note that $T^{\theta_0,\epsilon_0}=T$, as required for the result.   Let $S=\l\{i \in I: z_i\l(T\r)\in \l(\hat{z}_3,\hat{z}_5\r)\r\}$ and $O=\l\{i \in I: z_i\l(T\r) \in \l(\hat{z}_1,\hat{z}_2\r)\r\}$, so that, as described in Section \ref{proof sketch subsection}, starting at $\theta=\theta_0=0$ and $\epsilon=\epsilon_0=0$, as $\epsilon$ increases, taxes on the incomes earned by agents in $S$ rise and taxes on incomes earned by agents in $O$ fall.

Recalling that $T\l(z,\theta,\epsilon\r)= T^{\theta,\epsilon}\l(z\r)$, it follows from (\ref{T theta epsilon def 1}) that, for all $ i \in I,  \epsilon \in E, $ and $\theta' \in \l(\underline{\theta},\overline{\theta}\r)$, 
\begin{align}
\label{tax derivative 1} \l.\pdv{\theta}\r|_{\theta =\theta'} 
T\l(z_ i\l(\theta',\epsilon\r),\theta,\epsilon\r) =\mu_1\l(z_ i\l(\theta',\epsilon\r)\r)-\l.\pdv{\theta}\r|_{\theta = \theta'}\zeta\l(\theta',\epsilon\r)\mu_2\l(z_ i\l(\theta',\epsilon\r)\r),
\end{align}
and it follows from (\ref{T theta epsilon def 1}), (\ref{AXzeta equals zero}), and the fact that $\theta_0 =0$, that, for all $ i \in I$,
\begin{align}
\label{tax derivative 2} \l.\pdv{\epsilon}\r|_{\epsilon =\epsilon_0} T\l(z_ i\l(\theta_0,\epsilon_0\r),\theta_0,\epsilon\r) =\eta_1\l(z_ i\l(T\r)\r) + \eta_2\l(z_ i\l(T\r)\r). 
\end{align}
It follows from (\ref{AXintegral zeta condition}), (\ref{T theta epsilon def 2}), and (\ref{tax derivative 1}) that $\l(T^{\theta,\epsilon}\r)_{\theta \in \Theta,\epsilon \in E}$ satisfies (\ref{2}), and from (\ref{balancing condition}) and (\ref{tax derivative 2}) that $\l(T^{\theta,\epsilon}\r)_{\theta \in \Theta,\epsilon \in E}$ satisfies (\ref{3}).

Next I seek to establish that $\l(T^{\theta,\epsilon}\r)_{\theta \in \Theta,\epsilon \in E}$ satisfies (\ref{wanted key inequality}).  In the special case in which $\theta = \theta_0$ and $\epsilon =\epsilon_0$ (recall that $\theta_0=\epsilon_0=0$), the general statement in (\ref{AXintegral zeta condition}) reduces to 
\begin{align*}
\int g_ i\l(T\r) \l[\mu_1\l(z_ i\l(T\r)\r)-\l.\pdv{\theta}\r|_{\theta =\theta_0}\zeta\l(\theta,\epsilon_0\r)\mu_2\l(z_ i\l(T\r)\r)\r]  \dd i =0.
\end{align*}
Solving for $\l.\pdv{\theta}\r|_{\theta =\theta_0} \zeta\l(\theta, \epsilon_0\r)$ from the above equation, it follows that 
\begin{align}\label{zeta derivative equals one}
\l.\pdv{\theta}\r|_{\theta =\theta_0} \zeta\l(\theta, \epsilon_0\r)= \frac{\int g_ i\l(T\r) \mu_1\l(z_ i\l(T\r)\r)  \dd i}{\int g_ i\l(T\r) \mu_2\l(z_ i\l(T\r)\r)  \dd i}=1,
\end{align}
where the second equality follows from (\ref{Xi definition}).  

Consider the type $ i$ agent's optimization problem when facing tax policy $T^{\theta,\epsilon}$--that is, of choosing $z$ so as to maximize $z-v_ i\l(z\r)-T^{\theta,\epsilon}\l(z\r)$. It follows from the implicit function theorem applied to the first order condition for this optimization problem at $\l(\theta,\epsilon\r)=\l(\theta_0,\epsilon_0\r)$ that 
\begin{align}\label{income derivatives}
\begin{split}
\l.\pdv{\theta}\r|_{\theta =\theta_0} z_ i\l(\theta,\epsilon_0\r)=&- \frac{\mu'_1\l(z_ i\l(T\r)\r)-\l.\pdv{\theta}\r|_{\theta = \theta_0}\zeta\l(\theta,\epsilon_0\r)\mu'_2\l(z_ i\l(T\r)\r)}{T''\l(z_ i\l(T\r)\r) + v''_ i\l(z_ i\l(T\r)\r)},\\
=&- \frac{\mu'_1\l(z_ i\l(T\r)\r)-\mu'_2\l(z_ i\l(T\r)\r)}{T''\l(z_ i\l(T\r)\r) + v''_ i\l(z_ i\l(T\r)\r)}\\
\l.\pdv{\epsilon}\r|_{\epsilon =\epsilon_0} z_ i\l(\theta_0,\epsilon\r)=& -\frac{\eta'_1\l(z_ i\l(T\r)\r)+\eta'_2\l(z_ i\l(T\r)\r)}{T''\l(z_ i\l(T\r)\r) + v''_ i\l(z_ i\l(T\r)\r)},
\end{split}\;\;\;\;\;\;\;\;  \forall  i \in  \l[0,1\r],\end{align}
where the second equality for the term $\l.\pdv{\theta}\r|_{\theta =\theta_0} z_ i\l(\theta,\epsilon_0\r)$ uses (\ref{zeta derivative equals one}), and the equality for the term $\l.\pdv{\epsilon}\r|_{\epsilon =\epsilon_0} z_ i\l(\theta_0,\epsilon\r)$ uses the fact that $\l.\pdv{\epsilon}\r|_{\epsilon=\epsilon_0} \zeta\l(\theta_0,\epsilon\r)=0$, which follows from (\ref{AXzeta equals zero}) and the assumption that $\theta_0 = 0$.  These equations simplify when $ i \in \l[ i_3, i_5\r]$.  In particular,
\begin{align}\label{income derivatives simplified}
\begin{split}
\l.\pdv{\theta}\r|_{\theta =\theta_0} z_ i\l(\theta,\epsilon_0\r)=&- \frac{\mu'_1\l(z_ i\l(T\r)\r)}{T''\l(z_ i\l(T\r)\r) + v''_ i\l(z_ i\l(T\r)\r)},\\
\l.\pdv{\epsilon}\r|_{\epsilon =\epsilon_0} z_ i\l(\theta_0,\epsilon\r)=& -\frac{\eta'_1\l(z_ i\l(T\r)\r)}{T''\l(z_ i\l(T\r)\r) + v''_ i\l(z_ i\l(T\r)\r)},
\end{split} \;\;\;\;\;\;\;\;  \forall  i \in  \l[ i_3, i_5\r].\end{align}
This simplification is explained by the observations that (i) since $\mu_2\l(z_ i\l(T\r)\r)=1$ when $ i \in \l[ i_3, i_5\r]$, $\mu'_2\l(z_ i\l(T\r)\r)=0$ when $ i \in \l[ i_3, i_5\r]$, and (ii) the support of $\eta_2$ is $\l[\hat{z}_1,\hat{z}_2\r]$, so that $\eta'_2\l(z_ i\l(T\r)\r)=0$ when $ i \in \l[ i_3, i_5\r]$.  

When $\l(\theta,\epsilon\r)=\l(\theta_0,\epsilon_0\r)$ and $ i \in \l[ i_3, i_5\r]$, (\ref{tax derivative 1})-(\ref{tax derivative 2}) also simplify:
\begin{align}\label{tax derivatives simplified}
\begin{split}
 \l.\pdv{\theta}\r|_{\theta =\theta_0} T\l(z_ i\l(\theta_0,\epsilon_0\r),\theta,\epsilon_0\r) =\;&\mu_1\l(z_ i\l(T\r)\r)-1,\\ 
\l.\pdv{\epsilon}\r|_{\epsilon =\epsilon_0} T\l(z_ i\l(\theta_0,\epsilon_0\r),\theta_0,\epsilon\r) =\;&\eta_1\l(z_ i\l(T\r)\r), 
\end{split}  \;\;\;\;\;\;\;\;  \forall  i \in  \l[ i_3, i_5\r],
\end{align}
where the first equality uses (\ref{zeta derivative equals one})  and the fact that $\mu_2\l(z\r) =1$ when $z \in \l[\hat{z}_3,\hat{z}_5\r]$, and the second equality uses the fact that $\eta_2\l(z\r)=0$ when $z \in \l[\hat{z}_3,\hat{z}_5\r]$.

Recalling that $ i_a= i_3$ and $ i_b =  i_5$, it follows from (\ref{income derivatives simplified}) and (\ref{tax derivatives simplified}) that 
 \begin{align}\label{lemmas desired inequality}
\begin{split}
& \forall  i \in \l( i_a, i_b\r), \\
&\l.\pdv{\theta}\r|_{\theta =\theta_0} z_ i\l(\theta,\epsilon_0\r) \l.\pdv{\epsilon}\r|_{\epsilon =\epsilon_0} T\l(z_ i\l(\theta_0,\epsilon_0\r),\theta_0,\epsilon\r) -  \l.\pdv{\epsilon}\r|_{\epsilon =\epsilon_0} z_ i\l(\theta_0,\epsilon\r)\l.\pdv{\theta}\r|_{\theta =\theta_0} T\l(z_ i\l(\theta_0,\epsilon_0\r),\theta,\epsilon_0\r)\\
=
&\frac{-\overbrace{\mu'_1\l(z_ i\l(T\r)\r)}^+\overbrace{\eta_1\l(z_ i\l(T\r)\r)}^+
 +\l(\overbrace{\eta'_1\l(z_ i\l(T\r)\r)}^{+ \textup{ on }\l( i_3, i_4\r), -\textup{ on } \l( i_4, i_5\r)} \times \overbrace{\l[\mu_1\l(z_ i\l(T\r)\r)-1 \r]}^{- \textup{ on } \l( i_3, i_4\r), + \textup{ on }\l( i_4, i_5\r)}\r)}{\underbrace{T''\l(z_ i\l(T\r)\r) + v''_ i\l(z_ i\l(T\r)\r)}_+} <0.
\end{split}
\end{align}
where the signs are derived from the assumptions we made above about $\eta_1$ and $\mu_1$ -- in particular note that $\mu_1\l(z\r) >0$ is increasing on $\l(\hat{z}_3,\hat{z}_5\r)$ and, by Lemma \ref{intermediate point lemma}, $\mu_1\l(\hat{z}_4\r)=1$ -- as well as the fact that because $T$ is regular, $0> \l.\dv[2]{z_i}\r|_{z_i=z_i\l(T\r)}u\l(z_i-T\l(z_i\r)-v_i\l(z_i\r)\r)= -u'\l(z_i\l(T\r)-T_i\l(z_i\l(T\r)\r)-v_i\l(z_i\l(T\r)\r)\r) \times \l[v''_i\l(z_i\l(T\r)\r) + T''\l(z_i\l(T\r)\r)\r], \forall i \in I$\footnote{Observe that $\l.\dv[2]{z_i}\r|_{z_i=z_i\l(T\r)}u\l(z_i-T\l(z_i\r)-v_i\l(z_i\r)\r) =u''\l(z_i-T\l(z_i\r)-v_i\l(z_i\r)\r)\times\underbrace{\l[1-v'_i\l(z_i\l(T\r)\r) + T'\l(z_i\l(T\r)\r)\r]^2}_{=0} -u'\l(z_i\l(T\r)-T_i\l(z_i\l(T\r)\r)-v_i\l(z_i\l(T\r)\r)\r) \times \l[v''_i\l(z_i\l(T\r)\r) + T''\l(z_i\l(T\r)\r)\r]= -u'\l(z_i\l(T\r)-T_i\l(z_i\l(T\r)\r)-v_i\l(z_i\l(T\r)\r)\r) \times \l[v''_i\l(z_i\l(T\r)\r) + T''\l(z_i\l(T\r)\r)\r].$}   (see Section \ref{wb appendix non-individualized}), so that $v''_i\l(z_i\l(T\r)\r) + T''\l(z_i\l(T\r)\r)>0, \forall i \in I$.  Next observe that:
\begin{itemize}
\item The support of $ i \mapsto  \l.\pdv{\theta}\r|_{\theta =\theta_0} z_ i\l(\theta,\epsilon_0\r)  $ is $\l[ i_2, i_5\r]$.
\item The support of $ i \mapsto \l.\pdv{\epsilon}\r|_{\epsilon =\epsilon_0} T\l(z_ i\l(\theta_0,\epsilon_0\r),\theta_0,\epsilon\r) $ is $\l[ i_1, i_2\r] \cup \l[ i_3, i_5\r]$.
\end{itemize}

Recalling that $ i_a= i_3$ and $ i_b= i_5$, it follows that
\begin{align}\label{zT product 1}
\l.\pdv{\theta}\r|_{\theta =\theta_0} z_ i\l(\theta,\epsilon_0\r) \l.\pdv{\epsilon}\r|_{\epsilon =\epsilon_0} T\l(z_ i\l(\theta_0,\epsilon_0\r),\theta_0,\epsilon\r) = 0,\;\;\; \forall  i \not \in \l( i_a, i_b\r).
\end{align}
To understand why the above expression is equal to zero when $ i \in \l\{ i_2, i_3, i_5\r\}$, note that the expressions  $\l.\pdv{\theta}\r|_{\theta =\theta_0} z_ i\l(\theta,\epsilon_0\r)$ and $ \l.\pdv{\epsilon}\r|_{\epsilon =\epsilon_0} T\l(z_ i\l(\theta_0,\epsilon_0\r),\theta_0,\epsilon\r)$ are equal to zero on the boundaries of their supports.  
\begin{itemize}
\item The support of $ i \mapsto   \l.\pdv{\epsilon}\r|_{\epsilon =\epsilon_0} z_ i\l(\theta_0,\epsilon\r)$ is contained in $\l[ i_1, i_2\r] \cup \l[ i_3, i_5\r]$.
\item The support of $ i \mapsto \l.\pdv{\theta}\r|_{\theta =\theta_0} T\l(z_ i\l(\theta_0,\epsilon_0\r),\theta,\epsilon_0\r)$ is $\l[ i_2,1\r]$.
\end{itemize}
It follows that
\begin{align}\label{zT product 2}
\l.\pdv{\epsilon}\r|_{\epsilon =\epsilon_0} z_ i\l(\theta_0,\epsilon\r)\l.\pdv{\theta}\r|_{\theta =\theta_0} T\l(z_ i\l(\theta_0,\epsilon_0\r),\theta,\epsilon_0\r) =0, \;\;\; \forall  i \not \in \l( i_a, i_b\r).
\end{align}
Again, the above condition uses the fact that $\l.\pdv{\epsilon}\r|_{\epsilon =\epsilon_0} z_ i\l(\theta_0,\epsilon\r)$ and $\l.\pdv{\theta}\r|_{\theta =\theta_0} T\l(z_ i\l(\theta_0,\epsilon_0\r),\theta,\epsilon_0\r)$ are equal to zero on the boundaries of their supports.  Together (\ref{zT product 1}), (\ref{zT product 2}), and the inequality established in (\ref{lemmas desired inequality}) are equivalent to (\ref{wanted key inequality}).  We have now established that the family $\l(T^{\theta,\epsilon}\r) $ satisfies all of the conditions required by the lemma. $\square$

\subsubsection{\label{intermediate point section}Proof of Lemma \ref{intermediate point lemma}}
Assume, for contradiction, that, for all $ i \in \l( i_3, i_5\r), \mu_1\l(z_ i\l(T\r)\r) \neq 1$.  Then, since the function $ i \mapsto \mu_1\l(z_ i\l(T\r)\r)$ is smooth (this follows from the assumed smoothness of relevant functions and the implicit function theorem), $\mu_1\l(z_{ i_3}\l(T\r)\r) =0$, and $ i \mapsto \mu_1\l(z_ i\l(T\r)\r)$ is a constant function on $\l[ i_5,1\r]$, it follows from the intermediate value theorem that $\mu_1\l(z_ i\l(T\r)\r) < 1, \forall  i \in \l[ i_3,1\r]$.  So
\begin{align}\label{contradicting inequality}
\begin{split}
\int_{0}^{1} g_ i\l(T\r)\mu_1\l(z_ i\l(T\r)\r) \dd i =& \int_{ i_3}^{1}  g_ i\l(T\r)\mu_1\l(z_ i\l(T\r)\r) \dd i < \int_{ i_3}^{1}  g_ i\l(T\r) \dd i \\=&\int_{ i_3}^{1}  g_ i\l(T\r)\mu_2\l(z_ i\l(T\r)\r) \dd i < \int_{0}^{1}  g_ i\l(T\r)\mu_2\l(z_ i\l(T\r)\r) \dd i, 
\end{split}
\end{align}
where the first equality follows from the fact that the support of $\mu_1$ is $\l[\hat{z}_3,\bar{z}\r]$; the first inequality from the our conclusion that $\mu_1\l(z_ i\l(T\r)\r) < 1, \forall  i \in \l[ i_3,1\r]$ and the fact that $g_ i\l(T\r) >0, \forall  i \in\l[0,1\r]$; the second equality form the fact that $\mu_2\l(z\r)=1$ for all $z \in \l[\hat{z}_3,\bar{z}\r]$, and the last inequality from the fact that the $\mu_2$ is nonnegative everywhere and $\mu_2\l(z\r)> 0$ for $z \in \l(\hat{z}_2,\hat{z}_3\r)$.  However, (\ref{contradicting inequality}) contradicts (\ref{Xi definition}).  So the assumption that $\mu_1\l(z_ i\l(T\r)\r)$ is never equal to $1$ on $\l( i_3,  i_5\r)$ leads to a contradiction, completing the proof.  $\square$

\subsubsection{\label{lemma without assumptions on welfare weights variant section}A variant of Lemma \ref{lemma without assumptions on welfare weights}}

This section discusses the proof of a variant of Lemma \ref{lemma without assumptions on welfare weights}; I appeal to this variant in the proof of Lemma \ref{key lemma minus revenue}.
\begin{lem}\label{lemma without assumptions on welfare weights variant}Let $T$ be a regular tax policy and let $i_a, i_b \in \l(0,1\r)$ be such that $i_a < i_b$.  Then there exists a well-behaved family $\l(T^{\theta,\epsilon}\r)$ with $T^{\theta_0,\epsilon_0}=T$ for some interior parameter values $\theta_0,\epsilon_0$ and that satisfies (\ref{2}),  (\ref{3}), and 
\begin{align}\label{wanted key inequality variant}
\begin{split}
& \l.\pdv{\theta}\r|_{\theta =\theta_0} z_ i\l(\theta,\epsilon_0\r) \l.\pdv{\epsilon}\r|_{\epsilon =\epsilon_0} T\l(z_ i\l(\theta_0,\epsilon_0\r),\theta_0,\epsilon\r) \\& -  \l.\pdv{\epsilon}\r|_{\epsilon =\epsilon_0} z_ i\l(\theta_0,\epsilon\r)\l.\pdv{\theta}\r|_{\theta =\theta_0} T\l(z_ i\l(\theta_0,\epsilon_0\r),\theta,\epsilon_0\r)\end{split}\;\;\begin{cases} >0, &\textup{ if }  i \in \l( i_a, i_b\r),\\
= 0,&\textup{ if }  i \not\in \l( i_a, i_b\r).
\end{cases}
\end{align}
\end{lem}
This lemma differs from Lemma \ref{lemma without assumptions on welfare weights} only in that the inequality in (\ref{wanted key inequality variant}) points in the opposite direction to (\ref{wanted key inequality}).  If one modifies the construction in the proof of Lemma \ref{lemma without assumptions on welfare weights} only by assuming that $\eta_1$ is decreasing (rather than increasing) on $\l(\hat{z}_3,\hat{z}_4\r)$ and increasing (rather than decreasing) on $\l(\hat{z}_4,\hat{z}_5\r)$, so that $\eta_1\l(z\r) < 0$ (rather than $\eta_1\l(x\r) > 0$) on $\l(\hat{z}_3,\hat{z}_5\r)$, and correspondingly if one assumes that $\eta_2\l(z\r) > 0$ on $\l(\hat{z}_1,\hat{z}_2\r)$ (rather than $\eta_2\l(z\r) <0$), then one flips the inequality in (\ref{wanted key inequality}), and so attains (\ref{wanted key inequality variant}).  $\square$

\subsection{\label{key lemma with revenue section}Proof of Lemma \ref{key lemma with revenue}}

Assume that welfare weights $g$ are not structurally utilitarian.  By Lemma \ref{key lemma minus revenue}, in this case, we may choose a well behaved family $\l(T^{\theta,\epsilon}\r)_{\theta \in \Theta, \epsilon \in E} $ satisfying (\ref{2})-(\ref{4}).  The construction of this family is presented in the proofs of Lemmas \ref{lemma without assumptions on welfare weights} and \ref{key lemma minus revenue}.  Let us consider again the construction of $\l(T^{\theta,\epsilon}\r)$.  First, by Corollary \ref{convex tax policy corollary}, since $g$ is not structurally utilitarian we can select a regular tax policy $T$ such that for some such that for some $i_a, i_b \in \l(0,1\r)$ with $i_a < i_b$, either condition (\ref{negative derivative integral}) or (\ref{positive derivative integral}) is satisfied.  An examination of the construction of the proof of Corollary \ref{convex tax policy corollary} shows that it is possible to select $T$ such that
\begin{align}\label{marginal T nonzero}
T'\l(z_0\l(T\r)\r)\neq 0.
\end{align} 
We did not previously assume property (\ref{marginal T nonzero}) but let us assume henceforth that (\ref{marginal T nonzero}) is satisfied.  Next we a  use tax policy $T$ and $i_a$ and $i_b$ with the above properties to construct a family of tax polices, $\l(T^{\theta,\epsilon}\r)$, as in the proof of Lemma \ref{lemma without assumptions on welfare weights}, of the form $T^{\theta,\epsilon} = T + \l[\theta \times \mu_1\r] -\l[\zeta\l(\theta,\epsilon\r) \times \mu_2\r]+ \l[\epsilon \times \l(\eta_1 + \eta_2\r)\r]$ (see (\ref{T theta epsilon def 1})). The proof of Lemma \ref{key lemma minus revenue} shows that such a family satisfies (\ref{2})-(\ref{4}).  It follows from their construction in the proof of Lemma \ref{lemma without assumptions on welfare weights} that the supports of the functions $\mu_1, \mu_2, \eta_1$, and $\eta_2$ are all contained in the set $\l[\hat{z}_1,+\infty\r)$, where $\hat{z}_1=z_{i_1}\l(T\r)$ was defined in the beginning of the proof of Lemma \ref{lemma without assumptions on welfare weights}.  As $0 <i_1$, it follows from assumptions in Section \ref{additional structure section}, that 
\begin{align}\label{0 less i1}
z_0\l(T\r) < z_{i_1}\l(T\r)=\hat{z}_1.
\end{align}  
It follows that
\begin{align}\label{T constant low}
T^{\theta,\epsilon}\l(z\r) = T\l(z\r), \;\;\; \forall z \in \l[0,\hat{z}_1\r], \forall \theta \in \Theta, \forall \epsilon \in E.
\end{align}          
So $T^{\theta,\epsilon}\l(z\r)=T\l(z,\theta,\epsilon\r)$ does not depend on $\theta$ or $\epsilon$ for $z$ below $\hat{z}_1$.  Recall that in the construction of $\l(T^{\theta,\epsilon}\r)$, we assumed that $\theta_0=0$ and $\epsilon_0=0$, so that, by (\ref{T theta epsilon def 1}), $T^{\theta_0,\epsilon_0}=T$.

\begin{lem}\label{indifference and nonconstant revenue family of reforms theorem}
There exists a family of tax reforms $\l(\Delta T^{\xi}\r)_{\xi \in \Xi}$, where $\Xi =\l[\underline{\xi},\overline{\xi}\r]$ for real numbers $\underline{\xi},\overline{\xi}$ satisfying $\underline{\xi} < 0 < \overline{\xi}$, and such that $\Delta T^0 \equiv 0$, the support of $\Delta T^\xi$ is contained in $\l[0,\hat{z}_1\r]$ for all $\xi \in \Xi$, the map $\l(z,\xi\r)\mapsto \Delta T^{\xi}\l(z\r)$ is smooth, and for some sets $\Theta'=\l[\underline{\theta}',\overline{\theta}'\r] \subseteq \Theta$, $E'=\l[\underline{\epsilon}',\overline{\epsilon}'\r] \subseteq E$, with $\underline{\theta}' <0< \overline{\theta}'$ and $\underline{\epsilon}' < 0 <\overline{\epsilon}'$,
\begin{align}
\label{indifference3}  \int g_i\l(T^{\theta,\epsilon}+\Delta T^\xi\r) \l.\pdv{\xi}\r|_{\xi=\xi'}\Delta T\l(z_i\l(T^{\theta,\epsilon}+\Delta T^{\xi'}\r),\xi\r) \dd i =0, &\;\;\; \forall \theta \in \Theta', \forall \epsilon \in E', \forall \xi' \in \Xi, \\\
\label{nonconstant revenue3} \l.\dv{\xi}\r|_{\xi=\xi'} R\l(T^{\theta,\epsilon}+ \Delta T^\xi\r) \neq 0, &\;\;\; \forall \theta \in \Theta', \forall \epsilon \in E', \forall \xi' \in \Xi.
\end{align}
where in (\ref{indifference3}) we use the notation $\Delta T\l(z_i,\xi\r)=\Delta T^\xi\l(z_i\r)$.  Moreover, $\l(\Delta T^\xi\r)_{\xi \in \Xi}$ can be constructed so that $T^{\theta,\epsilon} + \Delta T^\xi$ is regular, for all $\xi \in \Xi, \theta \in \Theta',$ and $\epsilon \in E'$.     
\end{lem}  
To understand this lemma, first recall that $T^{\theta_0,\epsilon_0}=T$, and note that, by construction, all tax policies $T^{\theta,\epsilon}$ are equal to $T$ on the interval $\l[0,\hat{z}_1\r]$, which contains the support of all tax reforms $\Delta T^\xi$.  Lemma \ref{indifference and nonconstant revenue family of reforms theorem} says that the family of reforms $\l(\Delta T^\xi\r)$ is such that varying $\xi$ in $T^{\theta,\epsilon}+\Delta T^\xi$ has no effect on welfare according to welfare weights (see (\ref{indifference3})), but does have an effect on revenue (see (\ref{nonconstant revenue3})).  Obviously, if $T^{\theta,\epsilon}$ were an optimal tax policy, it would not be possible to do this.  However note that $T$, which coincides with all policies $T^{\theta,\epsilon}$ at the bottom of the income distribution, is such that marginal tax rate at at the income $z_0\l(T\r)$ at the bottom of the income distribution is non-zero, and, moreover, since $T$ is regular, $z_0\l(T\r) >0$ (see Section \ref{wb appendix non-individualized}), and hence, none of the tax policies $T^{\theta,\epsilon}$ are optimal.  As shown by \citeasnoun{saez2016generalized}, (see Section A.2 of their Online Appendix), at an optimal tax policy in the generalized social welfare weights framework, the marginal tax rate for the bottom earner is zero if the bottom earner has a positive income.  Lemma \ref{indifference and nonconstant revenue family of reforms theorem} is proven in Section \ref{indifference and nonconstant revenue family of reforms theorem section} of the Appendix.
 
 So let us assume that a family $\l(\Delta T^\xi\r)$ with the properties in Lemma \ref{indifference and nonconstant revenue family of reforms theorem} is chosen.  Noting that $\Delta T^0 \equiv 0$, it follows from (\ref{nonconstant revenue3}) and the implicit function theorem that there exists $\underline{\theta}'',\overline{\theta}'' \in \Theta'$ with $\underline{\theta}'' < 0 < \overline{\theta}''$ and $\underline{\epsilon}'',\overline{\epsilon}'' \in E'$ with $\underline{\epsilon}'' < 0 < \overline{\epsilon}''$ and a function $\hat{\xi}: \Theta'' \times E'' \rightarrow \Xi$, where $\Theta'' = \l[\underline{\theta}'',\overline{\theta}''\r]$ and $E''=\l[\underline{\epsilon}'',\overline{\epsilon}''\r]$, satisfying:
\begin{align}\label{fixed revenue condition 0}
\hat{\xi}\l(\theta_0,\epsilon_0\r) &=0,\\
R\l(T^{\theta,\epsilon} + \Delta T^{ \hat{\xi}\l(\theta,\epsilon\r)}\r) &= \label{fixed revenue condition}R\l(T^{\theta_0,\epsilon_0}\r), \;\;\; \forall \theta \in \Theta', \forall \epsilon \in E'',
\end{align}
where, in (\ref{fixed revenue condition 0}), $\Delta T^{ \hat{\xi}\l(\theta,\epsilon\r)}$ is $\Delta T^{\xi}$ evaluated at $\xi =\hat{\xi}\l(\theta,\epsilon\r)$.  Because the other functions occurring in (\ref{fixed revenue condition}) are smooth, it follows that $\hat{\xi}\l(\theta,\epsilon\r)$ is smooth. 
 
Define the doubly parameterized family of tax policies $\l(\hat{T}^{\theta,\epsilon}\r)_{\theta \in \Theta'', \epsilon \in E''}$ by 
\begin{align}\label{hat defined in terms of non-hat}
\hat{T}^{\theta,\epsilon} = T^{\theta,\epsilon} + \Delta T^{\hat{\xi}\l(\theta,\epsilon\r)}, \;\;\; \forall \theta \in \Theta'', \forall \epsilon \in E''.
\end{align}
It follows from Lemma \ref{indifference and nonconstant revenue family of reforms theorem}, the fact that $\l(T^{\theta,\epsilon}\r)$ is well-behaved, the fact that $\hat{\xi}\l(\theta,\epsilon\r)$ is smooth, and Lemma \ref{small perturbation still in T hat lemma} that, if above $\underline{\theta}'',\overline{\theta}'',\underline{\epsilon}''$, and $\overline{\epsilon}''$ are selected sufficiently close to zero, then $\l(\hat{T}^{\theta,\epsilon}\r)_{\theta \in \Theta'',\epsilon\in E''}$ is well-behaved.  The well-behavedness of $\l(\hat{T}^{\theta,\epsilon}\r)$ is elaborated in greater detail in Section \ref{well behaved section}, and specifically Section \ref{hat T theta epsilon well behaved section}.

\begin{lem}\label{lemma satisfying desired properties}
$\l(\hat{T}^{\theta,\epsilon}\r)_{\theta \in \Theta'',\epsilon\in E''} $ satisfies (\ref{2})-(\ref{4}).
\end{lem}
It is straightforward to verify that $\l(\hat{T}^{\theta,\epsilon}\r)_{\theta \in \Theta'',\epsilon\in E''} $ inherits properties (\ref{2})-(\ref{4}) from $\l(T^{\theta,\epsilon}\r)_{\theta \in \Theta,\epsilon\in E}$.  The calculations verifying Lemma \ref{lemma satisfying desired properties} are in Section \ref{lemma satisfying desired properties section}.  Moreover it follows from (\ref{hat defined in terms of non-hat}) and (\ref{fixed revenue condition}) that $\l(\hat{T}^{\theta,\epsilon}\r)_{\theta \in \Theta'',\epsilon\in E''} $ has constant revenue.  Thus, we have constructed a family of tax policies with the desired properties, which completes the proof. $\square$

\subsection{\label{well behaved section}Well-behavedness of families $\l(T^{\theta,\epsilon}\r)$ and $\l(\hat{T}^{\theta,\epsilon}\r)$ in the proof of Theorem \ref{main theorem}}
This section explains why the families of tax policies $\l(T^{\theta,\epsilon}\r)$ and $\l(\hat{T}^{\theta,\epsilon}\r)$ constructed in the proof of Theorem \ref{main theorem} are well-behaved.  Well-behavedness consists of conditions on agents' optimization problems when facing the tax policies as well as smoothness conditions. (See Section \ref{wb appendix non-individualized}.)  At a high level, the reason that $\l(T^{\theta,\epsilon}\r)$ and $\l(\hat{T}^{\theta,\epsilon}\r)$  satisfy the smoothness conditions is that these tax policies are constructed by combining functions that are assumed to be smooth in ways that preserve smoothness.  More specifically, smoothness follows because relevant functions are derived from the implicit function theorem applied to smooth functions, which preserves smoothness (see Theorem 1.37 on p. 30 of \citeasnoun{warner2013foundations}) or from the fact that solution to a parameterized initial value problem (whose existence and uniqueness are guaranteed by the Picard-Lindel\"{o}ff theorem) is smooth when the parameterized initial value problem is appropriately constructed out of smooth functions (see Corollary 4.1 on p. 101 of \citeasnoun{hartman1982ordinary}).  I give a more detailed argument below.

\subsubsection{\label{T theta epsilon well behaved section}The family $\l(T^{\theta,\epsilon}\r)$}
In the proof of Lemma \ref{lemma without assumptions on welfare weights} in Section \ref{lemma without assumptions on welfare weights main argument subsubsection}, I wrote that if $\underline{\theta},\overline{\theta},\underline{\epsilon},$ and $\overline{\epsilon}$, with $\underline{\theta} < 0 < \overline{\theta}, \underline{\epsilon} < 0 < \overline{\epsilon}$ are all chosen sufficiently close to 0, then the family $\l(T^{\theta,\epsilon}\r)_{\theta \in \l[\underline{\theta},\overline{\theta}\r],\epsilon \in \l[\underline{\epsilon},\underline{\epsilon}\r]}$ is well-behaved.  I now substantiate that claim.  First, for easy reference, recall definitions (\ref{hat T theta epsilon definition}) and (\ref{T theta epsilon def 2}):   
\begin{align}
\label{relabel hat T theta epsilon definition}T_*^{\theta,\epsilon} &= T + \theta \mu_1+\epsilon \l(\eta_1 + \eta_2\r),\\
\label{relabel T theta epsilon def 2}T^{\theta,\epsilon} & = T_*^{\theta,\epsilon} -\zeta\l(\theta,\epsilon\r) \mu_2.
\end{align} 
As stated in Section \ref{wb appendix non-individualized}, a (non-individualized) tax policy is regular if the tax policy is smooth in income, and, for each agent $i$, when facing the tax policy, there is a unique optimal income, and at this optimum, $i$'s income is non-negative and $i$'s second order condition holds with a strict inequality.   Recall that the tax policy $T$ in the definition of $T_*^{\theta,\epsilon}$ was assumed to be regular.  Now consider a tax policy of the form $T_*^{\theta,\epsilon} - \zeta \mu_2$, where $\zeta$ is a real number.  If $\theta=\epsilon=\zeta=0$, then $T_*^{\theta,\epsilon} - \zeta \mu_2 = T$. Since $\mu_1,\mu_2, \eta_1$, and $\eta_2$ are all assumed to be smooth in $z$, and $T_*^{\theta,\epsilon} - \zeta \mu_2$ varies smoothly in $\l(\theta,\epsilon,\zeta\r)$, it follows that if $\theta,\epsilon,$ and $\zeta$, are sufficiently close to zero, then $T_*^{\theta,\epsilon} - \zeta \mu_2$ continues to be regular: for each agent $i$, the optimum continues to be unique and positive, the second order condition continues to hold with a strict inequality, and the tax policy continues to be smooth in income.  To state this formally, Lemma \ref{small perturbation still in T hat lemma} implies that it is possible to choose $\theta^*>0,\epsilon^*>0,  \zeta^*>0$ sufficiently small that,
\begin{align}\label{lemma on regularity}
 \textup{for all } \theta, \epsilon,  \zeta, \textup{ if } \l|\theta\r| \leq \theta^*,\l|\epsilon\r| \leq \epsilon^*, \l|\zeta \r| \leq \zeta^*, \textup{ then } T_*^{\theta,\epsilon} - \zeta \mu_2 \textup{ is regular.} 
\end{align}

Next, I establish that the function $ \zeta\l(\theta,\epsilon\r)$ is smooth in its arguments. Define the functions $f_1, f_2: I\times \l[-\theta^*, \theta^*\r] \times \l[-\zeta^*,\zeta^*\r] \times  \l[-\epsilon^*,\epsilon^*\r] \rightarrow \mathbb{R}$ by
\begin{align*}
f_1\l(i,\theta,\zeta,\epsilon\r) =&\; g_ i\l(T_*^{\theta,\epsilon}- \zeta\mu_2\r) \mu_1\l(z_ i\l(T_*^{\theta,\epsilon}-\zeta\mu_2\r)\r), \\
f_2\l(i,\theta,\zeta,\epsilon\r) =&\;  g_ i\l(T_*^{\theta,\epsilon}- \zeta\mu_2\r)\mu_2\l(z_ i\l(T_*^{\theta,\epsilon}-\zeta\mu_2\r)\r).  
\end{align*}
For $\epsilon \in \l[-\epsilon^*,\epsilon^*\r]$, define $F_\epsilon\l(\theta,\zeta\r): \l[-\theta^*, \theta^*\r] \times \l[-\zeta^*,\zeta^*\r] \rightarrow \mathbb{R}$ by
\begin{align}\label{capital F epsilon theta zeta definition}
F_\epsilon\l(\theta,\zeta\r) = \frac{\int_0^1 f_1\l(i,\theta,\zeta,\epsilon\r) \dd i}{\int_0^1 f_2\l(i,\theta,\zeta,\epsilon\r) \dd i}.
\end{align}
Choose a real number $M$ satisfying $M \geq \l|F_{\epsilon}\l(\theta,\zeta\r)\r|$, for all $\epsilon \in \l[-\epsilon^*,\epsilon^*\r], \forall \theta \in \l[-\theta^*,\theta^*\r], \forall \zeta \in \l[ -\zeta^*,\zeta^*\r]$.  Let $\overline{\theta} = \min\l\{\theta^*,\frac{\zeta^*}{M}\r\}$. Since both the functions $f_1$ and $f_2$ are smooth in their arguments,\footnote{Note in particular that $z_ i\l(T_*^{\theta,\epsilon}-\zeta\mu_2\r)$ is smooth in $\l(i,\theta,\zeta,\epsilon\r)$ because, as it has been established above that $T_*^{\theta,\epsilon}-\zeta\mu_2$ is regular, $z_ i\l(T_*^{\theta,\epsilon}-\zeta\mu_2\r)$ is characterized via the implicit function theorem from the agent's first order condition, and the other functions featuring in this condition are smooth.} the integrals in both the numerator and the denominator of the right hand side of (\ref{capital F epsilon theta zeta definition}) are smooth in $\l(\theta,\epsilon,\zeta\r)$, and since the denominator is never equal to zero, it follows that $\l(\theta,\zeta,\epsilon\r)\mapsto F_\epsilon\l(\theta,\zeta\r)$ is smooth. The smoothness of $F_\epsilon\l(\theta,\zeta\r)$ implies that, in particular, $F_\epsilon\l(\theta,\zeta\r)$ is continuous in $\theta$ and uniformly Lipschitz continuous in $\zeta$.  It now follows from the Picard-Lindel\"{o}f theorem (see Theorem 1.1 on p. 8 of \citeasnoun{hartman1982ordinary}) that, for all $\epsilon \in \l[-\epsilon^*,\epsilon^*\r]$, there exists a unique function $\zeta_\epsilon\l(\theta\r): \l[-\overline{\theta},\overline{\theta}\r] \rightarrow \l[-\zeta^*,\zeta^*\r]$ satisfying  
\begin{align}\label{equivalent differential equation}
\begin{split}
\zeta_{\epsilon}\l(0\r)&=0,\\
\dv{\theta}\zeta_\epsilon\l(\theta\r) &= F_\epsilon\l(\theta,\zeta_{\epsilon}\l(\theta\r)\r), \;\;\; \forall \theta \in \l[-\overline{\theta},\overline{\theta}\r].
\end{split}
\end{align}
We can write $\zeta\l(\theta,\epsilon\r) = \zeta_\epsilon\l(\theta\r)$.  Note that (\ref{equivalent differential equation}) is equivalent to (\ref{AXzeta equals zero})-(\ref{AXintegral zeta condition}).  Corollary 4.1 on p. 101 of \citeasnoun{hartman1982ordinary} implies that, if, in a parameterized initial value problem, such as (\ref{equivalent differential equation}), the map $\l(\theta,\zeta,\epsilon\r)\mapsto F_\epsilon\l(\theta,\zeta\r)$ is smooth, then the parameterized solution $\l(\theta,\epsilon\r) \mapsto \zeta\l(\theta,\epsilon\r)$ is smooth as well, establishing the desired smoothness of $\zeta\l(\theta,\epsilon\r)$.

It now follows from the fact that the range of $\zeta\l(\theta,\epsilon\r)$, on $\l[-\underline{\theta},\overline{\theta}\r] \times \l[-\epsilon^*,\epsilon^*\r]$, is contained in $\l[-\zeta^*,\zeta^*\r]$ (see the preceding paragraph), combined with (\ref{relabel T theta epsilon def 2}) and (\ref{lemma on regularity}), that, for all $\l(\theta,\epsilon\r) \in \l[-\underline{\theta},\overline{\theta}\r] \times \l[-\epsilon^*,\epsilon^*\r]$, $T^{\theta,\epsilon}$ is regular.  Next, it follows from the smoothness of $\zeta\l(\theta,\epsilon\r)$, established above, together with (\ref{relabel hat T theta epsilon definition}) and (\ref{relabel T theta epsilon def 2}) (and the smoothness of the functions on the right hand side of (\ref{relabel hat T theta epsilon definition})) that $\l(z,\theta,\epsilon\r) \mapsto T\l(z,\theta,\epsilon\r)$ is smooth.  It now follows from Observation \ref{regularity observation}, which says that, for non-individualized tax policies, regularity of each $T^{\theta,\epsilon}$ and smoothness of $\l(z,\theta,\epsilon\r) \mapsto T\l(z,\theta,\epsilon\r)$ is equivalent to well-behavedness, that, if we set  $\underline{\theta}=-\overline{\theta}, \overline{\epsilon}= \epsilon^*, \underline{\epsilon}=-\epsilon^*$, then $\l(T^{\theta,\epsilon}\r)_{\theta \in \l[\underline{\theta},\overline{\theta}\r],\epsilon \in \l[\underline{\epsilon},\overline{\epsilon}\r]}$ is well-behaved.

\subsubsection{\label{hat T theta epsilon well behaved section}The family $\l(\hat{T}^{\theta,\epsilon}\r)$}
This section shows that if $\underline{\theta}'',\overline{\theta}'',\underline{\epsilon}'',$ and $\overline{\epsilon}''$, with $\underline{\theta}'' < 0 < \overline{\theta}'', \underline{\epsilon}'' < 0 < \overline{\epsilon}''$ are all chosen sufficiently close to 0, then the family $\l(T^{\theta,\epsilon}\r)_{\theta \in \l[\underline{\theta}'',\overline{\theta}''\r],\epsilon \in \l[\underline{\epsilon}'',\underline{\epsilon}''\r]}$ is well-behaved. Recall from (\ref{hat defined in terms of non-hat}) that 
\begin{align}\label{relabel hat defined in terms of non-hat}
\hat{T}^{\theta,\epsilon} = T^{\theta,\epsilon} + \Delta T^{\hat{\xi}\l(\theta,\epsilon\r)}. 
\end{align}
First I explain why $\l(z,\theta,\epsilon\r) \mapsto  \Delta T^{\hat{\xi}\l(\theta,\epsilon\r)}\l(z\r)$ is smooth.  Note that $\l(z,\theta,\epsilon\r) \mapsto  \Delta T^{\hat{\xi}\l(\theta,\epsilon\r)}\l(z\r)$ is the composition of the maps $\l(z,\xi\r) \mapsto \Delta T^\xi\l(z\r)$ and $\l(\theta,\epsilon\r) \mapsto \hat{\xi}\l(\theta,\epsilon\r)$.  The smoothness of $\l(z,\xi\r) \mapsto \Delta T^\xi\l(z\r)$ is established by Lemma \ref{indifference and nonconstant revenue family of reforms theorem}. (See in particular the discussion following (\ref{D T xi def}) in Section \ref{completing indifference and nonconstant revenue family of reforms theorem section}.)  The function $\l(\theta,\epsilon\r) \mapsto \hat{\xi}\l(\theta,\epsilon\r)$ is defined by (\ref{fixed revenue condition 0})-(\ref{fixed revenue condition}) via the implicit function theorem and the fact that it is smooth follows from the fact that the other functions in (\ref{fixed revenue condition}) are smooth.\footnote{We have already established the smoothness of $T^{\theta,\epsilon}$ and $\Delta T^\xi$ above, and, noting that each agent's optimal income varies smoothly in response to smooth changes in tax policy, tax revenue also varies smoothly in response to such smooth changes.}  This establishes the smoothness of $\l(z,\theta,\epsilon\r) \mapsto \Delta T^{\hat{\xi}\l(\theta,\epsilon\r)}\l(z\r)$.  The regularity of $\hat{T}^{\theta,\epsilon}$, for each $\theta$ and $\epsilon$, given that $\underline{\theta}'', \overline{\theta}'', \underline{\epsilon}'', \overline{\epsilon}''$ are selected sufficiently close to zero, now follows from a similar argument as that for the regularity of $T_*^{\theta,\epsilon}-\zeta \mu_2$ in the previous section, again appealing to and Lemma \ref{small perturbation still in T hat lemma} and the fact that $\hat{T}^{\theta_0,\epsilon_0} = T$.\footnote{That $\hat{T}^{\theta_0,\epsilon_0} = T$ follows from (\ref{relabel hat defined in terms of non-hat}) and the facts that $T^{\theta_0,\epsilon_0}=T$ and $\Delta T^{\hat{\xi}\l(\theta_0,\epsilon_0\r)} \equiv 0$; see Section \ref{key lemma with revenue section} for this last point.} 

The smoothness, established above, of $\l(z,\theta,\epsilon\r) \mapsto T\l(z,\theta,\epsilon\r)$ and $\l(z,\theta, \epsilon\r) \mapsto \Delta T^{\hat{\xi}\l(\theta,\epsilon\r)}\l(z\r)$, together with (\ref{relabel hat defined in terms of non-hat}) now implies the smoothness of $\l(z,\theta,\epsilon\r) \mapsto \hat{T}^{\theta,\epsilon}\l(z\r)$, which, appealing again to Observation \ref{regularity observation}, completes the argument that $\l(\hat{T}^{\theta,\epsilon}\r)$ is well-behaved.

\subsection{Proof of Proposition \ref{poverty cycle proposition}}

In the poverty alleviation model of Section \ref{poverty alleviation subsection}, condition (\ref{kappa differential equation}) is equivalent to condition (\ref{2}).  Given that $\kappa\l(\theta_0,\epsilon\r)=0,\forall \epsilon$, it follows that $\l.\pdv{\epsilon}\r|_{\epsilon=\epsilon_0} \kappa\l(\theta_0,\epsilon\r)=0$, which implies that the assumption that 
\begin{align}\label{epsilon alpha assumption}
\int g\l(\theta_0,\epsilon_0\r) \l[z_i\l(\theta_0,\epsilon_0\r)-\alpha\r] \dd i =0
\end{align}
is equivalent to (\ref{3}).  

I now establish some facts that will be useful for establishing (\ref{4}).  First, using (\ref{kappa differential equation}), we have   
\begin{align}\label{expression for the theta derivative of kappa}
\begin{split}
\l.\pdv{\theta}\r|_{\theta=\theta_0} \kappa\l(\theta,\epsilon_0\r) =\;& \int \frac{g_i\l(\theta_0,\epsilon_0\r)}{\int g_j \l(\theta_0,\epsilon_0\r) \dd j}\l[z_i\l(\theta_0,\epsilon_0\r)+ f\l(z_i\l(\theta_0,\epsilon_0\r)\r)\r] \dd i\\
=\;& \int \frac{g_i\l(\theta_0,\epsilon_0\r)}{\int g_j \l(\theta_0,\epsilon_0\r) \dd j}z_i\l(\theta_0,\epsilon_0\r) \dd i +\int \frac{g_i\l(\theta_0,\epsilon_0\r)}{\int g_j \l(\theta_0,\epsilon_0\r) \dd j} f\l(z_i\l(\theta_0,\epsilon_0\r)\r) \dd i\\
=\;&\alpha + \underbrace{\int \frac{g_i\l(\theta_0,\epsilon_0\r)}{\int g_j \l(\theta_0,\epsilon_0\r) \dd j} f\l(z_i\l(\theta_0,\epsilon_0\r)\r) \dd i}_\beta,
\end{split}
\end{align}
where the third equality follows from (\ref{epsilon alpha assumption}), and $\beta$ is a label for the last integral.  It follows from the assumptions of Section \ref{poverty alleviation subsection} that $\beta > 0$.  Let $\bar{i}$ be the unique agent satisfying  $z_{\bar{i}}\l(\theta_0,\epsilon_0\r)=\bar{z}$.  That such a $\bar{i}$ exists and is unique follows from Lemma \ref{interval optimum lemma}.  It also follows from Lemma \ref{interval optimum lemma} and the assumptions in Section \ref{poverty alleviation subsection} that all agents in the interval $\l[0,\bar{i}\r)$ earn an income less than $\bar{z}$ when facing tax policy $T^{\theta_0,\epsilon_0}$.  It follows from assumptions on $f$ in Section \ref{poverty alleviation subsection} that, for all incomes $z$ earned by agents in the interval $\l[0,\bar{i}\r]$, when facing tax policy $T^{\theta_0,\epsilon_0}$, $f\l(z\r)=0$, so that for all $i \in \l[0,\bar{i}\r]$, $f'\l(z_i\l(\theta_0,\epsilon_0\r)\r)=0$ and $f''\l(z_i\l(\theta_0,\epsilon_0\r)\r)=0$.  Taking this into account, and applying the implicit function theorem to the first order condition for agents' optimization problem when facing tax policy $T^{\theta_0,\epsilon_0}$, we have         
\begin{align}\label{income derivative poverty}
\begin{split}
\l.\pdv{\theta}\r|_{\theta =\theta_0} z_i\l(\theta,\epsilon_0\r) = \;&-\frac{1}{v''_i\l(z_i\l(\theta_0,\epsilon_0\r)\r)}, 
\\ \l.\pdv{\epsilon}\r|_{\epsilon =\epsilon_0} z_i\l(\theta_0,\epsilon\r) = \;& \frac{1}{v''_i\l(z_i\l(\theta_0,\epsilon_0\r)\r)},
\end{split}\;\;\;\;\; \forall i \in \l[0,\bar{i}\r]. 
\end{align}
Again, using the fact that $f'\l(z_i\l(\theta_0,\epsilon_0\r)\r)=0, \forall i \in \l[0,\bar{i}\r]$ and (\ref{expression for the theta derivative of kappa}), and the fact that $\l.\pdv{\epsilon}\r|_{\epsilon=\epsilon_0} \kappa\l(\theta_0,\epsilon\r)=0$, we have 
\begin{align}\label{tax derivative poverty}
\begin{split}
\l.\pdv{\theta}\r|_{\theta =\theta_0} T\l(z_i\l(\theta_0,\epsilon_0\r),\theta,\epsilon_0\r) =\;&z_i\l(\theta_0,\epsilon_0\r) -\l(\alpha +\beta\r),  \\
\l.\pdv{\epsilon}\r|_{\epsilon =\epsilon_0}T\l(z_i\l(\theta_0,\epsilon_0\r),\theta_0,\epsilon\r)=\;&-z_i\l(\theta_0,\epsilon_0\r) + \alpha,
\end{split} \;\;\; \forall i \in \l[0,\bar{i}\r].   
\end{align}
Using (\ref{income derivative poverty}) and (\ref{tax derivative poverty}), we have that, for all $i \in \l[0,\bar{i}\r]$,  
\begin{align*}
&\l.\pdv{\theta}\r|_{\theta =\theta_0} z_i\l(\theta,\epsilon_0\r) \l.\pdv{\epsilon}\r|_{\epsilon =\epsilon_0} T\l(z_i\l(\theta_0,\epsilon_0\r),\theta_0,\epsilon\r) -  \l.\pdv{\epsilon}\r|_{\epsilon =\epsilon_0} z_i\l(\theta_0,\epsilon\r)\l.\pdv{\theta}\r|_{\theta =\theta_0} T\l(z_i\l(\theta_0,\epsilon_0\r),\theta,\epsilon_0\r)\\
=\;& \l(-\frac{1}{v''_i\l(z_i\l(\theta_0,\epsilon_0\r)\r)}\r)\l(-z_i\l(\theta_0,\epsilon_0\r) + \alpha\r)-\l(\frac{1}{v''_i\l(z_i\l(\theta_0,\epsilon_0\r)\r)}\r)\l(z_i\l(\theta_0,\epsilon_0\r) -\l(\alpha +\beta\r)\r)\\
=\;& \frac{1}{v''_i\l(z_i\l(\theta_0,\epsilon_0\r)\r)}\beta > 0.
\end{align*}
Next recall the relationship between the variables $c_i, \hat{u}_i$ and $z_i$ from Section \ref{structural utilitarian section}:  $c_i=\hat{u}_i + v_i\l(z_i\r)$.  It follows that $\hat{g}_i\l(\hat{u}_i,z_i\r)=\tilde{g}\l(\hat{u}_i+v_i\l(z_i\r)\r)$, and hence $
\pdv{z_i}\hat{g}_i \l(\hat{u}_i, z_i\r) = \tilde{g}'\l(\hat{u}_i + v_i\l(z_i\r)\r) v'_i\l(z_i\r)=\tilde{g}'\l(c_i\r) v'_i\l(z_i\r)$.  It follows from the above that:
\begin{align}\label{simplified double derivative poverty alleviation}
\begin{split}
\int \pdv{z_i}\hat{g}_i\l(\hat{U}_i\l(\theta_0,\epsilon_0\r),z_i\l(\theta_0,\epsilon_0\r)\r)
&\l[\l.\pdv{\theta}\r|_{\theta =\theta_0} z_i\l(\theta,\epsilon_0\r) \l.\pdv{\epsilon}\r|_{\epsilon =\epsilon_0} T\l(z_i\l(\theta_0,\epsilon_0\r),\theta_0,\epsilon\r)\r.\\& \l. -  \l.\pdv{\epsilon}\r|_{\epsilon =\epsilon_0} z_i\l(\theta_0,\epsilon\r)\l.\pdv{\theta}\r|_{\theta =\theta_0} T\l(z_i\l(\theta_0,\epsilon_0\r),\theta,\epsilon_0\r)\r] \dd i\\
=\beta \int_0^{\bar{i}}\tilde{g}'\l(c_i\l(\theta_0,\epsilon_0\r)\r) \frac{v'_i\l(z_i\l(\theta_0,\epsilon_0\r)\r)}{v''_i\l(z_i\l(\theta_0,\epsilon_0\r)\r)} \dd i &< 0, 
\end{split}
\end{align} 
where the upper bound of integration in the second integral follows from that fact that all $i \in \l(\bar{i},1\r]$ are above the poverty line when facing tax policy $T^{\theta_0,\epsilon_0}$ and hence $\hat{g}'\l(c_i\l(\theta_0,\epsilon_0\r)\r)=0$ for all such agents.  The inequality follows from the fact that $v_i'\l(z_i\r) > 0$ and $v''_i\l(z_i\r) > 0$, for all $z_i$, $\tilde{g}'\l(c_i\r) \leq 0$, for all $c_i$, and, since a positive measure of agents in the interval $\l[0,\bar{i}\r]$ is beneath the poverty line at tax policy $T^{\theta_0,\epsilon_0}$,  
$\hat{g}'\l(c_i\l(\theta_0,\epsilon_0\r)\r) < 0$ for a positive measure of agents in $\l[0,\bar{i}\r]$.  It now follows from Lemma  \ref{alternative condition lemma} that the family $\l(T^{\theta,\epsilon}\r)$ in the poverty alleviation model of Section \ref{poverty alleviation subsection} satisfies (\ref{4}). (Note that the proof of Lemma \ref{alternative condition lemma} also establishes that the first integral in (\ref{simplified double derivative poverty alleviation}) is equal to $\l.\dv{\theta}\r|_{\theta = \theta_0} \int g_i\l(\theta,\epsilon_0\r) \l.\pdv{\epsilon}\r|_{\epsilon = \epsilon_0} T\l(z_i\l(\theta,\epsilon_0\r),\theta,\epsilon\r) $, which shows that (\ref{simplified double derivative poverty alleviation}) justifies the inequality that was said to be the key calculation corresponding to (\ref{4}) in Section \ref{poverty alleviation subsection} of the main text.)  $\square$

 \section{\label{supporting lemmas appendix}Lemmas supporting Lemma \ref{key lemma with revenue}}
This section proves Lemmas \ref{indifference and nonconstant revenue family of reforms theorem} and \ref{lemma satisfying desired properties}, to which I appealed in the proof of Lemma \ref{key lemma with revenue}.

\subsection{\label{indifference and nonconstant revenue family of reforms theorem section}Proof of Lemma \ref{indifference and nonconstant revenue family of reforms theorem}}
I begin the proof by establishing a pair of lemmas and then proceed to complete the proof. 

\subsubsection{Lemma \ref{linearity of marginal revenue lemma}}
The following lemma establishes the linearity of the function $f\l(\Delta T\r) = \l.\dv{\varepsilon}\r|_{\varepsilon=0} R\l(T + \varepsilon \Delta T\r)$.   

\begin{lem}\label{linearity of marginal revenue lemma}
Let $T$ be a regular tax policy.  Let $\Delta T_1$ and $\Delta T_2$ be smooth tax reforms and let $r_1$ and $r_2$ be real numbers.  Then $\l.\dv{\varepsilon}\r|_{\varepsilon=0} R\l(T+\varepsilon \l(r_1 \Delta T_1+r_2 \Delta T_2\r)\r)=  r_1\l.\dv{\varepsilon}\r|_{\varepsilon=0} R\l(T + \varepsilon\Delta T_1\r) + r_2 \l.\dv{\varepsilon}\r|_{\varepsilon=0} R\l(T+ \varepsilon \Delta T_1\r)$.  
\end{lem} 
Proof.  Let $\Delta T_1$ and $\Delta T_2$ be smooth tax reforms and let $r_1$ and $r_2$ be real numbers.  Then 
\begin{align}\label{linear effects of tax reforms}
\begin{split}
&\l.\dv{\varepsilon}\r|_{\varepsilon =0}z_i\l(T+ \varepsilon \l(r_1\Delta T^\gamma+r_2 \Delta T_2\r)\r)\\=\;& -\frac{r_1\Delta T'_1 \l(z_i\l(T\r)\r)+r_2 \Delta T'_2 \l(z_i\l(T\r)\r)}{T''\l(z_i\l(T\r)\r)+ v''_i\l(z_i\l(T\r)\r)}\\
=\;&r_1\l(-\frac{\Delta T'_1 \l(z_i\l(T\r)\r)}{T''\l(z_i\l(T\r)\r)+ v''_i\l(z_i\l(T\r)\r)}\r)+  r_2 \l(- \frac{\Delta T'_2 \l(z_i\l(T\r)\r)}{T''\l(z_i\l(T\r)\r)+ v''_i\l(z_i\l(T\r)\r)}\r)\\
=\;& r_1\l.\dv{\varepsilon}\r|_{\varepsilon=0} z_i\l(T+\varepsilon\Delta T_1\r) + r_2 \l.\dv{\varepsilon}\r|_{\varepsilon=0} z_i\l(T+\varepsilon \Delta T_2\r),
\end{split}
\end{align}
where the first and third equalities follow from applying the implicit function theorem to the first order conditions of agent $i$'s optimization problem when facing tax policies $T+ \varepsilon \l(r_1\Delta T_1+r_2 \Delta T_2\r)$, $T+\varepsilon\Delta T_1$, and $T+\varepsilon \Delta T_2$, and $\Delta T'_1\l(z\r)$ and $\Delta T'_2\l(z\r)$, are, respectively, the derivatives of $\Delta T_1\l(z\r)$ and $\Delta T_2\l(z\r)$.  Next, observe that 
\begin{align*}
&\l.\dv{\varepsilon}\r|_{\varepsilon =0} R\l(T+\varepsilon\l(r_1\Delta T_1 +r_2\Delta T_2\r)\r)\\=\;& \int \l[r_1\Delta T_1\l(z_i\l(T\r)\r)+r_2\Delta T_2\l(z_i\l(T\r)\r) \r]\dd i + \int  T'\l(z_i\l(T\r)\r) \l.\dv{\varepsilon}\r|_{\varepsilon =0}z_i\l(T+ \varepsilon \l(r_1\Delta T_1+r_2 \Delta T_2\r)\r)\dd i \\
=\;&\int \l[r_1\Delta T_1 \l(z_i\l(T\r)\r)+r_2\Delta T_2\l(z_i\l(T\r)\r) \r]\dd i \\&+ \int  T'\l(z_i\l(T\r)\r) \l[r_1\l.\dv{\varepsilon}\r|_{\varepsilon=0} z_i\l(T+\varepsilon\Delta T_1\r) + r_2 \l.\dv{\varepsilon}\r|_{\varepsilon=0} z_i\l(T+\varepsilon \Delta T_2\r)\r]\dd i \\
=\;& r_1\l[\int\Delta T_1\l(z_i\l(T\r)\r) \dd i +\int  T'\l(z_i\l(T\r)\r) \l.\dv{\varepsilon}\r|_{\varepsilon=0} z_i\l(T+\varepsilon\Delta T_1\r) \dd i\r]\\
&+ r_2 \l[ \int \Delta T_2\l(z_i\l(T\r)\r) \dd i +\int  T'\l(z_i\l(T\r)\r) \l.\dv{\varepsilon}\r|_{\varepsilon=0} z_i\l(T+\varepsilon \Delta T_2\r) \dd i\r]\\
=&r_1 \l.\dv{\varepsilon}\r|_{\varepsilon =0} R\l(T + \varepsilon\Delta T_1\r) +r_2  \l.\dv{\varepsilon}\r|_{\varepsilon =0} R\l(T + \varepsilon \Delta T_2\r),
\end{align*}
where the second equality follows from (\ref{linear effects of tax reforms}). $\square$

\subsubsection{\label{revenue unchanging welfare changing lemma section}Lemma \ref{revenue unchanging welfare changing lemma}}

Under the assumption that the lowest earned income is positive and the marginal tax rate at the bottom of the income distribution is nonzero, the following lemma establishes the existence of a desirable revenue-neutral tax reform in the generalized welfare weights framework.  This mirrors a standard result in traditional (utilitarian) optimal tax theory. \citeasnoun{saez2016generalized} present a very closely related result, namely an optimal tax formula that, as they observe in their Online Appendix, implies that when the lowest earned income is positive, the marginal tax rate at the bottom of the income distribution is zero in the generalized welfare weights framework, as in the traditional framework. Here, I prove a slightly stronger result than that there is a desirable reform when the bottom rate is nonzero: I also establish that the desirable reform can be assumed to have certain additional properties that are useful for our purposes. 

\begin{lem}\label{revenue unchanging welfare changing lemma}
Let $T$ be a regular tax policy (so that in particular $z_0\l(T\r)>0$), and suppose that  $T'\l(z_{0}\l(T\r)\r) \neq 0$.  Let $z_*$ be such that $z_{0}\l(T\r) < z_* \leq z_{1}\l(T\r)$.  Then there exists a desirable revenue neutral tax reform $\Delta T$ with support contained in $\l[0,z_*\r]$; formally, there exists a smooth tax reform $\Delta T$ with support contained in $\l[0,z_*\r]$ such that $\l.\dv{\varepsilon}\r|_{\varepsilon =0} R\l(T+ \varepsilon \Delta T\r) =0$ and $\int g_i\l(T\r) \Delta T\l(z_{i}\l(T\r)\r) \dd i <0$.  Moreover, there exist smooth tax reforms $\Delta T_1, \Delta T_2,$ with supports contained in $\l[0,z_*\r]$ such that $\Delta T = \Delta T_1 - \Delta T_2, \Delta T_2\l(z\r) \geq 0, \forall z,$ and $\l.\dv{\varepsilon}\r|_{\varepsilon=0} R\l(T + \varepsilon \Delta T_2\r) \neq 0$.  \end{lem}
\noindent Recall that the set of agents is $I=\l[0,1\r]$, $z_0\l(T\r)$ and $z_1\l(T\r)$ are the optimal responses to $T$ for agents $0$ and $1$ respectively.  By the assumptions of Section \ref{additional structure section}, $z_0\l(T\r)$ and $z_1\l(T\r)$ are respectively the bottom and top of the income distributions earned in response to $T$ (see also Lemma \ref{interval optimum lemma}).

I begin by stating some useful background facts and then proceed to prove the lemma. 
\paragraph{Background facts}
Choose a regular tax policy $T$, let $z_0=z_0\l(T\r)$, and $z_1=z_1\l(T\r)$.  Define the function $\zeta: I \rightarrow Z$ by $\zeta\l(i\r)=z_i\l(T\r)$, and let $\iota=\zeta^{-1}$ be the inverse of $\zeta$ so that $\iota\l(z\r)=i$ if and only if $z_i\l(T\r)=z$.  It follows from our assumptions in Section \ref{additional structure section} that $\zeta\l(0\r)=z_0\l(T\r) > 0$ and $\zeta\l(i\r)$ is strictly increasing in $i$.  Let $H$ be the cumulative distribution over incomes induced by tax policy $T$.  Then, recalling that agents are uniformly distributed on the interval $I=\l[0,1\r]$, it follows that $H\l(z\r) = 0$ for all $z \in Z$ such that $z < z_0$; $H\l(z\r) = \iota\l(z\r)$ for all $z \in Z$ with $z_0 \leq z \leq z_1$; and $H\l(z\r)= 1$ for all $z \in Z$ with $z_0 < z$.  So if $h$ is the density corresponding to the cumulative distribution $H$, we have $h\l(z\r)= H'\l(z\r) = \iota'\l(z\r)= \frac{1}{\zeta'\l(\iota\l(z\r)\r)}$ for all $z \in \l[z_0,z_1\r]$;\footnote{Strictly speaking, $h\l(z\r)$ is, respectively, the right- and left-derivative of $H\l(z\r)$ at $z=z_0$ and $z=z_1$, and we have $h\l(z_0\r)= \frac{1}{\zeta'\l(\iota\l(z_0\r)\r)}= \frac{1}{\zeta'\l(0\r)}$ and $h\l(z_1\r)=\frac{1}{\zeta'\l(1\r)}$.} and $h\l(z\r)=0$ for all $z \in Z$ with $z \not \in \l[z_0,z_1\r]$.  Observe, using a change of variables, that given a smooth tax reform $\Delta T$:
\begin{align}\label{change of variables derivative of revenue}
\begin{split}
\l.\dv{\varepsilon}\r|_{\varepsilon=0} R\l(T+ \varepsilon\Delta T\r) =\;& \int_0^1  \Delta T\l(z_i\l(T\r)\r) \dd i + \int_0^1 T'\l(z_i\l(T\r)\r) \l.\dv{\varepsilon}\r|_{\varepsilon=0} z_i\l(T + \varepsilon \Delta T\r) \dd i  \\
=\;&\int_0^1  \Delta T\l(z_i\l(T\r)\r) \dd i - \int_0^1 T'\l(z_i\l(T\r)\r) \frac{\Delta T'\l(z_i\l(T\r)\r)}{T''\l(z_i\l(T\r)\r) + v''_i\l(z_i\l(T\r)\r)} \dd i  \\
=\;&\int_{z_0}^{z_1}  \Delta T\l(z\r) h\l(z\r) \dd z - \int_{z_0}^{z_1} T'\l(z\r) \frac{\Delta T'\l(z\r)}{T''\l(z\r) + v''_{\iota\l(z\r)}\l(z\r)}h\l(z\r) \dd z\\
=\; & \int_{z_0}^{z_1}  \Delta T\l(z\r) h\l(z\r) \dd z - \int_{z_0}^{z_1} \Delta T'\l(z\r)  k_T\l(z\r) \dd z
\end{split}
\end{align}
where $\Delta T'$ is the derivative of $\Delta T$, $v''_{\iota\l(z\r)}\l(z\r)$ is $v''_i\l(z\r)$ evaluated at $i=\iota\l(z\r)$, and the second equality follows from applying the implicit function theorem to the first order condition for an agent's optimization problem when facing tax policy $T+\varepsilon \Delta T\l(z\r)$ around $\varepsilon =0$ and 
\begin{align}\label{definition k(z)}
k_T\l(z\r)=\frac{T'\l(z\r)h\l(z\r)}{T''\l(z\r) + v''_{\iota\l(z\r)}\l(z\r)}, \;\;\;\; \forall z \in \l[z_0,z_1\r]. 
\end{align}
Note that we include the subscript $T$ in $k_T$ to express the dependence of $k_T$ on the tax policy $T$ through the terms $T'\l(z\r)$ and $T''\l(z\r)$.  It follows from the fact that, for all regular $T$ and all $i \in I$, $\l.\dv[2]{z_i}\r|_{z_i=z_i\l(T\r)}u\l(z_i-T\l(z_i\r)-v_i\l(z_i\r)\r) < 0$ (see Section \ref{wb appendix non-individualized}), that the denominator on the right hand side of (\ref{definition k(z)}) is positive for all $z \in \l[z_0,z_1\r].$  Moreover the assumptions on $v_i$ and $y_i$ in Section \ref{additional structure section} imply that $\zeta'\l(i\r) > 0, \forall i \in I$, and hence that $h\l(z\r) > 0, \forall z \in \l[z_0,z_1\r]$.  It then follows that:
\begin{align}\label{sign of kT}
T'\l(z_0\r) \neq 0  \Rightarrow k_T\l(z_0\r) \neq 0.
\end{align}

\paragraph{Main argument}
Again, let $z_0 = z_0\l(T\r)$ and $z_1=z_1\l(T\r)$.  Choose $z_*$ such that $z_0 < z_* \leq z_1$.  Let $z_0  = z_0\l(T\r)$. As $T$ is regular, it follows that $z_0 > 0$ (see Section \ref{wb appendix non-individualized}).  Choose $z_-$ so that $0 < z_- < z_0$.   Consider a smooth tax reform $\Delta \hat{T}_1$ with $\Delta \hat{T}_1\l(z\r)= 2, \forall z \in \l[0,z_-\r],  \Delta \hat{T}'_1\l(z\r) < 0, \forall z \in \l(z_-,z_*\r), \Delta \hat{T}_1\l(z_0\r)=1$, and $\Delta \hat{T}_1\l(z\r)=0, \forall z \in \l[z_*,+\infty\r)$.  So the smooth tax reform $\Delta \hat{T}_1$ equal to $2$ until $z=z_-$, at which point it falls, passing through $\Delta \hat{T}_1=1$ when $z=z_0$, and reaching $\Delta \hat{T}_1=0$ at $z=z_*$ and remains at zero thereafter.  

For each $\gamma \in \l[1,+\infty\r)$, define $z^\gamma_-, z^\gamma_*$ by $\gamma \l(z^\gamma_--z_0\r) + z_0 = z_-, \gamma \l(z^\gamma_*-z_0\r)+z_0 = z_*$.  For $\gamma \geq 1$, we have $ z^\gamma_- < z_0
  < z^\gamma_*$; and $z^\gamma_- \uparrow z_0$ and $z^\gamma_* \downarrow z_0$ as $\gamma \uparrow +\infty$.   Define $i^\gamma$  by the condition $z_{i^\gamma}\l(T\r) = z^\gamma_*$; that such an $i_\gamma$ exists follows from Lemma \ref{interval optimum lemma}.  Using assumptions in Section \ref{additional structure section}, we have $i^\gamma \downarrow 0$ as $\gamma \uparrow +\infty$. 

Define the tax reform $\Delta T^\gamma_1$ by 
\begin{align}\label{Delta T gamma definitions}
\Delta T^\gamma_1 \l(z\r)= \begin{cases} 2, &\textup{if $z \in \l[0,z^\gamma_-\r]$},\\
\Delta \hat{T}_1\l(\gamma \l(z - z_0\r)+ z_0\r), &\textup{if $z \in \l(z^\gamma_-,z^\gamma_*\r)$},\\
0, & \textup{if $z \in \l[z^\gamma_*,+\infty\r)$ }.
\end{cases} 
\end{align}
Using the properties of $\Delta \hat{T}_1$, it is straightforward to verify that, for all $\gamma \geq 1$, $\Delta T^\gamma_1$ is a smooth function of $z$.  So $\Delta T^\gamma_1$ is similar to $\Delta \hat{T}_1$, except that in the former $z^\gamma_-$ and $z^\gamma_*$ play the roles of $z_-$ and $z_*$ in the latter. For $\gamma > 1, \Delta T^\gamma_1$ falls more steeply than $\Delta \hat{T}_1$ near $z=z_0$.  Observe that, for all $\gamma \geq 1$, $\l[0,z^\gamma_*\r]$ is the support of both $\Delta T^\gamma_1$, so that the support of $\Delta T^\gamma_1$ is contained in $\l[0,z_*\r]$.  

\begin{lem}\label{gamma limit nonzero revenue lemma}
Assume, as above, that $T'\l(z_0\r) \neq 0$.  Then $\lim_{\gamma \rightarrow \infty} \l.\dv{\varepsilon}\r|_{\varepsilon=0} R\l(T +\varepsilon \Delta T^\gamma_1\r) \neq 0$.  
\end{lem}  
Lemma \ref{gamma limit nonzero revenue lemma} is proven in Section \ref{gamma limit nonzero revenue lemma section}.    

Since we are assuming that $T'\l(z_0\r) \neq 0$, it follows from Lemma \ref{gamma limit nonzero revenue lemma} that there exists a tax reform $\Delta \hat{T}_2$ with support contained in $\l[0,z^*\r]$ such that $\Delta \hat{T}_2\l(z\r) \geq 0, \forall z \in Z$, and    
\begin{align}\label{Delta T 2 nonzero marginal revenue}
\l.\dv{\varepsilon}\r|_{\varepsilon=0} R\l(T+ \varepsilon\Delta \hat{T}_2\r) \neq 0.
\end{align}
In particular, we can choose $\Delta \hat{T}_2= \Delta \hat{T}^{\gamma_0}_1$ for some sufficiently large $\gamma_0$.  However, for our purposes, it is not important whether $\Delta \hat{T}_2= \Delta \hat{T}^{\gamma_0}_1$ for some sufficiently large (fixed) $\gamma_0$; it matters only that is has the properties we have just ascribed to it.         

It follows from Lemma \ref{linearity of marginal revenue lemma} and (\ref{Delta T 2 nonzero marginal revenue}) that, for each $\gamma >1$, there exists $r_\gamma$ such that  
\begin{align}\label{Revenue T r gamma Delta T 2 nonzero}
\l.\dv{\varepsilon}\r|_{\varepsilon =0} R\l(T+\varepsilon\l(\Delta T^\gamma_1-r_\gamma\Delta \hat{T}_2\r)\r)=0.
\end{align}
It follows from Lemma \ref{linearity of marginal revenue lemma}, (\ref{Delta T 2 nonzero marginal revenue}), and (\ref{Revenue T r gamma Delta T 2 nonzero}) that 
\begin{align}\label{r gamma ratio}
r_\gamma = \frac{\l.\dv{\varepsilon}\r|_{\varepsilon =0} R\l(T + \varepsilon \Delta T^\gamma_1\r)}{\l.\dv{\varepsilon}\r|_{\varepsilon =0} R\l(T + \varepsilon \Delta \hat{T}_2\r)}.
\end{align}
The negative of the marginal welfare effect of a small tax reform in direction $\Delta T^{\gamma}_1 - r_\gamma \Delta \hat{T}_2$ is
\begin{align*}
W_\gamma = \int g_i\l(T\r) \Delta T^\gamma\l(z_i\l(T\r)\r)  \dd i - r_\gamma \int g_i\l(T\r) \Delta \hat{T}_2\l(z_i\l(T\r)\r)  \dd i.
\end{align*}
Observe that because (i) the function $i \mapsto \Delta T^\gamma\l(z_i\l(T\r)\r)$, whose domain is $\l[0,1\r]$, has support $\l[0,i^\gamma\r]$ and $i^\gamma \downarrow 0$ as $\gamma$ approaches infinity and (ii) $\Delta T^\gamma\l(z\r)$ is bounded between $0$ and $2$, for all $z$, it follows that $\int g_i\l(T\r) \Delta T^\gamma\l(z_i\l(T\r)\r)  \dd i \rightarrow 0$ as $\gamma \rightarrow \infty$.  Note that because $\Delta \hat{T}_2\l(z\r) \geq 0, \forall z \in Z$, and $\Delta \hat{T}_2$ satisfies (\ref{Delta T 2 nonzero marginal revenue}), there must be a positive measure set of agents $i$ such that $\Delta \hat{T}_2\l(z_i\l(T\r)\r) > 0$.  It follows that  $\int g_i\l(T\r) \Delta \hat{T}_2\l(z_i\l(T\r)\r)  \dd i >0$. Lemma \ref{gamma limit nonzero revenue lemma} and (\ref{r gamma ratio}) imply that $\lim_{\gamma \rightarrow \infty} r_\gamma \neq 0$.\footnote{Observe that, from (\ref{change of variables derivative of revenue}), $\l.\dv{\varepsilon}\r|_{\varepsilon =0} R\l(T + \varepsilon \Delta \hat{T}_2\r)=\int_{z_0}^{z_1}  \Delta \hat{T}_2\l(z\r) h\l(z\r) \dd z - \int_{z_0}^{z_1} \Delta T'\l(z\r)  k_T\l(z\r) \dd z$, which is finite, so the denominator in (\ref{r gamma ratio}) is finite as well.}  It now follows from the results of the previous paragraph that if $\gamma$ is sufficiently large then $W_\gamma \neq 0$.  Then choose such a sufficiently large $\gamma$.  If $W_\gamma < 0$, then define $\Delta T_1= \Delta T^\gamma_1$ and $\Delta T_2 = r_\gamma \Delta \hat{T}_2$; and if $W_\gamma > 0$,  define $\Delta T_1=- \Delta T^\gamma_1$ and $\Delta T_2 = -r_\gamma \Delta \hat{T}_2$.  In either case define $\Delta T= \Delta T_1 - \Delta T_2$.  In both cases, we have $\int g_i\l(T\r) \Delta T\l(z_i\l(T\r)\r) \dd i < 0$ and, appealing to Lemma \ref{linearity of marginal revenue lemma}, (\ref{Delta T 2 nonzero marginal revenue}), and (\ref{Revenue T r gamma Delta T 2 nonzero}), $\l.\dv{\varepsilon}\r|_{\varepsilon=0}R\l(T + \varepsilon \Delta T\r) = 0$ and $\l.\dv{\varepsilon}\r|_{\varepsilon=0}R\l(T + \varepsilon \Delta T_2\r) \neq 0$. Finally note the support of $\Delta T$, $\Delta T_1$, and $\Delta T_2$ are all contained in $\l[0,z_*\r]$. We have now established all of the properties required by Lemma \ref{revenue unchanging welfare changing lemma}. $\square$

\paragraph{\label{gamma limit nonzero revenue lemma section}Proof of Lemma \ref{gamma limit nonzero revenue lemma}.}
It follows from (\ref{change of variables derivative of revenue}) and the fact that the support of $\Delta T^\gamma_1$ is $\l[0,z^\gamma_*\r]$ that 
  \begin{align}\label{marginal revenue in direction Delta T 1 gamma}
\l.\dv{\varepsilon}\r|_{\varepsilon=0} R\l(T +\varepsilon \Delta T_1^\gamma\r) = & \int_{z_0}^{z^\gamma_*}  \Delta T_1^\gamma\l(z\r) h\l(z\r) \dd z - \int_{z_0}^{z^\gamma_*} \dv{z}\Delta T_1^\gamma\l(z\r)  k_T\l(z\r) \dd z,
\end{align}
where $\dv{z} \Delta T^\gamma_1\l(z\r)$ is the derivative of $\Delta T^\gamma_1\l(z\r)$.  Because $z^\gamma_* \downarrow z_0$ as $\gamma \rightarrow \infty$ and $\Delta T^\gamma\l(z\r)$ is bounded between $0$ and $2$ for all $z$,  
\begin{align}\label{limit of first integral is zero}
\lim_{\gamma \rightarrow \infty}\int_{z_0}^{z^\gamma_*}  \Delta T_1^\gamma\l(z\r) h\l(z\r) \dd z=0.
\end{align}
Since $\dv{z}\Delta T^\gamma_1\l(z\r) \leq 0, \forall z \in Z$, it follows from the preceding that, if $\gamma$ is sufficiently large, we have:  
\begin{align}\label{bounds on the integral}
\begin{split}
\l(\max_{z\in \l[z_0,z^\gamma_*\r]} k_T\l(z\r)\r) \times \int_{z_0}^{z^\gamma_*} \dv{z}\Delta T_1^\gamma\l(z\r)   \dd z \leq\;& \int_{z_0}^{z_*^\gamma} \dv{z}\Delta T_1^\gamma\l(z\r)  k_T\l(z\r) \dd z\\ \leq\;& \l(\min_{z\in \l[z_0,z^\gamma_*\r]} k_T\l(z\r)\r) \times \int_{z_0}^{z^\gamma_*} \dv{z}\Delta T_1^\gamma\l(z\r)   \dd z.
\end{split}
\end{align}
Next observe that
\begin{align}\label{change of variables integral}
\int_{z_0}^{z^\gamma_*} \dv{z}\Delta T_1^\gamma\l(z\r)   \dd z=\int_{z_0}^{z^\gamma_*}\gamma\Delta \hat{T}'_1 \l(\gamma \l[z-z_0\r]+z_0\r)\dd z =  \int_{z_0}^{z_*}\Delta \hat{T}'_1 \l(\tilde{z}\r)\dd \tilde{z} = \Delta \hat{T}_1\l(z_*\r)-\hat{T}_1\l(z_0\r)=-1,
\end{align}
where $\Delta \hat{T}'_1\l(\tilde{z}\r)$ is the derivative of $\Delta \hat{T}_1\l(\tilde{z}\r)$ and the second equality uses the change of variables $z \mapsto \tilde{z}=\gamma\l[z-z_0\r] + z_0$.  Next observe that as $k$ is smooth and $z^\gamma \downarrow z_0$ and $\gamma \rightarrow \infty$,
\begin{align}\label{limits on the k T s}
\lim_{\gamma \rightarrow \infty} \max_{z\in \l[z_0,z^\gamma\r]} k_T\l(z\r) = k_T\l(z_0\r) \textup{ and }\lim_{\gamma \rightarrow \infty} \min_{z\in \l[z_0,z^\gamma\r]} k_T\l(z\r) =k_T\l(z_0\r).
\end{align}
It follows from (\ref{marginal revenue in direction Delta T 1 gamma}), (\ref{limit of first integral is zero}), (\ref{bounds on the integral}), (\ref{change of variables integral}), and (\ref{limits on the k T s}) that 
\begin{align*}
\lim_{\gamma \rightarrow \infty} \l.\dv{\varepsilon}\r|_{\varepsilon=0} R\l(T +\varepsilon \Delta T_1^\gamma\r) = k_T\l(z_0\r).
\end{align*}
It follows from (\ref{sign of kT}) and the assumption that $T'\l(z_0\r)\neq 0$ that $k_T\l(z_0\r) \neq 0$, which completes the proof. $\square$

\subsubsection{\label{completing indifference and nonconstant revenue family of reforms theorem section}Completion of the proof of Lemma \ref{indifference and nonconstant revenue family of reforms theorem}}

Lemma \ref{revenue unchanging welfare changing lemma} established that, under certain conditions, there exists a desirable revenue neutral tax reform.  It is intuitive that, starting from such a reform, and adding an appropriate lumpsum tax, one can attain a welfare-neutral reform that raises revenue. Lemma \ref{indifference and nonconstant revenue family of reforms theorem} establishes the existence of something similar: a parameterized family of tax reforms (at a subset of tax policies in the family $\l(T^{\theta,\epsilon}\r)$) such that varying the parameter affects revenue but is socially indifferent according to welfare weights.  This family is not constructed by modifying a desirable revenue neutral reform via a lumpsum tax, which would affect all taxpayers, but rather by local change in taxes that affects only taxpayers at the bottom of the income distribution.

I now use Lemmas \ref{linearity of marginal revenue lemma} and \ref{revenue unchanging welfare changing lemma} to prove Lemma \ref{indifference and nonconstant revenue family of reforms theorem}.  Recall from Section \ref{key lemma with revenue section} that $T^{\theta_0,\epsilon_0}=T$ for a regular tax policy $T$ satisfying $T'\l(z_0\l(T\r)\r) \neq 0$.  Recall also from Section \ref{key lemma with revenue section} that $\hat{z}_1=z_{i_1}\l(T\r)$, and moreover, $z_0\l(T\r)< \hat{z}_1 <z_1\l(T\r)$.  So letting $\hat{z}_1$ play the role of $z_*$ in Lemma \ref{revenue unchanging welfare changing lemma}, there exist tax reforms $\Delta T_1, \Delta T_2$, and $\Delta T$, all with supports contained in $\l[0,\hat{z}_1\r]$ and satisfying the properties in Lemma \ref{revenue unchanging welfare changing lemma} in relation to the tax policy $T=T^{\theta_0,\epsilon_0}$. 
Define the function\footnote{It follows from Lemma \ref{small perturbation still in T hat lemma} that if 
 if $\xi$ and $r$ are sufficiently close to $0$, then $T + \xi \Delta T_1-r\Delta T_2$ is regular, and hence $z_i\l(T + \xi \Delta T_1-r\Delta T_2 \r)$ is uniquely defined, and so $g_i\l(T + \xi \Delta T_1-r\Delta T_2 \r)$ is also uniquely defined.}
\begin{align}\label{F xi r definition}
 F\l(\xi,r\r) = \frac{\int g_i\l(T+\xi \Delta T_1-r \Delta T_2\r) \Delta T_1\l(z_i\l(T+\xi \Delta T_1-r \Delta T_2\r)\r) \dd i}{\int g_i\l(T+\xi \Delta T_1-r \Delta T_2\r) \Delta T_2\l(z_i\l(T+\xi \Delta T_1-r \Delta T_2\r)\r) \dd i}. 
\end{align}
It follows from the properties stated in Lemma \ref{revenue unchanging welfare changing lemma} (which apply to $T, \Delta T_1$ and $\Delta T_2$) and the smoothness of the relevant functions that if $\xi$ and $r$ are sufficiently close to zero, then the denominator in the above expression is nonzero.\footnote{\label{positive integral denominator footnote}In particular, the facts that $\Delta T_2\l(z\r) \geq 0, \forall z$, and $\l.\dv{\varepsilon}\r|_{\varepsilon=0} R\l(T+\varepsilon \Delta T_2\r) \neq 0$ imply that there exists a positive measure of agents $i$ such that $\Delta T_2\l(z_i\l(T\r)\r) >0$, hence, invoking again $\Delta T_2\l(z\r) \geq 0, \forall z$, it follows that from the fact that welfare weights are positive $\int g_i\l(T\r) \Delta T_2\l(z_i\l(T\r)\r) \dd i >0$.  That the denominator of (\ref{F xi r definition}) is positive now follows from our smoothness assumptions.} Note that $F$ is smooth in its arguments.  It follows from the Picard-Lindel\"{o}f  theorem that there exist real numbers $\underline{\xi}, \overline{\xi}$ with $\underline{\xi} < 0< \overline{\xi}$ and a function $s: \l[\underline{\xi},\overline{\xi}\r] \rightarrow \mathbb{R}$ satisfying
\begin{align}
\label{r initial condition} s\l(0\r)&=0,\\
\label{r evolution} s'\l(\xi\r)&= F\l(\xi,s\l(\xi\r)\r), \;\;\;\; \forall \xi \in \Xi,
\end{align}
where $\Xi=\l[\underline{\xi},\overline{\xi}\r]$.  Define the family of tax reforms $\l(\Delta T^\xi\r)_{\xi \in \Xi}$ by the condition
\begin{align}\label{D T xi def}
\Delta T^\xi =  \xi \Delta T_1 - s\l(\xi\r) \Delta T_2, \;\;\; \forall \xi \in \Xi. 
\end{align} 
Observe that $\Delta T^0 \equiv 0$ and, for all $\xi \in \Xi$, the support of $\Delta T^\xi$ is contained in $\l[0,\hat{z}_1\r]=\l[0,z_*\r]$ because the supports of $\Delta T_1$ and $\Delta T_2$ are contained in $\l[0,\hat{z}_1\r]$.  It follows from the smoothness of the function $F\l(\xi,r\r)$, and Corollary 4.1 on p. 101 of \citeasnoun{hartman1982ordinary} that the function $s\l(\xi\r)$ is smooth, and hence also, given the smoothness of $\Delta T_1$ and $\Delta T_2$, that the map $\l(z,\xi\r) \mapsto \Delta T^\xi\l(z\r)$ is smooth.  Lemma \ref{small perturbation still in T hat lemma} implies that it is possible to choose $\underline{\xi}$ and $\overline{\xi}$, and also $\underline{\theta}',\overline{\theta}' \in \Theta, \underline{\epsilon}',\overline{\epsilon}' \in E$ with $\underline{\theta}' < 0 < \overline{\theta}', \underline{\epsilon}' < 0< \overline{\epsilon}'$  so that $T^{\theta,\epsilon}+ \Delta T^\xi= T+ \l[\theta \times \mu_1\r] -\l[\zeta\l(\theta,\epsilon\r) \times \mu_2\r]+ \l[\epsilon \times \l(\eta_1 + \eta_2\r)\r]+ \Delta T^\xi$ is regular, for all $\theta \in \Theta'=\l[\underline{\theta}',\overline{\theta}'\r], \epsilon \in E'=\l[\underline{\epsilon}',\overline{\epsilon}'\r],$ and $\xi \in \Xi =\l[\underline{\xi},\overline{\xi}\r]$.  So let us assume that $\underline{\xi},\overline{\xi}, \underline{\theta}',\overline{\theta}', \underline{\epsilon}'$, and $\overline{\epsilon}'$ are so chosen.     

Next observe that 
 \begin{align}\label{relation of two reforms}
\pdv{\xi}\Delta T \l(z,\xi\r)=\Delta T_1\l(z\r) - s'\l(\xi\r) \Delta T_2\l(z\r),  \;\;\; \forall \xi \in \Xi, \forall z \in Z,   
\end{align}
where $\Delta T\l(z,\xi\r)=\Delta T^\xi\l(z\r)$.  Recalling that $T^{\theta_0,\epsilon_0}=T$, and using (\ref{F xi r definition}), (\ref{D T xi def}), and (\ref{relation of two reforms}), it follows that (\ref{r evolution}) is equivalent to 
\begin{align}\label{indifference 3 simpler}
\int g_i\l(T+\Delta T^\xi\r) \l.\pdv{\xi}\r|_{\xi=\xi'}\Delta T\l(z_i\l(T+\Delta T^{\xi'}\r),\xi\r) \dd i =0, \;\;\; \forall \xi' \in \Xi.
\end{align}
It follows that the family $\l(\Delta T^\xi\r)$ satisfies (\ref{indifference 3 simpler}). 

Taking the derivative of the relation (\ref{relation of two reforms}) with respect to $z$ yeilds:
 \begin{align}\label{relation of two reforms derivative}
\pdv[2]{}{\xi}{z}\Delta T \l(z,\xi\r)=\Delta T'_1\l(z\r) - s'\l(\xi\r) \Delta T'_2\l(z\r),  \;\;\; \forall \xi \in \Xi, \forall z \in Z,   
\end{align}
Now consider the tax reform $\Delta T_1-s'\l(0\r) \Delta T_2$.  This is just the tax reform $\Delta T_1 - r \Delta T_2$ in the special case in which $r=s'\l(0\r)$.  When facing the tax policies $T+\varepsilon\l(\Delta T_1 - s'\l(0\r) \Delta T_2\r)$ and $T + \Delta T^\xi$, agent $i$ faces, respectively, optimization problems $\max_{z_i} \l[z_i - T\l(z_i\r) - \varepsilon \l(\Delta T_1\l(z_i\r) - s'\l(0\r)\Delta T_2\l(z_i\r)\r) - v_i\l(z_i\r)\r]$ and $\max_{z_i} \l[z_i - T\l(z_i\r) - \Delta T^\xi\l(z_i\r)-v_i\l(z_i\r)\r]$.  Note that, because $T$ is regular, when $\varepsilon$ and $\xi$ are sufficiently small, all agents select an interior income (see also Lemma \ref{small perturbation still in T hat lemma}).  Applying the implicit function theorem to the agent's first order conditions for these two problems, we have:
\begin{align}\label{relation of two optimal incomes}
\begin{split}
&\l.\dv{\varepsilon}\r|_{\varepsilon=0} z_i\l(T+\varepsilon \l(\Delta T_1 - s'\l(0\r) \Delta T_2\r)\r)= -\frac{\Delta  T'_1\l(z_i\l(T\r)\r)-s'\l(0\r)\Delta  T'_2\l(z_i\l(T\r)\r) }{T''\l(z_i\l(T\r)\r)+v''_i\l(z_i\l(T\r)\r)}\\ &=- \frac{\l.\pdv[2]{}{\xi}{z}\r|_{\xi=0,z=z_i\l(T\r)} \Delta T\l(z,\xi\r)}{T''\l(z_i\l(T\r)\r)+v''_i\l(z_i\l(T\r)\r)}= \l.\dv{\xi}\r|_{\xi=0} z_i\l(T+\Delta T^\xi\r),
\end{split}
\end{align}
where the second equality follows from (\ref{relation of two reforms derivative}).  This, in turn, implies that 
\begin{align}\label{marginal revenue equivalence}
\begin{split}
&\l.\dv{\varepsilon}\r|_{\varepsilon=0} R\l(T+ \varepsilon \l(\Delta T_1-s'\l(0\r) \Delta T_2\r)\r) \\=& \int \l[ \Delta T_1\l(z_i\l(T\r)\r) - s'\l(0\r) \Delta T_2\l(z_i\l(T\r)\r) \r]\dd i + \int T'\l(z_i\l(T\r)\r) \l.\dv{\varepsilon}\r|_{\varepsilon=0} z_i\l(T+\varepsilon \l(\Delta T_1 - s'\l(0\r) \Delta T_2 \r)\r)\dd i\\
=& \int \l.\pdv{\xi}\r|_{\xi=0} \Delta T \l(z_i\l(T\r),\xi\r)\dd i + \int T'\l(z_i\l(T\r)\r) \l.\dv{\xi}\r|_{\xi=0} z_i\l(T+\Delta T^\xi\r)  \dd i\\
=&\l.\dv{\xi}\r|_{\xi=0} R\l(T+ \Delta T^\xi\r),
\end{split} 
\end{align}
where the second equality uses (\ref{relation of two reforms}) and (\ref{relation of two optimal incomes}).

Next observe that, by the properties implied by Lemma \ref{revenue unchanging welfare changing lemma}, $0 > \int g_i\l(T\r) \Delta T\l(z_i\l(T\r)\r) \dd i =  \int g_i\l(T\r) \Delta T_1\l(z_i\l(T\r)\r) \dd i - \int g_i\l(T\r) \Delta T_2\l(z_i\l(
T\r)\r) \dd i$.  So, since, again by the properties in Lemma \ref{revenue unchanging welfare changing lemma}, $  \int g_i\l(T\r) \Delta T_2\l(z_i\l(T\r)\r) \dd i>0$ (see footnote \ref{positive integral denominator footnote} of the Appendix), it follows that  $F\l(0,0\r) = \frac{\int g_i\l(T\r) \Delta T_1\l(z_i\l(T\r)\r) \dd i}{\int g_i\l(T\r) \Delta T_2\l(z_i\l(T\r)\r) \dd i} < 1$.  So, by (\ref{F xi r definition}) and (\ref{r evolution}), $s'\l(0\r) < 1$.  It follows from Lemma \ref{linearity of marginal revenue lemma} and the properties of Lemma \ref{revenue unchanging welfare changing lemma} that
\begin{align*}
0 =\;&\l.\dv{\varepsilon}\r|_{\varepsilon =0} R\l(T+ \varepsilon \Delta T\r) = \l.\dv{\varepsilon}\r|_{\varepsilon =0} R\l(T+ \varepsilon \l(\Delta T_1-\Delta T_2\r)\r)\\=\;& \l.\dv{\varepsilon}\r|_{\varepsilon =0} R\l(T+ \varepsilon \Delta T_1\r)-\l.\dv{\varepsilon}\r|_{\varepsilon =0} R\l(T+ \varepsilon \Delta T_2\r) \\
\neq\;&  \l.\dv{\varepsilon}\r|_{\varepsilon =0} R\l(T+ \varepsilon \Delta T_1\r)-s'\l(0\r)\l.\dv{\varepsilon}\r|_{\varepsilon =0} R\l(T+ \varepsilon \Delta T_2\r)\\ 
 =\;&\l.\dv{\varepsilon}\r|_{\varepsilon =0} R\l(T+ \varepsilon \l(\Delta T_1-s'\l(0\r)\Delta T_2\r)\r)\\
 =\; & \l.\dv{\xi}\r|_{\xi=0} R\l(T+ \Delta T^\xi\r),
\end{align*}  
where the non-equality $\neq$ in the above derivation follows from the facts that $s'\l(0\r) \neq 1$ and $\l.\dv{\varepsilon}\r|_{\varepsilon =0} R\l(T+ \varepsilon \Delta T_2\r) \neq 0$ (see Lemma \ref{revenue unchanging welfare changing lemma} for the latter).  So, to summarize,  $\l.\dv{\xi}\r|_{\xi=0} R\l(T+ \Delta T^\xi\r) \neq 0$.  By our smoothness assumptions, if $\underline{\xi}$ and $\overline{\xi}$ in $\Xi=\l[\underline{\xi},\overline{\xi}\r]$ are selected so as to be sufficiently close to $0$,    
\begin{align}\label{nonconstant revenue3 simpler}
 \l.\dv{\xi}\r|_{\xi=\xi'} R\l(T+ \Delta T^\xi\r) \neq 0, \;\;\; \forall \xi' \in \Xi.
\end{align}
Let us assume that $\underline{\xi}$ and $\overline{\xi}$ are so chosen.

Using the facts that, by the construction of $\l(T^{\theta,\epsilon}\r)$, $T^{\theta,\epsilon}\l(z\r) + \Delta T^{\xi}\l(z\r) =T\l(z\r) +  \Delta T^{\xi}\l(z\r), \forall z \in \l[0,\hat{z}_1\r], \forall \theta \in \Theta', \forall \epsilon \in E', \forall \xi \in \Xi$, and that, for all $\xi \in \Xi$, the support of $\Delta T^\xi$ is contained in $\l[0,\hat{z}_1\r]$ (see in particular Lemma \ref{extended tax policy support lemma} and (\ref{z g equivalence lower interval 1})-(\ref{z g equivalence lower interval 2}) of Lemma \ref{useful but simple lemma}\footnote{The proof of Lemmas \ref{extended tax policy support lemma} and \ref{useful but simple lemma} depend on the fact that, for all $\xi \in \Xi$, the support of $\Delta T^\xi$ is contained in $\l[0,\hat{z}_1\r]$, but not the more detailed properties established in the current lemma, Lemma \ref{indifference and nonconstant revenue family of reforms theorem}.} and note that $T^{\theta_0,\epsilon_0}=T$), it follows that, for all $\theta \in \Theta'$, for all $\epsilon \in E'$, and for all $\xi' \in \Xi$, 
\begin{align}\label{equivalent utility and revenue 1}
\begin{split}
&\int g_i\l(T+\Delta T^\xi\r) \l.\pdv{\xi}\r|_{\xi=\xi'}\Delta T\l(z_i\l(T+\Delta T^{\xi'}\r),\xi\r) \dd i\\ =\;&\int g_i\l(T^{\theta,\epsilon}+\Delta T^\xi\r) \l.\pdv{\xi}\r|_{\xi=\xi'}\Delta T\l(z_i\l(T^{\theta,\epsilon}+\Delta T^{\xi'}\r),\xi\r) \dd i,\end{split}\\
\label{equivalent utility and revenue 2}&\l.\dv{\xi}\r|_{\xi=\xi'} R\l(T+ \Delta T^\xi\r)=\l.\dv{\xi}\r|_{\xi=\xi'} R\l(T^{\theta,\epsilon}+ \Delta T^\xi\r)
\end{align}
Conditions (\ref{indifference3}) and (\ref{nonconstant revenue3}) follow from (\ref{indifference 3 simpler}), (\ref{nonconstant revenue3 simpler}), (\ref{equivalent utility and revenue 1}), and (\ref{equivalent utility and revenue 2}).  We have now proven all the properties required by Lemma \ref{indifference and nonconstant revenue family of reforms theorem}.  $\square$

\subsection{\label{lemma satisfying desired properties section}Proof of Lemma \ref{lemma satisfying desired properties}}

I now prove a pair of supporting lemmas, and then proceed to verify the properties required by Lemma \ref{lemma satisfying desired properties}.  The following sections appeal to the notation and terminology used in Section \ref{key lemma with revenue section}. 

\subsubsection{\label{proofs of several lemmas}Supporting lemmas}

Here I establish a pair of lemmas that  collect some properties that follow fairly immediately from above definitions.

\begin{lem}\label{extended tax policy support lemma}
For all $\theta \in \Theta', \epsilon \in E',$ and $\xi \in \Xi$, 
\begin{align*}
z_i\l(T^{\theta,\epsilon} + \Delta T^\xi\r) \begin{cases} \in \l[0,\hat{z}_1\r], & \textup{ if } i \in \l[0,i_1\r], \\ 
\in \l[\hat{z}_1,+\infty\r), & \textup{ if } i \in \l[i_1,1\r].
\end{cases}
\end{align*}
\end{lem}

\begin{lem}\label{useful but simple lemma} For all $i \in \l[0,i_1\r], \theta \in \Theta', \epsilon \in E'$, and $\xi \in \Xi$, 
\begin{align}\label{z g equivalence lower interval 1}
z_i\l(T^{\theta_0,\epsilon_0} + \Delta T^\xi\r) & = z_i\l(T^{\theta,\epsilon} + \Delta T^\xi\r),\\
\label{z g equivalence lower interval 2}g_i\l(T^{\theta_0,\epsilon_0} + \Delta T^\xi\r) &= g_i\l(T^{\theta,\epsilon} + \Delta T^\xi\r),
\end{align}
For all $i \in \l[i_1,1\r],  \theta \in \Theta'',$ and $\epsilon \in E''$, 
\begin{align}
\label{equivalence of hat and not hat 1}
z_i\l(\hat{T}^{\theta,\epsilon}\r) &= z_i\l(T^{\theta,\epsilon}\r),\\
\label{equivalence of hat and not hat 2}
g_i\l(\hat{T}^{\theta,\epsilon}\r) &= g_i\l(T^{\theta,\epsilon}\r).
\end{align}
\end{lem}
I now proceed to prove both lemmas.  It is useful to state a pair of facts, which follow from, respectively, the construction of $T^{\theta,\epsilon}$ (Fact \ref{more specialized condition fact}) and the characterization of regular tax policies in Section \ref{wb appendix non-individualized} (Fact \ref{optimality first order characterized fact}). For Fact \ref{optimality first order characterized fact} and the remainder of this section, it is also convenient to introduce the following notation: For any tax policy $T$ and agent $i$ and income $z_i$, let $U^T_i\l(z_i\r)= u\l(z_i-T\l(z_i\r)-v_i\l(z_i\r)\r)$, be $i$'s utility when facing tax policy $T$ and choosing income level $z_i$.
\begin{fact}\label{more specialized condition fact}
$T^{\theta,\epsilon}\l(z\r)$ does not depend on $\theta$ and $\epsilon$ when $z \in \l[0,\hat{z}_1\r]$; that is $T^{\theta,\epsilon}\l(z\r)=T^{\theta_0,\epsilon_0}\l(z\r), \forall z \in \l[0,\hat{z}_1\r], \forall \theta \in \Theta, \forall \epsilon \in E$.
\end{fact} 
\begin{fact}\label{optimality first order characterized fact}
For all regular tax policies $T$, and for all agents $i$, there exists a unique optimal income $z_i\l(T\r)$ for $i$ when facing $T$ and $z_i\l(T\r)$ is characterized by $i$'s first order condition in the sense that if $\dv{z}  U_i^T\l(z\r)=0$, then $z=z_i\l(T\r)$.
\end{fact}
To simplify notation, I write $\bar{T}^{\theta,\epsilon,\xi}=  T^{\theta,\epsilon} + \Delta T^\xi$.  Fix some $\theta' \in \Theta', \epsilon' \in E'$, and $\xi' \in \Xi$.  Recall that $\hat{z}_1 \in \l(z_0\l(T^{\theta_0,\epsilon_0}\r),z_1\l(T^{\theta_0,\epsilon_0}\r)\r)$, and that $i_1$ is the unique agent in $I$ such that $z_{i_1}\l(T^{\theta_0,\epsilon_0}\r) = \hat{z}_1$.  Let $I_0 := \l[0,i_1\r)$ and $I_1 := \l(i_1,1\r]$.

By Fact \ref{more specialized condition fact} and because $T^{\theta_0,\epsilon_0}$ is smooth, $\dv{z}T^{\theta_0,\epsilon_0}\l(\hat{z}_1\r)=\dv{z}T^{\theta',\epsilon'}\l(\hat{z}_1\r)$.  So, because $\hat{z}_1=z_{i_1}\l(T^{\theta_0,\epsilon_0}\r)$, 
\begin{align}\label{second hat U i derivative equality}
\dv{z} U_{i_1}^{T^{\theta',\epsilon'}}\l(\hat{z}_1\r)=\dv{z} U_{i_1}^{T^{\theta_0,\epsilon_0}}\l(\hat{z}_1\r) = 0.
\end{align} 
Again, by Fact \ref{more specialized condition fact}, and the fact that the support of $\Delta T^\xi$ is contained in $\l[0,\hat{z}_1\r]$, it follows that 
\begin{align}\label{first complicated tax policy equivalence}
\forall z \leq \hat{z}_1,\;\;\;& \bar{T}^{\theta',\epsilon',\xi'}\l(z\r)=\bar{T}^{\theta_0,\epsilon_0,\xi'}\l(z\r), \\
\label{second complicated tax policy equivalence}\forall z \geq \hat{z}_1,\;\;\;& \bar{T}^{\theta',\epsilon',\xi'}\l(z\r) =  T^{\theta',\epsilon'}\l(z\r).
\end{align}
Using (\ref{second complicated tax policy equivalence}) and the smoothness of the tax policies $\bar{T}^{\theta',\epsilon',\xi'}$ and $T^{\theta',\epsilon'}$, it follows that     
\begin{align}\label{first hat U i derivative equality} 
\dv{z} U_{i_1}^{T^{\theta',\epsilon'}}\l(\hat{z}_1\r) = \dv{z} U_{i_1}^{\bar{T}^{\theta',\epsilon',\xi'}}\l(\hat{z}_1\r).
\end{align}
It follows from Fact \ref{optimality first order characterized fact}, (\ref{first hat U i derivative equality}), and (\ref{second hat U i derivative equality}), and the fact that $T^{\theta_0,\epsilon_0}$ and $\bar{T}^{\theta',\epsilon',\xi'}$ are regular (the latter was established in Section \ref{completing indifference and nonconstant revenue family of reforms theorem section}) that
\begin{align}\label{optima for i 1}
\hat{z}_1=z_{i_1}\l(T^{\theta_0,\epsilon_0}\r) = z_{i_1}\l(\bar{T}^{\theta',\epsilon',\xi'}\r).   
\end{align} 
Because $\bar{T}^{\theta',\epsilon',\xi'}$ is regular, (\ref{optima for i 1}) and the fact that, for all regular tax policies $T$, the map $i \mapsto z_i\l(T\r)$ is strictly increasing in $i$ (see Lemma \ref{interval optimum lemma}) together establish Lemma \ref{extended tax policy support lemma}.  

It follows from (\ref{first complicated tax policy equivalence}) and Lemma \ref{extended tax policy support lemma} that, for all $i \in \l[0,i_1\r]$, $\dv{z} U_i^{\bar{T}^{\theta',\epsilon',\xi'}}\l(z_i\l(\bar{T}^{\theta_0,\epsilon_0,\xi'}\r)\r)=\dv{z} U_i^{\bar{T}^{\theta_0,\epsilon_0,\xi'}}\l(z_i\l(\bar{T}^{\theta_0,\epsilon_0,\xi'}\r)\r)=0.$ So, using the fact that $\bar{T}^{\theta',\epsilon',\xi'}$ and $\bar{T}^{\theta_0,\epsilon_0,\xi'}$ are regular, it follows from Fact \ref{optimality first order characterized fact} that (\ref{z g equivalence lower interval 1}) holds.  Similarly, it follows from (\ref{second complicated tax policy equivalence}) and Lemma \ref{extended tax policy support lemma}\footnote{Observe that $\Delta T^0 \equiv 0$, so that, setting $\xi=0$, Lemma \ref{extended tax policy support lemma} implies that, when $i \in \l[i_1,1\r]$, $z_i\l(T^{\theta,\epsilon}\r) \geq \hat{z}_1$.} that, for all $i \in \l[i_1,1\r]$, $\dv{z} U_i^{\bar{T}^{\theta',\epsilon',\xi'}}\l(z_i\l(T^{\theta',\epsilon'}\r)\r)=\dv{z} U_i^{T^{\theta',\epsilon'}}\l(z_i\l(T^{\theta',\epsilon'}\r)\r)=0.$ So, using the fact that $\bar{T}^{\theta',\epsilon',\xi'}$ and $\bar{T}^{\theta',\epsilon'}$ are regular, it follows from Fact \ref{optimality first order characterized fact} that (\ref{equivalence of hat and not hat 1}) holds.  If follows immediately from the definition of $g_i\l(T\r)$ (see Sections \ref{standard aspects section}-\ref{generalized welfare weights introduction subsection}), (\ref{z g equivalence lower interval 1}), and (\ref{equivalence of hat and not hat 1}) that (\ref{z g equivalence lower interval 2}) and (\ref{equivalence of hat and not hat 2}) hold.  We have now established Lemma \ref{useful but simple lemma}. $\square$

\subsubsection{Verification of properties required by Lemma \ref{lemma satisfying desired properties}}

I now proceed with the proof of Lemma \ref{lemma satisfying desired properties}.  Choose $\epsilon \in \l(\underline{\epsilon}'',\overline{\epsilon}''\r)$ and $\theta' \in \l(\underline{\theta}'',\overline{\theta}''\r)$.  We have:
\begin{align}\label{first zero two}
\begin{split}
&\int_{0}^{1} g_i\l(\hat{T}^{\theta',\epsilon}\r) \l.\pdv{\theta}\r|_{\theta = \theta'} \hat{T}\l(z_i\l(\hat{T}^{\theta',\epsilon}\r),\theta,\epsilon\r) \dd i \\
=&\underbrace{\int_{0}^{i_1} g_i\l(\hat{T}^{\theta',\epsilon}\r) \l.\pdv{\theta}\r|_{\theta = \theta'} \hat{T}\l(z_i\l(\hat{T}^{\theta',\epsilon}\r),\theta,\epsilon\r) \dd i}_A \\& +  
\underbrace{\int_{i_1}^{1} g_i\l(\hat{T}^{\theta',\epsilon}\r) \l.\pdv{\theta}\r|_{\theta = \theta'} \hat{T}\l(z_i\l(\hat{T}^{\theta',\epsilon}\r),\theta,\epsilon\r) \dd i}_B,
\end{split}
\end{align}
where $A$ and $B$ are simply labels for the two integrals on the right-hand side, and we use the notation $\hat{T}\l(z_i,\theta,\epsilon\r)=\hat{T}^{\theta,\epsilon}\l(z_i\r)$, as in Section \ref{parameterized families subsection}.  Next observe that
\begin{align}\label{second zero two}
\begin{split}
A=& \int_{0}^{i_1} g_i\l(\hat{T}^{\theta',\epsilon}\r) \l[ \l.\pdv{\theta}\r|_{\theta = \theta'}T\l(z_i\l(\hat{T}^{\theta',\epsilon}\r),\theta,\epsilon\r)\r.\\&\l.+\l.\pdv{\xi}\r|_{\xi=\hat{\xi}\l(\theta',\epsilon\r)}\Delta T\l(z_i\l(\hat{T}^{\theta',\epsilon}\r),\xi\r) \l.\pdv{\theta}\r|_{\theta=\theta'} \hat{\xi}\l(\theta,\epsilon\r)\r] \dd i\\
=&\l[\l.\pdv{\theta}\r|_{\theta = \theta'}\hat{\xi}\l(\theta,\epsilon\r)\r]\int_{0}^{i_1} g_i\l(\hat{T}^{\theta',\epsilon}\r) \l.\pdv{\xi}\r|_{\xi=\hat{\xi}\l(\theta',\epsilon\r)}\Delta T\l(z_i\l(\hat{T}^{\theta',\epsilon}\r),\xi\r) \dd i\\
=&\l[\l.\pdv{\theta}\r|_{\theta = \theta'}\hat{\xi}\l(\theta,\epsilon\r)\r]\int_{0}^{i_1} g_i\l(T^{\theta',\epsilon}+\Delta T^{\hat{\xi}\l(\theta',\epsilon\r)}\r) \l.\pdv{\xi}\r|_{\xi=\hat{\xi}\l(\theta',\epsilon\r)}\Delta T^\xi\l(z_i\l(T^{\theta',\epsilon}+\Delta T^{\hat{\xi}\l(\theta',\epsilon\r)}\r),\xi\r) \dd i\\
=&\l[\l.\pdv{\theta}\r|_{\theta = \theta'}\hat{\xi}\l(\theta,\epsilon\r)\r]\int_{0}^{i_1} g_i\l(T^{\theta_0,\epsilon_0}+\Delta T^{\hat{\xi}\l(\theta',\epsilon\r)}\r) \l.\pdv{\xi}\r|_{\xi=\hat{\xi}\l(\theta',\epsilon\r)}\Delta T^\xi\l(z_i\l(T^{\theta_0,\epsilon_0}+\Delta T^{\hat{\xi}\l(\theta',\epsilon\r)}\r),\xi\r) \dd i\\
=&\l[\l.\pdv{\theta}\r|_{\theta = \theta'}\hat{\xi}\l(\theta,\epsilon\r)\r]\int_{0}^{1} g_i\l(T^{\theta_0,\epsilon_0}+\Delta T^{\hat{\xi}\l(\theta',\epsilon\r)}\r) \l.\pdv{\xi}\r|_{\xi=\hat{\xi}\l(\theta',\epsilon\r)}\Delta T^\xi\l(z_i\l(T^{\theta_0,\epsilon_0}+\Delta T^{\hat{\xi}\l(\theta',\epsilon\r)}\r),\xi\r) \dd i\\
=&\;0,
\end{split}
\end{align}
where the first equality follows from the definition (\ref{hat defined in terms of non-hat}) of $\hat{T}^{\theta,\epsilon}$; the second equality follows from Fact \ref{more specialized condition fact} and Lemma \ref{extended tax policy support lemma}, which imply that, when $i \in \l[0,i_1\r]$,  $\l.\pdv{\theta}\r|_{\theta = \theta'}T\l(z_i\l(\hat{T}^{\theta',\epsilon}\r),\theta,\epsilon\r)=0$;  the third equality follows again follows from (\ref{hat defined in terms of non-hat}); the fourth equality follows from (\ref{z g equivalence lower interval 1})-(\ref{z g equivalence lower interval 2}); the fifth equality follows from the fact that, for all $\xi \in \Xi$, the support of $\Delta T^\xi$ is contained in $\l[0,\hat{z}_1\r]$ and Lemma \ref{extended tax policy support lemma}, so that the integrand in the expression following the third equality is equal to zero when $i \in \l[i_1,1\r]$; and the last equality follows from (\ref{indifference3}).  

Next, observe that 
\begin{align}\label{third zero two}
\begin{split}
B&= \int_{i_1}^{1} g_i\l(\hat{T}^{\theta',\epsilon}\r) \l.\pdv{\theta}\r|_{\theta = \theta'} T\l(z_i\l(\hat{T}^{\theta',\epsilon}\r),\theta,\epsilon\r) \dd i\\
&= \int_{i_1}^{1} g_i\l(T^{\theta',\epsilon}\r) \l.\pdv{\theta}\r|_{\theta = \theta'} T\l(z_i\l(T^{\theta',\epsilon}\r),\theta,\epsilon\r) \dd i\\
&= \int_{0}^{1} g_i\l(T^{\theta',\epsilon}\r) \l.\pdv{\theta}\r|_{\theta = \theta'} T\l(z_i\l(\hat{T}^{\theta',\epsilon}\r),\theta,\epsilon\r) \dd i\\
&=0,
\end{split}
\end{align}
where the first equality follows from Lemma \ref{extended tax policy support lemma}, (\ref{hat defined in terms of non-hat}), and the fact that, for all $\xi \in \Xi$, the support of  $\Delta T^\xi$ is contained in $\l[0,\hat{z}_1\r]$, so that $T\l(z_i\l(\hat{T}^{\theta',\epsilon}\r),\theta,\epsilon\r)=\hat{T}\l(z_i\l(\hat{T}^{\theta',\epsilon}\r),\theta,\epsilon\r)$ when $i \in \l[i_1,1\r]$; the second equality follows from (\ref{equivalence of hat and not hat 1})-(\ref{equivalence of hat and not hat 2}); the third equality follows from the fact that, by Fact \ref{more specialized condition fact}, $T\l(z_i\l(T^{\theta',\epsilon}\r),\theta,\epsilon\r)$ does not depend on $\theta$ when $i \in \l[0,i_1\r]$, so the integrand in the expression following the second equality is zero when $i \in \l[0,i_1\r]$; and the last equality follows from the fact that $\l(T^{\theta,\epsilon}\r)$ satisfies (\ref{2}).  

Putting together (\ref{first zero two}), (\ref{second zero two}), and (\ref{third zero two}), it follows that $\l(\hat{T}^{\theta,\epsilon}\r)_{\theta \in \Theta'',\epsilon \in E''}$ satisfies (\ref{2}).  

Next observe that:
\begin{align}\label{conditions this C D 1}
\begin{split}
&\int_{0}^{1} g_i\l(\hat{T}^{\theta_0,\epsilon_0}\r) \l.\pdv{\epsilon}\r|_{\epsilon = \epsilon_0} \hat{T}\l(z_i\l(\hat{T}^{\theta_0,\epsilon_0}\r),\theta_0,\epsilon\r) \dd i\\
=& \underbrace{\int_{0}^{i_1} g_i\l(\hat{T}^{\theta_0,\epsilon_0}\r) \l.\pdv{\epsilon}\r|_{\epsilon = \epsilon_0} \hat{T}\l(z_i\l(\hat{T}^{\theta_0,\epsilon_0}\r),\theta_0,\epsilon\r) \dd i}_C\\
&+ \underbrace{\int_{i_1}^{1} g_i\l(\hat{T}^{\theta_0,\epsilon_0}\r) \l.\pdv{\epsilon}\r|_{\epsilon = \epsilon_0} \hat{T}\l(z_i\l(\hat{T}^{\theta_0,\epsilon_0}\r),\theta_0,\epsilon\r) \dd i}_D.
\end{split}
\end{align}
Analyzing the first term:
\begin{align}\label{conditions this C D 2}
\begin{split}
C = &  \int_{0}^{i_1} g_i\l(\hat{T}^{\theta_0,\epsilon_0}\r)\l[ \l.\pdv{\epsilon}\r|_{\epsilon = \epsilon_0} T\l(z_i\l(\hat{T}^{\theta_0,\epsilon_0}\r),\theta_0,\epsilon\r)\r.\\&\l.+\l.\pdv{\xi}\r|_{\xi=\hat{\xi}\l(\theta_0,\epsilon_0\r)}\Delta T^{\xi}\l(z_i\l(\hat{T}^{\theta_0,\epsilon_0}\r)\r)\l.\pdv{\epsilon}\r|_{\epsilon=\epsilon_0}\hat{\xi}\l(\theta_0,\epsilon\r)\r] \dd i \\
= &\l[\l.\pdv{\epsilon}\r|_{\epsilon=\epsilon_0}\hat{\xi}\l(\theta_0,\epsilon\r)\r]  \int_{0}^{i_1} g_i\l(\hat{T}^{\theta_0,\epsilon_0}\r)\l.\pdv{\xi}\r|_{\xi=\hat{\xi}\l(\theta_0,\epsilon_0\r)}\Delta T^{\xi}\l(z_i\l(\hat{T}^{\theta_0,\epsilon_0}\r),\xi\r) \dd i \\
= &\l[\l.\pdv{\epsilon}\r|_{\epsilon=\epsilon_0}\hat{\xi}\l(\theta_0,\epsilon\r)\r]  \int_{0}^{i_1} g_i\l(T^{\theta_0,\epsilon_0}+ \Delta T^{\hat{\xi}\l(\theta_0,\epsilon_0\r)}\r)\l.\pdv{\xi}\r|_{\xi=\hat{\xi}\l(\theta_0,\epsilon_0\r)}\Delta T^{\xi}\l(z_i\l(T^{\theta_0,\epsilon_0}+ \Delta^{\hat{\xi}\l(\theta_0,\epsilon_0\r)}\r),\xi\r) \dd i \\
= &\l[\l.\pdv{\epsilon}\r|_{\epsilon=\epsilon_0}\hat{\xi}\l(\theta_0,\epsilon\r)\r]  \int_{0}^{1} g_i\l(T^{\theta_0,\epsilon_0}+ \Delta T^{\hat{\xi}\l(\theta_0,\epsilon_0\r)}\r)\l.\pdv{\xi}\r|_{\xi=\hat{\xi}\l(\theta_0,\epsilon_0\r)}\Delta T^{\xi}\l(z_i\l(T^{\theta_0,\epsilon_0}+ \Delta T^{\hat{\xi}\l(\theta_0,\epsilon_0\r)}\r),\xi\r) \dd i \\
=&\;0,
\end{split}
\end{align}
where the first equality follows from (\ref{hat defined in terms of non-hat}); the second equality from Fact \ref{more specialized condition fact} and Lemma \ref{extended tax policy support lemma}, which imply that, when $i \in \l[0,i_1\r]$, $\l.\pdv{\epsilon}\r|_{\epsilon = \epsilon_0} T\l(z_i\l(\hat{T}^{\theta_0,\epsilon_0}\r),\theta_0,\epsilon\r)=0$; the third equality follows from (\ref{hat defined in terms of non-hat}); the fourth equality follows from Lemma \ref{extended tax policy support lemma} and the fact that, for all $\xi \in \Xi$, the support of $\Delta T^\xi$ is contained in $\l[0,\hat{z}_1\r]$, so that the integrand in the expression following the fourth equality is zero when $i \in \l[i_1,1\r]$; and the last equality follows from (\ref{indifference3}).  

Analyzing the second term:
\begin{align}\label{conditions C D 3}
\begin{split}
D&= \int_{i_1}^{1} g_i\l(\hat{T}^{\theta_0,\epsilon_0}\r) \l.\pdv{\epsilon}\r|_{\epsilon = \epsilon_0} T\l(z_i\l(\hat{T}^{\theta_0,\epsilon_0}\r),\theta_0,\epsilon\r) \dd i\\
&= \int_{i_1}^{1} g_i\l(T^{\theta_0,\epsilon_0}\r) \l.\pdv{\epsilon}\r|_{\epsilon = \epsilon_0} T\l(z_i\l(T^{\theta_0,\epsilon_0}\r),\theta_0,\epsilon\r) \dd i\\
&= \int_{0}^{1} g_i\l(T^{\theta_0,\epsilon_0}\r) \l.\pdv{\epsilon}\r|_{\epsilon = \epsilon_0} T\l(z_i\l(T^{\theta_0,\epsilon_0}\r),\theta_0,\epsilon\r) \dd i\\
&=0,
\end{split}
\end{align}
where the first equality follows from Lemma \ref{extended tax policy support lemma}, (\ref{hat defined in terms of non-hat}), and the fact that, for all $\xi \in \Xi$, the support of  $\Delta T^\xi$ is contained in $\l[0,\hat{z}_1\r]$, so that  $T\l(z_i\l(\hat{T}^{\theta_0,\epsilon_0}\r),\theta_0,\epsilon\r)=\hat{T}\l(z_i\l(\hat{T}^{\theta_0,\epsilon_0}\r),\theta_0,\epsilon\r)$ when $i \in \l[i_1,1\r]$; the second follows from the fact that, by (\ref{fixed revenue condition 0}) and $\Delta T^0 \equiv 0$, $T^{\theta_0,\epsilon_0} = \hat{T}^{\theta_0,\epsilon_0}$; the third equality follows from the fact that, by Fact \ref{more specialized condition fact} and Lemma \ref{extended tax policy support lemma}, $\l.\pdv{\epsilon}\r|_{\epsilon = \epsilon_0} T\l(z_i\l(T^{\theta_0,\epsilon_0}\r),\theta_0,\epsilon\r)=0$ when $i \in \l[0,i_1\r]$; and the last equality follows from the fact that $\l(T^{\theta,\epsilon}\r)$ satisfies (\ref{3}).  

Putting together (\ref{conditions this C D 1}), (\ref{conditions this C D 2}), and (\ref{conditions C D 3}), we see that $\l(\hat{T}^{\theta,\epsilon}\r)_{\theta \in \Theta'', \epsilon \in E''}$ satisfies (\ref{3}).

Next observe that:
\begin{align}\label{inequality condition E G}
\begin{split}
&\l.\dv{\theta}\r|_{\theta = \theta_0} \int_{0}^{1} g_i\l(\hat{T}^{\theta,\epsilon_0}\r) \l.\pdv{\epsilon}\r|_{\epsilon = \epsilon_0} \hat{T}\l(z_i\l(\hat{T}^{\theta,\epsilon_0}\r),\theta,\epsilon\r) \dd i \\
=&\underbrace{\l.\dv{\theta}\r|_{\theta = \theta_0} \int_{0}^{i_1} g_i\l(\hat{T}^{\theta,\epsilon_0}\r) \l.\pdv{\epsilon}\r|_{\epsilon = \epsilon_0} \hat{T}\l(z_i\l(\hat{T}^{\theta,\epsilon_0}\r),\theta,\epsilon\r) \dd i}_E\\
&+\underbrace{\l.\dv{\theta}\r|_{\theta = \theta_0} \int_{i_1}^{1} g_i\l(\hat{T}^{\theta,\epsilon_0}\r) \l.\pdv{\epsilon}\r|_{\epsilon = \epsilon_0} \hat{T}\l(z_i\l(\hat{T}^{\theta,\epsilon_0}\r),\theta,\epsilon\r) \dd i}_F.
\end{split}
\end{align}
Analyzing the first term:
\begin{align}\label{inequality condition E}
\begin{split}
E=&\l.\dv{\theta}\r|_{\theta = \theta_0} \int_{0}^{i_1} g_i\l(\hat{T}^{\theta,\epsilon_0}\r) \l.\pdv{\epsilon}\r|_{\epsilon = \epsilon_0} T\l(z_i\l(\hat{T}^{\theta,\epsilon_0}\r),\theta,\epsilon\r) \dd i \\&+\l.\dv{\theta}\r|_{\theta = \theta_0}
\l[\l( \l.\pdv{\epsilon}\r|_{\epsilon = \epsilon_0} \hat{\xi}\l(\theta,\epsilon\r) \r)\int_{0}^{i_1}g_i\l(\hat{T}^{\theta,\epsilon_0}\r)\l.\pdv{\xi}\r|_{\xi=\hat{\xi}\l(\theta,\epsilon_0\r)}\Delta T^\xi\l(z_i\l(\hat{T}^{\theta,\epsilon_0}\r) ,\xi\r) \dd i \r]\\
=&\l.\dv{\theta}\r|_{\theta = \theta_0}
\l[\l( \l.\pdv{\epsilon}\r|_{\epsilon = \epsilon_0} \hat{\xi}\l(\theta,\epsilon\r) \r)\int_{0}^{i_1}g_i\l(\hat{T}^{\theta,\epsilon_0}\r)\l.\pdv{\xi}\r|_{\xi=\hat{\xi}\l(\theta,\epsilon_0\r)}\Delta T^\xi\l(z_i\l(\hat{T}^{\theta,\epsilon_0}\r) ,\xi\r) \dd i \r]\\
=&\l.\dv{\theta}\r|_{\theta = \theta_0}
\l[\l( \l.\pdv{\epsilon}\r|_{\epsilon = \epsilon_0} \hat{\xi}\l(\theta,\epsilon\r) \r)\int_{0}^{i_1}g_i\l(T^{\theta,\epsilon_0}+\Delta T^{\hat{\xi}\l(\theta,\epsilon_0\r)}\r)\r.\\ &\l.\times \l.\pdv{\xi}\r|_{\xi=\hat{\xi}\l(\theta,\epsilon_0\r)}\Delta T^\xi\l(z_i\l(T^{\theta,\epsilon_0}+\Delta T^{\hat{\xi}\l(\theta,\epsilon_0\r)}\r) ,\xi\r) \dd i \r]\\
=&\l.\dv{\theta}\r|_{\theta = \theta_0}
\l[\l( \l.\pdv{\epsilon}\r|_{\epsilon = \epsilon_0} \hat{\xi}\l(\theta,\epsilon\r) \r)\int_{0}^{1}g_i\l(T^{\theta,\epsilon_0}+\Delta T^{\hat{\xi}\l(\theta,\epsilon_0\r)}\r)\r.\\ &\l.\times \l.\pdv{\xi}\r|_{\xi=\hat{\xi}\l(\theta,\epsilon_0\r)}\Delta T^\xi\l(z_i\l(T^{\theta,\epsilon_0}+\Delta T^{\hat{\xi}\l(\theta,\epsilon_0\r)}\r) ,\xi\r) \dd i \r]\\
&=0,
\end{split}
\end{align}
where the first equality follows from (\ref{hat defined in terms of non-hat}); the second equality from Fact \ref{more specialized condition fact} and Lemma \ref{extended tax policy support lemma}, which imply that $\l.\pdv{\epsilon}\r|_{\epsilon = \epsilon_0} T\l(z_i\l(\hat{T}^{\theta,\epsilon_0}\r),\theta,\epsilon\r)=0, \forall \theta \in \Theta''$, when $i \in \l[0,i_1\r]$; the third equality follows from (\ref{hat defined in terms of non-hat}); the fourth equality follows from Lemma \ref{extended tax policy support lemma} and the fact that, for all $\xi \in \Xi$, the support of $\Delta T^\xi$ is contained in $\l[0,\hat{z}_1\r]$, so that the integrand in the expression following the fourth equality is zero, for all values of $\theta$ in $\Theta''$, when $i \in \l[i_1,1\r]$; and the last equality follows from (\ref{indifference3}).  

Analyzing the second term:
\begin{align}\label{inequality condition G}
\begin{split}
F=&\l.\dv{\theta}\r|_{\theta = \theta_0} \int_{i_1}^{1} g_i\l(\hat{T}^{\theta,\epsilon_0}\r) \l.\pdv{\epsilon}\r|_{\epsilon = \epsilon_0} T\l(z_i\l(\hat{T}^{\theta,\epsilon_0}\r),\theta,\epsilon\r)  \dd i \\&+\l.\dv{\theta}\r|_{\theta = \theta_0}
\l[\l( \l.\pdv{\epsilon}\r|_{\epsilon = \epsilon_0} \hat{\xi}\l(\theta,\epsilon\r) \r)\int_{i_1}^{1}g_i\l(\hat{T}^{\theta,\epsilon_0}\r)\l.\pdv{\xi}\r|_{\xi=\hat{\xi}\l(\theta_0,\epsilon_0\r)}\Delta T^\xi\l(z_i\l(\hat{T}^{\theta,\epsilon_0}\r),\xi\r)  \dd i \r]\\ 
=& \l.\dv{\theta}\r|_{\theta = \theta_0} \int_{i_1}^{1} g_i\l(\hat{T}^{\theta,\epsilon_0}\r) \l.\pdv{\epsilon}\r|_{\epsilon = \epsilon_0} T\l(z_i\l(\hat{T}^{\theta,\epsilon_0}\r),\theta,\epsilon\r)  \dd i\\
=& \l.\dv{\theta}\r|_{\theta = \theta_0} \int_{i_1}^{1} g_i\l(T^{\theta,\epsilon_0}\r) \l.\pdv{\epsilon}\r|_{\epsilon = \epsilon_0} T\l(z_i\l(T^{\theta,\epsilon_0}\r),\theta,\epsilon\r)  \dd i \\
=& \l.\dv{\theta}\r|_{\theta = \theta_0} \int_{0}^{1} g_i\l(T^{\theta,\epsilon_0}\r) \l.\pdv{\epsilon}\r|_{\epsilon = \epsilon_0} T\l(z_i\l(T^{\theta,\epsilon_0}\r),\theta,\epsilon\r)  \dd i \\
<&\; 0,
\end{split}
\end{align}
where the first equality follows from (\ref{hat defined in terms of non-hat}); the second equality follows from Lemma \ref{extended tax policy support lemma}, (\ref{hat defined in terms of non-hat}), and the fact that, for all $\xi \in \Xi$, the support of  $\Delta T^\xi$ is contained in $\l[0,\hat{z}_1\r]$, so that\\ $\l.\pdv{\xi}\r|_{\xi=\hat{\xi}\l(\theta,\epsilon_0\r)}\Delta T^\xi\l(z_i\l(\hat{T}^{\theta,\epsilon_0}\r),\xi\r)=0, \forall \theta \in \Theta''$, when $i \in \l[i_1,1\r]$; the third equality follows from Lemma \ref{extended tax policy support lemma}, (\ref{hat defined in terms of non-hat}) and (\ref{equivalence of hat and not hat 1})-(\ref{equivalence of hat and not hat 2}); the fourth equality follows from the fact that, by Fact \ref{more specialized condition fact},\\ $\l.\pdv{\epsilon}\r|_{\epsilon = \epsilon_0} T\l(z_i\l(T^{\theta,\epsilon_0}\r),\theta,\epsilon\r)=0, \forall \theta \in \Theta''$, when $i \in \l[0,i_1\r]$; and the inequality follows from the fact that $\l(T^{\theta,\epsilon}\r)$ satisfies (\ref{4}).

Putting together (\ref{inequality condition E G}), (\ref{inequality condition E}), and (\ref{inequality condition G}), it follows that $\l(\hat{T}^{\theta,\epsilon}\r)_{\theta \in \Theta'', \epsilon \in E''}$ satisfies (\ref{4}).

We have now established that $\l(\hat{T}^{\theta,\epsilon}\r)_{\theta \in \Theta'', \epsilon \in E''}$  satisfies (\ref{2})-(\ref{4}), completing the proof of Lemma \ref{lemma satisfying desired properties}.  $\square$

\section{\label{additional lemmas appendix section}Additional lemmas}
The lemmas in this section apply to tax policies that are not individualized.

\begin{lem}\label{interval optimum lemma}
For all regular tax policies $T$, $\l\{z_i\l(T\r): i \in I\r\}=\l[z_0\l(T\r),z_1\l(T\r)\r]$, and the map $i \mapsto z_i\l(T\r)$ is strictly increasing. 
\end{lem}
Proof.  Let $T$ be a regular tax policy.  It follows from our assumptions (see Section \ref{wb appendix non-individualized}) that $z_i\l(T\r)$ is characterized by the first order condition $1-T'\l(z_i\l(T\r)\r) -v'_i\l(z_i\l(T\r)\r)=0$.  The smoothness of $T$ and $\l(z,i\r)\mapsto v_i\l(z\r)$ imply that the function $i \mapsto z_i\l(T\r)$ is smooth. It follows from the facts that (i) $v_i\l(z\r)=v\l(z,y_i\r) \forall i, \forall z$, (ii) $ \pdv[2]{}{z}{y} v\l(z,y\r) < 0, \forall z, \forall y$, and (iii) $\dv{i} y_i > 0, \forall i$, that the map $i \mapsto z_i\l(T\r)$ is strictly increasing (see Section \ref{additional structure section} for the preceding assumptions).  Since $i \mapsto z_i\l(T\r)$ is continuous and strictly increasing on $I=\l[0,1\r]$, $\l\{z_i\l(T\r): i \in I\r\}=\l[z_0\l(T\r),z_1\l(T\r)\r]$.  $\square$ 

\begin{lem}\label{small perturbation still in T hat lemma}
Let $T $ be a regular tax policy.  For $j=1,\ldots, n$, let $\Theta_j = \l[-\bar{\theta}_j,\bar{\theta}_j\r] \subseteq \mathbb{R}$, where $\theta_j  > 0$, and let $\bar{\Theta} = \times_{j=1}^n \Theta_j$.  Write $\bar{\theta}=\l(\theta_1,\ldots,\theta_j,\ldots,\theta_n\r)$.  Let $\l(\Delta T^{\bar{\theta}}\r)_{\bar{\theta} \in \bar{\Theta}}$ be a family of tax reforms such that the map $\l(z,\bar{\theta}\r) \mapsto \Delta T^{\bar{\theta}}\l(z\r)$ is smooth and $\Delta T^{\l(0,0,\ldots,0\r)} \equiv 0$.  Then there exist $\theta^*_j \in \Theta_j$ with $\theta^*_j > 0$ for $j=1,\ldots,n$ such that, for all $\bar{\theta}=\l(\theta_1,\ldots,\theta_j, \ldots, \theta_n\r) \in \bar{\Theta},$ if $\l|\theta_j\r| \leq \theta^*_j$ for $j=1,\ldots, n$, then $T + \Delta T^{\bar{\theta}} $ is regular.     
\end{lem} 
Proof.  I use the following notation: for any tax policy $T$ and income $z_i$, define $U^T_i\l(z_i\r)= u\l(z_i-T\l(z_i\r)-v_i\l(z_i\r)\r)$.  Now let $T$ be a regular tax policy. It follows that, for all agents $i$, $\dv[2]{z_i} U^T_i\l(z_i\l(T\r)\r) <0$ (see Section \ref{wb appendix non-individualized}). Also, since $T$ is regular, $z_i\l(T\r) > 0$, for all agents $i$.  By the smoothness of the primitives and $T$, it follows that there is a neighborhood $N_i$ of the income $z_i\l(T\r)$ such that, for all $z_i \in N_i$, $\dv[2]{z_i} U^T_i\l(z_i\r) <0$ and $z_i >0$.  For each $i$, let 
\begin{align*}
\delta_i = \sup\l\{\delta > 0: z_i\l(T\r)-\delta > 0, \forall z_i \in \l(z_i\l(T\r)-\delta, z_i\l(T\r) + \delta\r),  \dv[2]{z_i} U^T_i\l(z_i\r) <0\r\}.
\end{align*}  We have $\delta_i > 0, \forall i$, and, moreover, the smoothness of the primitives and of $T$ implies that $i \mapsto \delta_i$ is smooth.  Since a continuous function attains its minimum on a compact set, it follows that $\delta^* = \min\l\{ \delta_i: i \in \l[0,1\r]\r\}$ exists and $\delta^* > 0$.  For each $i$, define the neighborhood $N'_i = \l(z_i\l(T\r) - \frac{1}{2} \delta^*, z_i\l(T\r) + \frac{1}{2} \delta^*\r)$ and let $\bar{N}'_i$ be the closure of $N'_i$.  For each $\bar{\theta} \in \bar{\Theta}$, define $T^{\bar{\theta}}= T+\Delta T^{\bar{\theta}}$.  Define $\gamma^{\bar{\theta}}_i = U^{T^{\bar{\theta}}}
\l(z_i\l(T\r)\r) - \max_{z_i \in Z \setminus N'_i}  U^{T^{\bar{\theta}}}_i\l(z_i\r)$ and $\gamma^{\bar{\theta}}= \min_{i \in \l[0,1\r]} \gamma^{\bar{\theta}}_i$.  As $T^{\l(0,0,\ldots,0\r)}=T+\Delta T^{\l(0,0,\ldots,0\r)}=T$, and, as $T$ is regular, $U_i^{T^{\l(0,0,\ldots,0\r)}}\l(z_i\r)$ has a unique maximizer $z_i\l(T\r)$, it follows that, for all $i$, $\gamma_i^{\l(0,0,\ldots,0\r)} >0$, and hence, again because a continuous function attains its minimum on a compact set, $\gamma^{\l(0,0,\ldots,0\r)} >0$.  It follows from our smoothness assumptions that there exist $\theta'_j \in \Theta_j$ with $\theta'_j > 0$ for $j=1,\ldots,n$ such for all $\bar{\theta}=\l(\theta_1,\ldots,\theta_j, \ldots, \theta_n\r) \in \bar{\Theta},$ if $\l|\theta_j\r| \leq \theta'_j$ for $j=1,\ldots, n$, $\gamma^{\bar{\theta}} >0$, so that, for all such $\bar{\theta}$, $U_i^{T^{\bar{\theta}}}\l(z_i\r)$ does not have any maximizers $z_i$ outside of $N'_i$.  Note that we have: $\forall i \in I, \forall z_i \in \bar{N}'_i, \dv[2]{z_i} U^{T^{\l(0,0,\ldots,0\r)}}_i\l(z_i\r) <0$.  So $\max_{i \in \l[0,1\r], z_i \in \bar{N}'_i}  \dv[2]{z_i} U^{T^{\l(0,0,\ldots,0\r)}}_i <0$.  As the map $\bar{\theta} \mapsto \max_{i \in \l[0,1\r], z_i \in \bar{N}'_i}  \dv[2]{z_i} U^{T^{\bar{\theta}}}_i\l(z_i\r)$ is continuous, it follows that there exist $\theta''_j \in \Theta_j$ with $\theta''_j > 0$ for $j=1,\ldots,n$ such for all $\bar{\theta}=\l(\theta_1,\ldots,\theta_j, \ldots, \theta_n\r) \in \bar{\Theta},$ if $\l|\theta_j\r| \leq \theta''_j$ for $j=1,\ldots, n$, then, for all agents $i$ and all $z_i \in \bar{N}'_i,\dv[2]{z_i} U^{T^{\bar{\theta}}}_i\l(z_i\r) <0$, so that $U^{T^{\bar{\theta}}}_i\l(z_i\r)$ is strictly convex on $\bar{N}'_i$, implying that $U^{T^{\bar{\theta}}}_i\l(z_i\r)$ has a unique maximizer on $\bar{N}'_i$.  It follows that if $\theta^*_j = \min\l\{\theta_j',\theta_j''\r\}$ for $j=1,\ldots, n$, then, then for all $\bar{\theta}=\l(\theta_1,\ldots,\theta_j, \ldots, \theta_n\r) \in \bar{\Theta}$, if $\l|\theta_j\r| \leq \theta^*_j$ for $j=1,\ldots, n$, then for, all agents $i$, $U^{T^{\bar{\theta}}}_i\l(z_i\r)$ has a unique maximizer $z_i\l(T^{\bar{\theta}}\r)>0$, and, moreover, $\dv[2]{z_i} U^{T^{\bar{\theta}}}_i\l(z_i\l(T^{\bar{\theta}}\r)\r) <0$, so that $T^{\bar{\theta}}$ is regular.  $\square$

\section{\label{nonquasilinear appendix}Theorems \ref{necessity theorem} and \ref{main theorem} without quasilinear utility}

\subsection{\label{preliminaries h section}Preliminaries}

This Appendix explains how Theorems \ref{necessity theorem} and \ref{main theorem} are still valid without the assumption of quasilinearity.  In particular, I describe how the proofs of the theorems must be modified if the assumption of quasilinearity is removed.  I assume that utility takes the form $U_i\l(c_i,z_i\r)=U\l(c_i,z_i;x_i,y_i\r)$, where $U\l(c_i,z_i;x_i,y_i\r)$ is smooth in $\l(c_i,z_i; x_i,y_i\r)$ unless $\l(x_i,y_i\r)$ are discrete, in which case $U\l(c_i,z_i;x_i,y_i\r)$ is smooth in $\l(c_i,z_i\r)$.  I assume that  $U_i\l(c_i,z_i\r)$ is strictly increasing in $c_i$ (with a strictly positive partial derivative everywhere), strictly decreasing in $z_i$, and strictly concave in $\l(c_i,z_i\r)$.   I assume for simplicity
that $c_i$ can take on any real value and that for any income level $z_i$, the range of $c_i \mapsto U_i\l(c_i,z_i\r)$ is the entire real line.  I continue to assume that, in the absence of taxes, all agents earn a positive income.  

In what follows it will be useful to define the function $\tilde{c}_i\l(u_i,z_i\r)$ by the following condition:
\begin{align}\label{definition of tilde c}
U_i\l(\tilde{c}_i\l(u_i,z_i\r),z_i\r)=u_i, \;\;\; \forall z_i \in Z, \forall u_i \in \mathbb{R}. \end{align}
So, $\tilde{c}_i\l(u_i,z_i\r)$ is the level of consumption that is necessary to give $i$ utility $u_i$ given income level $z_i$; $\tilde{c}_i\l(u_i,z_i\r)$ is well-defined because $U_i$ is strictly increasing in $c_i$.  It follows from the implicit function theorem that:
\begin{align}\label{c i derivative MRS}
\pdv{z_i} \tilde{c}_i\l(u_i,z_i\r) = - \frac{\pdv{z_i} U_i\l(\tilde{c}_i\l(u_i,z_i\r),z_i\r)}{\pdv{c_i} U_i\l(\tilde{c}_i\l(u_i,z_i\r),z_i\r)}, \;\;\; \forall u_i, \forall z_i.
\end{align} 
I assume that along any $\l(c_i,z_i\r)$-indifference curve, the marginal rate of substitution of consumption for avoiding the effort of earning income exceeds $1$ as $z$ becomes large: 
\begin{align*}
\forall u_i, \lim_{z_i \rightarrow+ \infty} \pdv{z_i} \tilde{c}_i\l(u_i,z_i\r) >1.
\end{align*}  In other words, as one increases both income and consumption along an indifference curve, it is eventually necessary to compensate an agent by more than a dollar in order to bear the cost of earning another dollar of income.   This has the consequence that, whenever facing a tax policy under which marginal tax rates become nonnegative once income is sufficiently large, the agent optimally selects some finite income and does not want to increase their income without bounds.

Note that when $U_i\l(c_i,z_i\r)=u\l(c_i-v_i\l(z_i\r)\r)$ where $u' >0$ and $u'' < 0$ everywhere, and all the other assumptions of Section \ref{standard aspects section} are satisfied, then all of the above assumptions are satisfied, so the assumption here in essence generalize the assumptions made for the quasilinear case. I also carry over other assumptions (or analogous assumptions) and notation from the quasilinear case.   

The key preliminary definitions and results supporting the main results continue to hold in this more general framework.  Observe first that, even without quasilinear preferences, the envelope theorem still implies that for any well behaved parameterized family of tax policies $\l(T^\theta\r)$, $\dv{\theta} U_i\l(T^\theta\r) =  -\pdv{c_i}U_i\l(c_i\l(\theta\r),z_i\l(\theta\r)\r)\pdv{\theta} T_i\l(z_i\l(T\r),\theta\r)$.  It follows that the local and global improvement principles are still valid when welfare weights are utilitarian.  So the justification for the global and local improvement principles that was given in Section \ref{improvement and indifference principles subsection}, on analogy with the utilitarian case, still applies without quasilinearity.  Likewise, the supporting Proposition \ref{indifference Pareto coro} on Pareto indifference and weak Pareto along paths is unchanged, and so the result continues to hold. 

We can no longer define $\hat{g}_i\l(\hat{u}_i,z_i\r)$ as we did in (\ref{g g hat relation}) in Section \ref{structural utilitarian section} because that definition depended on the assumption of quasilinear utility.  Instead we define $\tilde{g}_i\l(u_i,z_i\r)$ which is a function of the variable $u_i=U_i\l(c_i,z_i\r)$ and $z_i$, as follows: 
\begin{align}\label{g g tilde relation}
\tilde{g}_i\l(u_i,z_i\r) = g_i\l(\tilde{c}_i\l(u_i,z_i\r), z_i\r), \;\;\; \forall z_i \in Z, \forall u_i \in \mathbb{R}.  
\end{align}
Next define:
\begin{align}
\label{k d U relation} k_i\l(u_i,z_i\r) &= \pdv{c_i} U_i\l(\tilde{c}_i\l(u_i,z_i\r),z_i\r),\\
\label{h g d U relation}h_i\l(u_i,z_i\r) &= \frac{\tilde{g}_i\l(u_i,z_i\r)}{\pdv{c_i} U_i\l(\tilde{c}_i\l(u_i,z_i\r),z_i\r)}.
\end{align}
Then observe that
\begin{align}\label{g = hk}
\tilde{g}_i\l(u_i,z_i\r)= k_i\l(u_i,z_i\r) h_i\l(u_i,z_i\r).
\end{align}
Now choose $z_i, z'_i$ and $u_i$ and observe that it follows from (\ref{definition of tilde c}) that 
\begin{align}\label{c i U i equality}
U_i\l(\tilde{c}_i\l(u_i,z_i\r),z_i\r)=U_i\l(\tilde{c}_i\l(u_i,z'_i\r),z'_i\r).
\end{align} 
Then if $g_i$ is structurally utilitarian, $h_i\l(u_i,z_i\r) = \frac{\tilde{g}_i\l(u_i,z_i\r)}{\pdv{c_i} U_i\l(\tilde{c}_i\l(u_i,z_i\r),z_i\r)}= \frac{\tilde{g}_i\l(u_i,z'_i\r)}{\pdv{c_i} U_i\l(\tilde{c}_i\l(u_i,z'_i\r),z'_i\r)} = h_i\l(u_i,z'_i\r)$, where the second equality follows from (\ref{c i U i equality}) and the definition of structural utilitarianism without quasilinearity (Definition \ref{nonquasilinear welfare weights definition}).   So for structurally utilitarian weights, $h_i\l(u_i,z_i\r)$ does not depend on $z_i$.  It is also easy to see that, if $h_i\l(u_i,z_i\r)$ does not depend on $z_i$, then the corresponding welfare weights are structurally utilitarian.  This can be summarized in a form a proposition which is the non-quasilinear analog of Proposition \ref{g observation}.   
\begin{prop}\label{h observation}
Let $g$ and $\tilde{g}$ be related as in (\ref{g g tilde relation}) and let $h$ be defined in terms of $\tilde{g}$ as in (\ref{h g d U relation}).  Then welfare weights $g$ are structurally utilitarian if and only if $\forall i \in I, \forall u_i \in \mathbb{R},\forall z_i \in Z,  \pdv{z_i}h_i\l(u_i,z_i\r) =0.$
\end{prop}    

\subsection{Proof of Theorem \ref{necessity theorem} without quasilinearity}

I now present the proof of Theorem \ref{necessity theorem} without assuming quasilinearity using the more general definition of structural utilitarianism (Definition \ref{nonquasilinear welfare weights definition}). First assume welfare weights are generalized utilitarian.  This means that welfare weights are of the form $g_i\l(c_i,z_i\r)= F'_i\l(U_i\l(c_i,z_i\r)\r) \pdv{c_i} U\l(c_i,z_i\r)$.  Assume that $U_i\l(c_i,z_i\r)=U_i\l(c'_i,z'_i\r)$.  Then  $F'_i\l(U_i\l(c_i,z_i\r)\r)=F'_i\l(U_i\l(c'_i,z'_i\r)\r)$.  So 
\begin{align*}  
\frac{\pdv{c_i}U_i\l(c_i,z_i\r)}{\pdv{c_i}U_i\l(c'_i,z'_i\r)}=\frac{F'_i\l(U_i\l(c_i,z_i\r)\r)\pdv{c_i}U_i\l(c_i,z_i\r)}{F'_i\l(U_i\l(c'_i,z'_i\r)\r)\pdv{c_i}U_i\l(c'_i,z'_i\r)}=\frac{g_i\l(c_i,z_i\r)}{g_i\l(c'_i,z'_i\r)}. 
\end{align*}  
So, welfare weights are structurally utilitarian.  

Going in the other direction, assume that welfare weights $g$ are structurally utilitarian.  Let $h_i$ be defined from $g_i$ via (\ref{g g tilde relation}) and (\ref{h g d U relation}).  It follows from Proposition \ref{h observation} that $h_i\l(u_i,z_i\r)$ does not depend on $z_i$, and hence we can write this function as 
$h_i\l(u_i\r)$, without the argument $z_i$.  Now define $F_i\l(u_i\r)=\int_0^{u_i}h_i\l(u_i\r) \dd u_i$.  Appealing to (\ref{definition of tilde c}), (\ref{g g tilde relation}) and (\ref{h g d U relation}), note that because $g_i\l(c_i,z_i\r)=g\l(c_i,z_i;x_i,y_i\r)$ and $U_i\l(c_i,z_i\r)=U\l(c_i,z_i;x_i,y_i\r)$, we can write $F_i\l(u_i\r)=F\l(u_i;x_i,y_i\r)$ and $F$ inherits the appropriate smoothness properties from $g$ and $U$.  If follows from the above construction that $F'_i\l(U_i\l(c_i,z_i\r)\r) \pdv{c_i} U_i\l(c_i,z_i\r)= h_i\l(U_i\l(c_i,z_i\r)\r)\pdv{c_i} U_i\l(c_i,z_i\r) = \frac{g_i\l(c_i,z_i\r)}{\pdv{c_i}U_i\l(c_i,z_i\r)}\pdv{c_i} U_i\l(c_i,z_i\r)= g_i\l(c_i,z_i\r)$.  So welfare weights are generalized utilitarian. $\square$

\subsection{An example}

In this section, I present as informal example with individualized taxes, similar to the examples in Section \ref{preview section}, that illustrates that, in the non-quasilinear case, if structural utilitarianism in the sense of Definition \ref{nonquasilinear welfare weights definition} is violated, then it is possible to construct a social preference cycle.  In this example, I will not be concerned with holding revenue constant because that can be achieved with a modification of the example by means similar to that presented in Section \ref{individualized version section}.  The purpose of this section is to provide the reader with intuition and an understanding of the essence of the argument that a failure of structural utilitarianism leads to a social preference cycle in the non-quasilinear case.       

Suppose that there is just a single observable binary characteristic $x_i$ such that $x_i=A$ if $i \in \l[0,\frac{1}{2}\r]$ and $x_i=B$ if $i \in \l(\frac{1}{2},1\r]$, and taxes are conditioned on this characteristic.  There are no unobservable characteristics.  All agents of type $A$ are identical with one another and all agents of type $B$ are identical with one another as well.  I write $U_A\l(c,z\r)=U\l(c,z,A\r)$ and $U_B\l(c,z\r)=U\l(c,z,B\r)$ for the utility functions of agents with characteristics $A$ and $B$ respectively.  Likewise, I write $g_A\l(c,z\r)=g\l(c,z,A\r)$ and $g_B\l(c,z\r)=g\l(c,z,B\r)$ for the welfare weights of types $A$ and $B$ respectively.

Suppose that welfare weights for type $A$ agents are not structurally utilitarian.  It follows from Definition \ref{nonquasilinear welfare weights definition} that there exist allocations $\l(c_0,z_0\r)$, $\l(c_1,z_1\r)$ such that \begin{align}\label{u star definition} U_A\l(c_0,z_0\r)=U_A\l(c_1,z_1\r)=u^*\end{align} but $\frac{g_A\l(c_0,z_0\r)}{\pdv{c}U_A\l(c_0,z_0\r)} \neq \frac{g_A\l(c_1,z_1\r)}{\pdv{c}U_A\l(c_1,z_1\r)}$.  
Assume without loss of generality that $\frac{g_A\l(c_0,z_0\r)}{\pdv{c}U_A\l(c_0,z_0\r)} < \frac{g_A\l(c_1,z_1\r)}{\pdv{c}U_A\l(c_1,z_1\r)}$.  Then there exists a number $b$ such that 
 \begin{align}\label{b inequalities}
 \frac{g_A\l(c_0,z_0\r)}{\pdv{c}U_A\l(c_0,z_0\r)} < b< \frac{g_A\l(c_1,z_1\r)}{\pdv{c}U_A\l(c_1,z_1\r)}. 
 \end{align}
Because utility functions are strictly concave, and hence upper contour sets are strictly convex, for any consumption-income bundle $\l(c^*,z^*\r)$, it is possible to construct a linear tax policy (linear in $z$) $\bar{T}^{c^*,z^*}\l(z\r) = \tau\l(c^*,z^*\r)z + \kappa\l(c^*,z^*\r)$ such that type $A$'s optimal consumption and income in response to $\bar{T}^{c^*,z^*}$ is $\l(c^*,z^*\r)$.  As above, let $\tilde{c}_A\l(u,z\r)$ be the level of consumption that gives agents of type $A$ a utility of $u$ when their income is $z$.  Let $\l(\hat{c},\hat{z}\r)$ be type $B$'s optimal consumption and income in the absence of taxes.  (Of course $\hat{c}=\hat{z}$).

Now consider a family of tax policies $\l(T^{\zeta,u}\r)$ parameterized by real numbers $\zeta$ and $u$, where $\zeta \geq 0$, defined by
\begin{align*}
T^{\zeta,u}\l(z,x\r) = \begin{cases} \bar{T}^{\tilde{c},\zeta}\l(z\r) \textup{ with }\tilde{c}=\tilde{c}_A\l(u,\zeta\r), & \textup{if } x =A,\\
\frac{b}{g_B\l(\hat{c},\hat{z}\r)}\l(u-u^*\r), & \textup{if } x=B.
\end{cases}
\end{align*} 
I now explain this tax policy.  First consider type $A$ agents.  At $T^{\zeta,u}$, type $A$ agents face linear tax policy of the form $\bar{T}^{c^*,z^*}$ where $c^*= \tilde{c}_A\l(u,\zeta\r)$ and $z^*=\zeta$.  As explained above, this leads type $A$ agents to select consumption $\tilde{c}_A\l(u,\zeta\r)$ and income $\zeta$, and hence to attain utility $u$.  Type $B$ agents face only a lumpsum tax $\frac{b}{g_B\l(\hat{c},\hat{z}\r)}\l(u-u^*\r)$, where $u^*$ is defined by (\ref{u star definition}) and $b$ satisfies (\ref{b inequalities}).  

By construction, holding fixed $u$ and varying $\zeta$ in $T^{\zeta,u}$, type $A$ agents' utilities remain constant at $u$ when facing $T^{\zeta,u}$.  The taxes faced by type $B$ agents do not depend on $\zeta$.  Hence, all agents are indifferent when facing $T^{\zeta,u}$ as $\zeta$ varies while $u$ is held fixed, and so by Pareto indifference along paths (Proposition \ref{indifference Pareto coro}), which, as explained above, continues to hold in the non-quasilinear case, we have:
\begin{align}\label{nonq indifference}
T^{\zeta,u} \sim^g T^{\zeta',u}, \;\;\;\; \forall \zeta, \zeta', \forall u.
\end{align} 
Let $T_A\l(z,\zeta,u\r)$ be the taxes paid by type $A$ agents under $T^{\zeta,u}$ when earning income $z$.  ($T_B\l(z,\zeta,u\r)$ is defined similarly for type $B$ agents.)  Let $U_A\l(\zeta,u\r)$ be type $A$ agents' utility when facing tax policy $T^{\zeta,u}$ and let $c_A\l(T^{\zeta,u}\r)$ and $z_A\l(T^{\zeta,u}\r)$ be respectively the optimal consumption and income for type $A$ agents when facing tax policy $T^{\zeta,u}$.  It follows form the envelope theorem that
\begin{align}\label{envelope type A apx}
\pdv{u}U_A\l(\zeta,u\r) = -\pdv{c}U_A\l(c_A\l(T^{\zeta,u}\r),z_A\l(T^{\zeta,u}\r)\r) \l.\pdv{u'}\r|_{u'=u}T_A\l(z\l(T^{\zeta,u}\r),\zeta,u'\r).
\end{align}
On the other hand because, for all $u$ and $\zeta$, $U_A\l(\zeta,u\r)=u$, it follows that
\begin{align}\label{1 apx 1}
\pdv{u}U_A\l(\zeta,u\r) =1, \;\;\; \forall \zeta,\forall u.
\end{align} 
Putting (\ref{envelope type A apx}) and (\ref{1 apx 1}) together, we have 
\begin{align*}
\l.\pdv{u}\r|_{u=u'}T_A\l(z_A\l(T^{\zeta,u}\r),\zeta,u'\r)= -\frac{1}{\pdv{c}U_A\l(c_A\l(T^{\zeta,u}\r),z_A\l(T^{\zeta,u}\r)\r)}.  
\end{align*}
By construction we have:
\begin{align*}
\l.\pdv{u'}\r|_{u'=u}T_B\l(z_B\l(T^{\zeta,u}\r),\zeta,u'\r) = \frac{b}{g_B\l(\hat{c},\hat{z}\r)}
\end{align*}
Note that when $\zeta = z_0$ and $u=u^*$, $z_A\l(T^{\zeta,u}\r) =z_0$ and $c_A\l(T^{\zeta,u}\r)= \tilde{c}_A\l(u^*,z_0\r)=c_0$.  Also, when $u=u^*$, type $B$ agents face no taxes under $T^{\zeta,u}$, and hence $c_B\l(T^{\zeta,u}\r)=\hat{c}$ and $z_B\l(T^{\zeta,u}\r)=\hat{z}$.  
It follows that
\begin{align*}
&\int g_i\l(T^{z_0,u^*}\r) \l.\pdv{u}\r|_{u=u^*} T_i\l(z_i\l(T^{z_0,u^*}\r),z_0,u\r) \dd i \\=&  \underbrace{\int_0^{\frac{1}{2}} g_A\l(c_0,z_0\r)\l( -\frac{1}{\pdv{c}U_A\l(c_0,z_0\r)}\r)\dd i}_{\textup{type $A$ agents}}+ \underbrace{\int_{\frac{1}{2}} g\l(\hat{c},\hat{z}\r) \frac{b}{g_B\l(\hat{c},\hat{z}\r)} \dd i}_{\textup{type $B$ agents}}\\
=& \frac{1}{2} \l(-\frac{g_A\l(c_0,z_0\r)}{\pdv{c}U_A\l(c_0,z_0\r)}+b \r) > 0,
\end{align*}
where the inequality follows from (\ref{b inequalities}).  Similarly, 
\begin{align*}
\int g_i\l(T^{z_1,u^*}\r) \l.\pdv{u}\r|_{u=u^*} T_i\l(z_i\l(T^{z_1,u^*}\r),z_1,u\r) \dd i  = \frac{1}{2}  \l(-\frac{g_A\l(c_1,z_1\r)}{\pdv{c}U_A\l(c_1,z_1\r)}+b \r) < 0.
\end{align*}
It follows from the local improvement principle (Proposition \ref{local improvement principle proposition}), which also continues to hold in the non-quasilinear case, that for sufficiently small $\epsilon >0$,   
\begin{align}\label{comparisons nonq}
\begin{split}
T^{z_0,u^*} &\succ_g T^{z_0,u^*+\epsilon}, \textup{ and} \\
T^{z_1,u^*} &\prec_g T^{z_1,u^*+\epsilon}. 
\end{split}
\end{align}
Putting together (\ref{nonq indifference}) and (\ref{comparisons nonq}), we have the social preference cycle:
\begin{align*}
T^{z_1,u^*} \prec_g T^{z_1,u^*+\epsilon}\sim_g T^{z_0,u^*+\epsilon}\prec_g  T^{z_0,u^*} \sim^g T^{z_1,u^*}.
\end{align*}
So, on the assumption that welfare weights are not structurally utilitarian, we have derived a cycle.

\subsection{\label{main thm apx without quasilinearity}Proof of Theorem \ref{main theorem} without quasilinearity}
In this section, I explain how to modify the proof of Theorem \ref{main theorem} when quasilinearity is no longer assumed.  (The statement of the theorem must be modified to appeal to the assumptions of Sections \ref{preliminaries h section} and \ref{additional structure h section} rather than Section \ref{additional structure section}.)    
\subsubsection{\label{additional structure h section}Additional structure for the non-quasilinear version of Theorem \ref{main theorem}}
I now assume, as in Section \ref{additional structure section}, that there are no observable characteristics, but there is a single one-dimensional real valued unobservable characteristic $y$, so that we can write $U_i\l(c_i,z_i\r)=U\l(c_i,z_i;y_i\r)$, and that the function $i \mapsto y_i$ is smooth and that the derivative of $y_i$ with respect to $i$ is positive at all values of $i$ in $I=\l[0,1\r]$.  Moreover, I assume the single-crossing condition that, for all $\l(c,z,y\r) \in \mathbb{R} \times Z \times Y$, $\dv{y} \frac{\pdv{z}U\l(c,z,y\r)}{\pdv{c} U\l(c,z,y\r)}>0$.  This single crossing condition implies that for every regular tax policy $T$, $i \mapsto z_i\l(T\r)$ is strictly increasing in $i$.  Note that in Section \ref{additional structure section}, we assumed that$\pdv[2]{}{y}{z}v\l(z,y\r) < 0$, so that when $U\l(c,z;y\r) = u\l(c-v\l(z,y\r)\r)$, $\dv{y}\frac{\pdv{z}U\l(c,z,y\r)}{\pdv{c} U\l(c,z,y\r)}=-\pdv[2]{}{y}{z}v\l(z,y\r)>0$.  So the above single-crossing condition generalizes the assumption we made in the quasilinear case.  

\subsubsection{Modifications of the main lemmas}
Theorem \ref{main theorem} is proven by means of a series of lemmas, and in this section I will discuss how these lemmas must be altered when we drop the assumption of quasilinearity and revert to the weaker assumptions of Sections \ref{preliminaries h section} and \ref{additional structure h section} above.  

\paragraph{Lemma \ref{cycle structure lemma}}
Lemma \ref{cycle structure lemma} is unaltered relative to the quasilinear case and the proof is identical.  

\paragraph{Corollary \ref{convex tax policy corollary}}

The following result is the non-quasilinear analog of Corollary \ref{convex tax policy corollary}.

\begin{cor}\label{convex tax policy corollary h} Let $h$ be related to $g$ as specified by (\ref{g g tilde relation}) and (\ref{h g d U relation}).  If $g$ is not structurally utilitarian, then there exists a regular tax policy $T $ for which there exist agents $i_a, i_b \in \l(0,1\r)$ with $i_a < i_b$ such that either
\begin{align}\label{negative derivative integral h}
\forall i \in \l(i_a,i_b\r),\;\;\;\pdv{z_i}h_i\l(U_i\l(T\r),z_i\l(T\r)\r)<0
\end{align}
or 
\begin{align}\label{positive derivative integral h}
\forall i \in \l(i_a,i_b\r),\;\;\; \pdv{z_i}h_i\l(U_i\l(T\r),z_i\l(T\r)\r)>0.
\end{align}
\end{cor}
In the proof of Corollary \ref{convex tax policy corollary h}, Proposition \ref{h observation} plays the role  that Proposition \ref{g observation} plays in the proof of Corollary \ref{convex tax policy corollary}; moreover the proof of Corollary \ref{convex tax policy corollary h} is a bit more involved than that of Corollary \ref{convex tax policy corollary} because one cannot rely on the convenient properties of quasilinear preferences.   
\paragraph{Lemma \ref{alternative condition lemma}}
We also have the following lemma, which is an analog of Lemma \ref{alternative condition lemma}.  
\begin{lem}\label{alternative condition lemma h}
\label{simplification contradiction lemma}Assume that $\l(T^{\theta,\epsilon}\r)$ is well-behaved and satisfies (\ref{2}).  Then (\ref{4}) holds if and only if 
\begin{align}\label{alternative essential condition h}
\begin{split}
\int m_i\l(\theta_0,\epsilon_0\r) &\l[\l.\pdv[2]{}{\theta}{z_i}\r|_{\theta =\theta_0,z_i=z_i\l(\theta_0,\epsilon_0\r)} T\l(z_i,\theta,\epsilon_0\r) \l.\pdv{\epsilon}\r|_{\epsilon =\epsilon_0} T\l(z_i\l(\theta_0,\epsilon_0\r),\theta_0,\epsilon\r)\r.\\& \l. 
- \l.\pdv[2]{}{\epsilon}{z_i}\r|_{\epsilon =\epsilon_0,z_i=z_i\l(\theta_0,\epsilon_0\r)} T\l(z_i,\theta_0,\epsilon\r)  \l.\pdv{\theta}\r|_{\theta =\theta_0} T\l(z_i\l(\theta_0,\epsilon_0\r),\theta,\epsilon_0\r)\r] \dd i < 0.
\end{split}
\end{align}
where 
\begin{align*}
m_i\l(\theta_0,\epsilon_0\r)&= \frac{\l[\pdv{c_i} U_i\l(c_i\l(\theta_0,\epsilon_0\r),z_i\l(\theta_0,\epsilon_0\r)\r)\r]^2\pdv{z_i}
h_i\l(U_i\l(\theta_0,\epsilon_0\r),z_i\l(\theta_0,\epsilon_0\r)\r)}{\l.\dv[2]{z_i}\r|_{z_i=z_i\l(\theta_0,\epsilon_0\r)}U_i\l(z_i-T\l(z_i\r),z_i\r)}.
\end{align*}
\end{lem}
The structure of the proof is similar to the structure of the proof of Lemma \ref{alternative condition lemma}, and features terms $A$, $B$, and $C$, which play the same role as the terms $A$, $B$, and $C$ in Lemma \ref{alternative condition lemma}. However precise details of these terms differ in the two lemmas.  In Lemma \ref{alternative condition lemma h}, 
\begin{align}\label{A B C revised definitions}
\begin{split}
A=& \int m_i\l(\theta_0,\epsilon_0\r)\l.\pdv[2]{}{\epsilon}{z_i}\r|_{\epsilon =\epsilon_0,z_i=z_i\l(\theta_0,\epsilon_0\r)} T\l(z_i,\theta_0,\epsilon\r)  \l.\pdv{\theta}\r|_{\theta =\theta_0} T\l(z_i\l(\theta_0,\epsilon_0\r),\theta,\epsilon_0\r) \dd i\\
B=& \int m_i\l(\theta_0,\epsilon_0\r)  \l.\pdv[2]{}{\theta}{z_i}\r|_{\theta =\theta_0,z_i=z_i\l(\theta_0,\epsilon_0\r)} T\l(z_i,\theta,\epsilon_0\r) \l.\pdv{\epsilon}\r|_{\epsilon =\epsilon_0} T\l(z_i\l(\theta_0,\epsilon_0\r),\theta_0,\epsilon\r) \dd i\\
C=&\int g_i \l(\pdv[2]{T_i}{\theta}{\epsilon} + \frac{\l[\pdv{T_i}{\theta}{z_i}\pdv{T_i}{\epsilon}+\pdv{T_i}{\epsilon}{z_i}\pdv{T_i}{\theta}\r]\l[\pdv[2]{U_i}{c_i}\l(1-\pdv{T_i}{z_i}\r)+\pdv[2]{U_i}{c_i}{z_i}\r]+\pdv{U_i}{c_i}\pdv{T_i}{\theta}{z_i}\pdv{T_i}{\epsilon}{z_i}}{\dv[2]{U_i}{z_i}}\r) \dd i\\
&- \int \pdv{\tilde{g}_i}{u_i}\pdv{U_i}{c_i}\pdv{T_i}{\theta}\pdv{T_i}{\epsilon} \dd i +\int \pdv{\tilde{g}_i}{z_i}\frac{\l[\pdv[2]{U_i}{c_i}\l(1-\pdv{T_i}{z_i}\r)+\pdv[2]{U_i}{c_i}{z_i}\r]\pdv{T_i}{\theta} \pdv{T_i}{\epsilon}}{\dv[2]{U_i}{z_i}}\dd i
\end{split}
\end{align}
The following table explicitly defines the shorthand terms in the in the expression for $C$.
\begin{align}\label{abbreviations table}
\begin{tabular}{ll}
$g_i = g_i\l(\theta_0,\epsilon_0\r)$&$\pdv{\tilde{g}_i}{u_i}= \pdv{u_i}\tilde{g}_i\l(U_i\l(\theta_0,\epsilon_0\r),z_i\l(\theta_0,\epsilon_0\r)\r)$\\
$\pdv{\tilde{g}_i}{z_i}=\pdv{z_i}\tilde{g}_i\l(U_i\l(\theta_0,\epsilon_0\r),z_i\l(\theta_0,\epsilon_0\r)\r)$ & $\pdv{U_i}{c_i}=\pdv{c_i}U_i\l(c_i\l(\theta_0,\epsilon_0\r),z_i\l(\theta_0,\epsilon_0\r)\r)$\\
$\pdv[2]{U_i}{c_i}=\pdv[2]{c_i}U_i\l(c_i\l(\theta_0,\epsilon_0\r),z_i\l(\theta_0,\epsilon_0\r)\r)$ & $\pdv[2]{U_i}{c_i}{z_i}= \pdv[2]{}{c_i}{z_i}U_i\l(c_i\l(\theta_0,\epsilon_0\r),z_i\l(\theta_0,\epsilon_0\r)\r)$\\
 $\dv[2]{U_i}{z_i}=\l.\dv[2]{z_i}\r|_{z_i=z_i\l(\theta_0,\epsilon_0\r)}U_i\l(z_i-T\l(z_i\r),z_i\r)$ & $\pdv{T_i}{z_i}=\pdv{z_i}T\l(z_i\l(\theta_0,\epsilon_0\r),\theta_0,\epsilon_0\r)$\\
 $\pdv{T_i}{\theta}=\l.\pdv{\theta}\r|_{\theta =\theta_0} T\l(z_i\l(\theta_0,\epsilon_0\r),\theta,\epsilon_0\r)$& $\pdv{T_i}{\epsilon}= \l.\pdv{\epsilon}\r|_{\epsilon =\epsilon_0} T\l(z_i\l(\theta_0,\epsilon_0\r),\theta_0,\epsilon\r)$ \\
 $\pdv{T_i}{\theta}{z_i}=\l.\pdv{}{\theta}{z_i}\r|_{\theta=\theta_0,z_i=z_i\l(\theta_0,\epsilon_0\r)}T\l(z_i,\theta,\epsilon_0\r)$ & $\pdv{T_i}{\epsilon}{z_i}=\l. \pdv{}{\epsilon}{z_i}\r|_{\epsilon=\epsilon_0,z_i=z_i\l(\theta_0,\epsilon_0\r)}T\l(z_i,\theta_0,\epsilon\r)$\\
 $\pdv[2]{T_i}{\theta}{\epsilon}=\l.\pdv[2]{}{\theta}{\epsilon}\r|_{\theta=\theta_0,\epsilon=\epsilon_0} T\l(z_i\l(\theta_0,\epsilon_0\r),\theta,\epsilon\r)$ &
\end{tabular}
\end{align}
The proof of Lemma \ref{alternative condition lemma h} relies on several facts.  First observe that, by (\ref{k d U relation}), 
\begin{align*} \pdv{z_i} k_i\l(u_i,z_i\r) = \pdv[2]{c_i} U_i\l(\tilde{c}_i\l(u_i,z_i\r),z_i\r)\pdv{z_i}\tilde{c_i}\l(u_i,z_i\r) + \pdv[2]{}{z_i}{c_i}U_i\l(\tilde{c}_i\l(u_i,z_i\r),z_i\r).
 \end{align*}
It follows from agent $i$'s first order condition that $1-T'\l(z_i\l(\theta_0,\epsilon_0\r)\r) =-\frac{\pdv{z_i}U_i\l(c_i\l(\theta_0,\epsilon_0\r),z_i\l(\theta_0,\epsilon_0\r)\r)}{\pdv{c_i}U_i\l(c_i\l(\theta_0,\epsilon_0\r),z_i\l(\theta_0,\epsilon_0\r)\r)}.$ Using (\ref{c i derivative MRS}) and the fact that $\tilde{c}_i\l(U_i\l(\theta_0,\epsilon_0\r),z_i\l(\theta_0,\epsilon_0\r)\r)=c_i\l(\theta_0,\epsilon_0\r)$ and the abbreviations in (\ref{abbreviations table}), we have
\begin{align}\label{pdv z k}
 \pdv{z_i}k_i\l(U_i\l(\theta_0,\epsilon_0\r),z_i\l(\theta_0,\epsilon_0\r)\r)= \pdv[2]{U_i}{c_i}\l(1-\pdv{T_i}{z_i}\r)+\pdv[2]{U_i}{c_i}{z_i}.\end{align} 
Moreover, applying the implicit function theorem to the agent's first order conditions, using the abbreviations in (\ref{abbreviations table}), we have 
\begin{align}\label{implicit function theorem derivatives of z}
\begin{split}
\pdv{}{\theta}z\l(\theta_0,\epsilon_0\r) =&\frac{\l[\pdv[2]{U_i}{c_i}\l(1-\pdv{T_i}{z_i}\r)+\pdv[2]{U_i}{c_i}{z_i}\r]\pdv{T_i}{\theta}+\pdv{U_i}{c_i}\pdv[2]{T_i}{\theta}{z_i}}{\dv[2]{U_i}{z_i}},\\
\pdv{z_i}{\epsilon}\l(\theta_0,\epsilon_0\r) =&\frac{\l[\pdv[2]{U_i}{c_i}\l(1-\pdv{T_i}{z_i}\r)+\pdv[2]{U_i}{c_i}{z_i}\r]\pdv{T_i}{\epsilon}+\pdv{U_i}{c_i}\pdv[2]{T_i}{\epsilon}{z_i}}{\dv[2]{U_i}{z_i}}.
\end{split} 
\end{align}
Using the envelope theorem, (\ref{g = hk}), (\ref{pdv z k}), and (\ref{implicit function theorem derivatives of z}), it is staightforward to show that when $A, B,$ and $C$ are defined as in (\ref{A B C revised definitions}), then (\ref{A+C}) and (\ref{B+C}) hold, and then the argument for Lemma \ref{alternative condition lemma h} proceeds similarly to the argument for Lemma \ref{alternative condition lemma}. 

\paragraph{Lemma \ref{lemma without assumptions on welfare weights}}
Lemma \ref{lemma without assumptions on welfare weights} needs to be modified as follows for the non-quasilinear case:

\begin{lem}\label{lemma without assumptions on welfare weights h}
Let $T$ be a regular tax policy and let $i_a, i_b \in \l(0,1\r)$ be such that $i_a < i_b$.  Then there exists a well-behaved family $\l(T^{\theta,\epsilon}\r)$ with $T^{\theta_0,\epsilon_0}=T$ for some interior parameter values $\theta_0,\epsilon_0$ and that satisfies (\ref{2}),  (\ref{3}), and 
\begin{align}\label{wanted key inequality h}
\begin{split}
&\l.\pdv[2]{}{\theta}{z_i}\r|_{\theta =\theta_0,z_i=z_i\l(\theta_0,\epsilon_0\r)} T\l(z_i,\theta,\epsilon_0\r) \l.\pdv{\epsilon}\r|_{\epsilon =\epsilon_0} T\l(z_i\l(\theta_0,\epsilon_0\r),\theta_0,\epsilon\r)\\& 
- \l.\pdv[2]{}{\epsilon}{z_i}\r|_{\epsilon =\epsilon_0,z_i=z_i\l(\theta_0,\epsilon_0\r)} T\l(z_i,\theta_0,\epsilon\r)  \l.\pdv{\theta}\r|_{\theta =\theta_0} T\l(z_i\l(\theta_0,\epsilon_0\r),\theta,\epsilon_0\r)\end{split}\;\;\begin{cases} > 0, &\textup{ if } i \in \l(i_a,i_b\r),\\
= 0,&\textup{ if } i \not\in \l(i_a,i_b\r).
\end{cases}
\end{align}
\end{lem}
The reason that the inequality points in opposite directions in Lemmas \ref{lemma without assumptions on welfare weights} and \ref{lemma without assumptions on welfare weights h} is that, in the lemma preceding Lemma \ref{lemma without assumptions on welfare weights h}, namely,  Lemma \ref{alternative condition lemma h}, the term $\frac{1}{\l.\dv[2]{z_i}\r|_{z_i=z_i\l(\theta_0,\epsilon_0\r)}U_i\l(z_i-T\l(z_i\r),z_i\r)}$, which is negative, has been absorbed into $m_i\l(\theta_0,\epsilon_0\r)$, whereas, in Lemma \ref{alternative condition lemma}, the corresponding term was part of $\l.\pdv{\theta}\r|_{\theta =\theta_0} z_i\l(\theta,\epsilon_0\r)$ and $ \l.\pdv{\epsilon}\r|_{\epsilon =\epsilon_0} z_i\l(\theta_0,\epsilon\r)$.\footnote{In particular, in the quasilinear case, using the fact that, by construction, $T^{\theta_0,\epsilon_0}=T$, we have $\l.\pdv{\theta}\r|_{\theta =\theta_0} z_i\l(\theta,\epsilon_0\r)=-\frac{\l.\dv{\theta}\r|_{\theta =\theta_0}\l.\dv{z_i}\r|_{z_i=z_i\l(\theta_0,\epsilon_0\r)}U_i\l(z_i-T\l(z_i,\theta,\epsilon_0\r), z_i\r)}{\l.\dv[2]{z_i}\r|_{z_i=z_i\l(\theta_0,\epsilon_0\r)}U_i\l(z_i-T\l(z_i\r),z_i\r)}$ and $\l.\pdv{\theta}\r|_{\epsilon =\epsilon_0} z_i\l(\theta,\epsilon_0\r)$ is similar.  Note that, in the non-quasilinear case, the term $\l.\pdv[2]{}{\theta}{z_i}\r|_{\theta =\theta_0,z_i=z_i\l(\theta_0,\epsilon_0\r)} T\l(z_i,\theta,\epsilon_0\r)$ differs from $\l.\pdv{\theta}\r|_{\theta =\theta_0} z_i\l(\theta,\epsilon_0\r)$ in a number of ways, and not just in omitting the denominator $\l.\dv[2]{z_i}\r|_{z_i=z_i\l(\theta_0,\epsilon_0\r)}U_i\l(z_i-T\l(z_i\r),z_i\r)$.}  
In any event, just as in the quasilinear case, it was possible, with a slight modification in the construction to flip the inequality in (\ref{wanted key inequality}) (see Lemma \ref{lemma without assumptions on welfare weights variant}), it is also possible to do the same for (\ref{wanted key inequality h}).

The proof of Lemma \ref{lemma without assumptions on welfare weights h} is similar to the poof of Lemma \ref{lemma without assumptions on welfare weights}.  The construction of the family $\l(T^{\theta,\epsilon}\r)$ is the same as in Lemma \ref{lemma without assumptions on welfare weights}; the fact that $i \mapsto z_i\l(T\r)$ is increasing, which is used in the construction, now follows from the single-crossing condition.  Many other aspects of the argument are unchanged.  As (\ref{alternative essential condition h}) and (\ref{wanted key inequality h}), unlike (\ref{alternative essential condition}) and (\ref{wanted key inequality}), do not feature the terms $\l.\pdv{\theta}\r|_{\theta =\theta_0} z_i\l(\theta,\epsilon_0\r)$ and $ \l.\pdv{\epsilon}\r|_{\epsilon =\epsilon_0} z_i\l(\theta_0,\epsilon\r)$, we no longer have to appeal to the conditions (\ref{income derivatives}) and (\ref{income derivatives simplified}).  In place of (\ref{lemmas desired inequality}), we now derive the condition,  
\begin{align}\label{lemmas desired inequality h}\begin{split}
& \forall  i \in \l( i_a, i_b\r), \\
&\l.\pdv[2]{}{\theta}{z_i}\r|_{\theta =\theta_0,z_i=z_i\l(\theta_0,\epsilon_0\r)} T\l(z_i,\theta,\epsilon_0\r) \l.\pdv{\epsilon}\r|_{\epsilon =\epsilon_0} T\l(z_i\l(\theta_0,\epsilon_0\r),\theta_0,\epsilon\r)\\& 
- \l.\pdv[2]{}{\epsilon}{z_i}\r|_{\epsilon =\epsilon_0,z_i=z_i\l(\theta_0,\epsilon_0\r)} T\l(z_i,\theta_0,\epsilon\r)  \l.\pdv{\theta}\r|_{\theta =\theta_0} T\l(z_i\l(\theta_0,\epsilon_0\r),\theta,\epsilon_0\r)\\
=
&\overbrace{\mu'_1\l(z_ i\l(T\r)\r)}^+\overbrace{\eta_1\l(z_ i\l(T\r)\r)}^+
 -\l(\overbrace{\eta'_1\l(z_ i\l(T\r)\r)}^{+ \textup{ on }\l( i_3, i_4\r), -\textup{ on } \l( i_4, i_5\r)} \times \overbrace{\l[\mu_1\l(z_ i\l(T\r)\r)-1 \r]}^{- \textup{ on } \l( i_3, i_4\r), + \textup{ on }\l( i_4, i_5\r)}\r) >0.
\end{split}
\end{align}
The equality (\ref{lemmas desired inequality h}) appeals to similar facts as (\ref{lemmas desired inequality}) to derive and sign the relevant terms on the right hand side of the equality.  The argument that the expression on the right hand side of (\ref{wanted key inequality h}) is equal to zero outside of $\l(i_a,i_b\r)$ is similar to the corresponding argument in Lemma \ref{lemma without assumptions on welfare weights}.  This completes the summary of how the proof of Lemma \ref{lemma without assumptions on welfare weights h} differs from that of Lemma \ref{lemma without assumptions on welfare weights}.

\paragraph{Lemma \ref{key lemma minus revenue}}
Lemma \ref{key lemma minus revenue} continues to hold in the non-quasilinear case, and its proof in the non-quasilinear case is very similar to its proof in the quasilinear case.  In particular, note that, for each $i$, the terms $m_i\l(\theta_0,\epsilon_0\r)$ and $\pdv{z_i}
h_i\l(U_i\l(\theta_0,\epsilon_0\r),z_i\l(\theta_0,\epsilon_0\r)\r)$ always have opposite signs when nonzero, and one term is equal to zero if and only if the other is equal to zero as well.   

\paragraph{Lemma \ref{key lemma with revenue}}

The basic structure of the argument for Lemma \ref{key lemma with revenue}, as explained in Section \ref{key lemma with revenue section} of the appendix, is unchanged.  However, some of the lemmas supporting Lemma \ref{key lemma with revenue} must be modified.  In the proof of Lemma \ref{indifference and nonconstant revenue family of reforms theorem}, the specific expressions in (\ref{relation of two optimal incomes}) must be modified because they depend on the assumption of quasilinearity, but the equality $\l.\dv{\varepsilon}\r|_{\varepsilon=0} z_i\l(T+\varepsilon \l(\Delta T_1 - s'\l(0\r) \Delta T_2\r)\r)= \l.\dv{\xi}\r|_{\xi=0} z_i\l(T+\Delta T^\xi\r)$ continues to hold, so the proof can proceed as before.  Similarly, in in the proof of Lemma \ref{linearity of marginal revenue lemma} the specific terms in (\ref{linear effects of tax reforms}) depend on quasilinearity but $\l.\dv{\varepsilon}\r|_{\varepsilon =0}z_i\l(T+ \varepsilon \l(r_1\Delta T^\gamma+r_2 \Delta T_2\r)\r)=  r_1\l.\dv{\varepsilon}\r|_{\varepsilon=0} z_i\l(T+\varepsilon\Delta T_1\r) + r_2 \l.\dv{\varepsilon}\r|_{\varepsilon=0} z_i\l(T+\varepsilon \Delta T_2\r)$ still holds, and so again the proof can proceed as before.  In Lemma \ref{revenue unchanging welfare changing lemma}, (\ref{change of variables derivative of revenue}) becomes
\begin{align*}
\l.\dv{\varepsilon}\r|_{\varepsilon=0} R\l(T+ \varepsilon\Delta T\r) = \int_{z_0}^{z_1}  \Delta T\l(z\r) \ell_T\l(z\r) \dd z - \int_{z_0}^{z_1} \Delta T'\l(z\r)  k_T\l(z\r) \dd z,
\end{align*}
where
\begin{align*}
\ell_T\l(z\r) = \l[1- \frac{\l[\pdv[2]{c}U_{\iota\l(z\r)}\l(z-T\l(z\r),z\r)\l(1-T'\l(z\r)\r)+\pdv[2]{}{c}{z}U_{\iota\l(z\r)}\l(z-T\l(z\r),z\r)\r]T'\l(z\r)}{\l.\dv[2]{}{\tilde{z}}\r|_{\tilde{z}=z}U_{\iota\l(z\r)}\l(\tilde{z}-T\l(\tilde{z}\r),\tilde{z}\r)}
\r]&h\l(z\r),\\ &\forall z \in \l[z_0,z_1\r],
\end{align*}
and $k_T\l(z\r)$ is modified to become:
\begin{align*}
k_T\l(z\r) =\frac{\pdv{c}U_{\iota\l(z\r)}\l(z-T\l(z\r),z\r)T'\l(z\r)}{\l.\dv[2]{}{\tilde{z}}\r|_{\tilde{z}=z}U_{\iota\l(z\r)}\l(\tilde{z}-T\l(\tilde{z}\r),\tilde{z}\r)}h\l(z\r), \;\;\; \forall z \in \l[z_0,z_1\r].
\end{align*}
Accordingly, in the proof of Lemma \ref{gamma limit nonzero revenue lemma}, (\ref{marginal revenue in direction Delta T 1 gamma}) becomes
 \begin{align*}
\l.\dv{\varepsilon}\r|_{\varepsilon=0} R\l(T +\varepsilon \Delta T_1^\gamma\r) = & \int_{z_0}^{z^\gamma_*}  \Delta T_1^\gamma\l(z\r) \ell_T\l(z\r) \dd z - \int_{z_0}^{z^\gamma_*} \dv{z}\Delta T_1^\gamma\l(z\r)  k_T\l(z\r) \dd z,
\end{align*}
and (\ref{limit of first integral is zero}) becomes
\begin{align*}
\lim_{\gamma \rightarrow \infty}\int_{z_0}^{z^\gamma_*}  \Delta T_1^\gamma\l(z\r) \ell_T\l(z\r) \dd z=0.
\end{align*}
Otherwise the proofs of Lemmas \ref{revenue unchanging welfare changing lemma} and \ref{gamma limit nonzero revenue lemma} remain the same.  Some of the precise details of Lemma \ref{interval optimum lemma} need to be changed, but the basic structure of the argument, which relies on the single-crossing property, remains the same.   The proofs of Lemmas \ref{lemma satisfying desired properties}, \ref{extended tax policy support lemma}, \ref{useful but simple lemma} and \ref{small perturbation still in T hat lemma} are unchanged.

\end{document}